\documentclass[12pt,a4paper]{article}

\RequirePackage[OT1]{fontenc}
\RequirePackage{amsthm}
\RequirePackage[cmex10]{amsmath}
\RequirePackage{natbib}
\RequirePackage[colorlinks,citecolor=blue,urlcolor=blue, linkcolor=blue, bookmarksopen=true]{hyperref}

\usepackage[margin=1in]{geometry}
\usepackage[latin1]{inputenc}
\usepackage{amsmath}
\usepackage{amsfonts}
\usepackage{amssymb}
\usepackage{comment}
\usepackage{bbm}
\usepackage{graphicx}
\usepackage[usenames,dvipsnames]{color}
\usepackage{multirow}
\usepackage[labelfont=bf]{caption}
\usepackage{placeins}
\usepackage{array}
\usepackage{enumitem}
\usepackage{rotating}
\usepackage{adjustbox}
\usepackage{mathrsfs}
\usepackage{gensymb}
\usepackage{float}
\usepackage{subcaption}
\usepackage{stmaryrd}
\usepackage{yfonts}
\usepackage{xurl}

\newcolumntype{R}[2]{%
    >{\adjustbox{angle=#1,lap=\width-(#2)}\bgroup}%
    l%
    <{\egroup}%
}

\newcommand*\elide{\textup{[\,\dots]}}

\newcommand{\qmin}{ \mathsf{q}^{\min} }
\newcommand{\qmax}{ \mathsf{q}^{\max} }
\newcommand{\qcap}{ \mathsf{q}^{\text{cap}} }
\newcommand{\pmin}{ \mathsf{p}^{\min} }
\newcommand{\pmax}{ \mathsf{p}^{\max} }
\newcommand{\pgro}{ \mathsf{p}^\gamma }
\newcommand{\piavg}{ \pi^{\text{avg}} }
\newcommand{\pminup}{ \mathsf{p}^{\min\uparrow} }
\newcommand{\pmaxdown}{ \mathsf{p}^{\max\downarrow} }

\DeclareMathOperator*{\argmax}{arg\,max}

\title{\textsc{Endogenous shareholding auctions}\protect}

\author{\textsc{Andrew Mackenzie\thanks{%
Department of Economics, Rutgers University, New Brunswick, NJ, United States of America. Email: andrew.k.mackenzie@rutgers.edu} \hspace{0.01mm} and Christian Trudeau\thanks{%
Department of Economics, University of Windsor 401 Sunset Avenue, Windsor, Ontario, Canada. Email:
trudeauc@uwindsor.ca}} \thanks{Christian Trudeau acknowledges financial support by the Social Sciences and Humanities Research Council of Canada, grant number 435-2019-0141. We thank Ata Atay, Steven Chen, Piotr Dworczak, Justin Leroux, Rich McLean, Will Phan, Constantine Sorokin, and Alex Teytelboym, as well as seminar participants at University of Liverpool, Georgetown University, and multiple conferences.}}
\date{First draft: July 2, 2026 \\ This draft: \today}

\begin{document}

\maketitle

\begin{abstract}
We introduce {\it endogenous shareholding auctions} for production economies where a designer must elicit demand in order to determine price and quantity. Each of these auctions has the property that the auction's profit is distributed across the designer and the bidders in accordance with ownership shares that are determined over the course of the auction. We characterize this class, and a larger class, on the basis of standard axioms. Finally, we investigate optimal auctions according to both prior-free domination and subjective expected welfare. Applications include public housing, voluntary separation, and monopoly regulation.
\end{abstract}

\hypertarget{Section1}{}
\section{Introduction}

\hypertarget{Section1.1}{}
\subsection{Overview}

Auctions are profitable. For example, the 2017 incentive auction of the US Federal Communications Commission involved purchasing broadcast rights at a cost of about \$10.1 billion and selling broadcast rights for a revenue of about \$19.8 billion \citep{Milgrom-Segal2020}. The question of how to share an auction's profit with the bidders is subtle, because (i)~the question would hardly occur to an unconstrained designer who seeks to maximize his own profit, and (ii)~even when the designer cares about bidder welfare, there are many settings for which any profit-sharing with the bidders is incompatible with incentive compatibility and standard normative requirements. In this article, we show that for a setting where a costly production decision will be made after bids are gathered, profit-sharing is compatible with some of the literature's most demanding requirements of incentive compatibility, efficiency, fairness, and voluntary participation, and we argue that this setting is relevant for public housing, voluntary separation, and monopoly regulation (see \hyperlink{Section1.2}{Section~1.2}).

More precisely, we introduce the novel class of {\it endogenous shareholding auctions}, and show that for our setting this is precisely the class of strategy-proof mechanisms whose allocations are (ex-post) efficient, envy-free among the bidders (\citealp{Tinbergen1930}; \citealp{Foley1967}),\footnote{Often attributed to Tinbergen via \cite{Tinbergen1946}; see \cite{Heilmann-Wintein2021}.} and voluntary. These auctions are like VCG mechanisms (\citealp{Vickrey1961}; \citealp{Clarke1971}; \citealp{Groves1973}) in that they select envy-free competitive equilibria in our setting, but unlike VCG mechanisms in that their competitive equilibria may involve subsidies for the bidders that are funded out of the auction's profit. In the context of public housing, these subsidies make housing more affordable for winners while providing financial support to losers; in the context of voluntary separation, they share downsizing benefits by improving exit packages while rewarding employees who stay; in the context of monopoly regulation, they transfer surplus from the monopolist to the consumers as dividends from the regulated auction. The trick is that the bidder shareholding must be determined endogenously by the bids, for the only way to set it exogenously without compromising strategy-proofness is to set it to zero and thereby stick to VCG.

Production is essential for our main message. In our model, production costs are convex and known in advance to the designer, any produced objects are identical, and bidders have unit demand. This includes as a special case the classic problem of assigning $q$ identical objects {\it without} production, because we can set the first $q$ marginal costs to zero and the rest to infinity, but there are only endogenous shareholding auctions distinct from VCG when there are at least three distinct marginal costs. Without production, it is well-known that all Groves mechanisms `burn money' in the sense that they do not distribute all profit to the bidders \citep{Green-Laffont1979}, and it is possible to `burn less money' than VCG via profit-sharing if we require anonymity instead of no-envy (\citealp{Bailey1997}; \citealp{Cavallo2006});\footnote{For the sale of one object to $n$ bidders, the Bailey-Cavallo mechanism is such that (i)~the highest bidder wins and pays the second-highest bid, (ii)~each of the two highest bidders receives as subsidy $\frac{1}{n}$ times the third-highest bid, and (iii)~every other bidder receives as subsidy $\frac{1}{n}$ times the second-highest bid. In this case, the second-highest bidder may envy the third-highest bidder.} the novelty is that when there is production, there are profit-sharing auctions that provide equal opportunities not only ex-ante but also ex-post.

First, we characterize the mechanisms that are efficient, strategy-proof, and envy-free as the {\it endogenous subsidy auctions} (\hyperlink{Theorem1}{Theorem~1} and \hyperlink{Theorem2}{Theorem~2}). Equivalently, these are the envy-free Groves mechanisms \citep{Holmstrom1979}, a rich class of mechanisms with considerable structure that is revealed through analysis of {\it option sets} in the context of our particular model.\footnote{For a given agent and a given profile of preference reports of the others, the {\it option set} is the set of outcomes that the agent can obtain by varying his own report. The analysis of option sets has proven useful for the study of strategy-proofness across a variety of models (\citealp{Sprumont1995}; \citealp{Barbera2011}).} Our result differs from previous results about envy-free Groves mechanisms (\citealp{Papai2003}; \citealp{Ohseto2006}; \citealp{Yengin2012}; \citealp{Yengin2017}) in that it concerns a model that may have production, and in fact our class only includes interesting alternatives to VCG when there is production. We remark that our assumptions about costs cannot be weakened: in the more general model of costly inclusion, where costs may depend on the set of agents who win, there are envy-free Groves mechanisms if and only if costs are convex and only depend on the number of winners, as we assume here \citep{Mackenzie-Trudeau2023}. We describe the endogenous subsidy auctions in \hyperlink{Section2}{Section~2}, and we provide a proof sketch for the key lemma that drives this result in \hyperlink{Section6}{Section~6}.

Second, we characterize the endogenous subsidy auctions that are voluntary, or ex-post individually rational for all parties, as the {\it endogenous shareholding auctions} (\hyperlink{Theorem3}{Theorem~3}). As their name suggests, these auctions select competitive equilibria that only involve redistribution of the auction firm's ownership shares, with the common shareholding of each bidder determined endogenously by the preference reports. These competitive equilibria are evocative of competitive equilibria from equal incomes (CEEI), a classic notion from general equilibrium theory (\citealp{Foley1967}; \citealp{Kolm1971}; \citealp{Schmeidler-Vind1972}; \citealp{Varian1974}) that has been widely celebrated (see \citealp{Budish2011}); the difference is that our competitive equilibria need only provide equal income to the bidders. We also describe these auctions in \hyperlink{Section2}{Section~2}.

Third, we investigate the endogenous shareholding auctions that are optimal in a prior-free sense, both from the perspective of a profit-maximizing designer and from the perspective of the bidders. For the former, reinforcing earlier findings for models without production,\footnote{In particular, (i)~when selling identical objects to buyers with downward sloping demand curves, the Vickrey auction maximizes the expected revenue among all efficient and individually rational auctions that are Bayesian incentive compatible \citep{Krishna-Perry2000}, and (ii)~when selling distinct objects to buyers with unit demand and non-quasilinear preferences, the minimum price Walrasian mechanisms are ex-post revenue maximizing among mechanisms that satisfy strategy-proofness, individual rationality, no subsidy, and weak versions of efficiency and fairness (\citealp{Kazumura-Mishra-Serizawa2020}; \citealp{Sakai-Serizawa2023}).} we find that the VCG mechanisms are optimal for the designer (\hyperlink{Theorem4}{Theorem~4}). This result is robust in the sense that it does not depend at all on the designer's beliefs, which is notable because when the designer is {\it not} constrained by our axioms, his favorite auction does generally depend on his beliefs \citep{Myerson1981}. By contrast, the class of bidder-optimal\footnote{As worst-case performance (\citealp{Moulin2009}; \citealp{Guo-Conitzer2009}; \citealp{Carroll2015}; \citealp{Guo-Shmaya2023}) does not allow us to distinguish between endogenous shareholding auctions, we instead describe the undominated members of this class. This approach has also been used to compare anonymous Groves mechanisms for selling identical objects \citep{Guo-Markakis-Apt-Conitzer2013}, Groves mechanisms for making public decisions \citep{Athanasiou-Valletta2021}, `canonical' mechanisms for selling one object \citep{Borgers-Li-Wang2025}, and regulatory mechanisms when the monopolist is privately informed about marginal costs \citep{Mishra-Patil2025}.} endogenous shareholding auctions is the novel class of {\it valvular auctions} (\hyperlink{Theorem5}{Theorem~5}), whose name suggests a valve-based water system that can be run alongside a clock auction in order to determine the ownership shares. In \hyperlink{Section1.2}{Section~2}, we provide a concrete example of a valvular auction (\hyperlink{Figure1}{Figure~1}) and an illustration of the valve-based water system (\hyperlink{Figure2}{Figure~2}).

Finally, we investigate the endogenous shareholding auctions that are optimal for a rational designer who maximizes subjective expected welfare. Rather than focus on a particular social welfare function, we prove a general result: for any subjective prior with compact support and any continuous social welfare function, there is a mechanism that is optimal in the sense that it maximizes expected welfare (\hyperlink{Theorem6}{Theorem~6}). For emphasis, the social welfare function need not be linear. At a high level, the proof involves exploiting the rich mathematical structure of the endogenous shareholding auctions to establish the required compactness and continuity; see \hyperlink{Section6}{Section~6} for a proof sketch.

\hypertarget{Section1.2}{}
\subsection{Applications}

In this section, we discuss our model and results in the context of public housing, voluntary separation, and monopoly regulation.

\vspace{\baselineskip} \noindent \textsc{Example 1.} {\it Public housing.} Globally, public housing is one of the major forms of in-kind transfers from governments \citep{Currie-Gahvari2008}, and many houses today are purchased {\it presale}: before construction has been completed or even begun \citep{Chen-Hu-Jin-Zhou2026}. There is a notable intersection of public housing with presale purchase in Singapore, where over 80\% of resident households live in public housing, with construction commencing after properties are allocated and typically completed within three years \citep{Lee-Ferdowsian-Yap2023}. It is sometimes a reasonable approximation that public housing is identical, particularly when restricting attention to a given development, and unit demand captures the requirement that nobody receives multiple units. Convex costs are reasonable when land is scarce, as Cobb-Douglas is a good first-order approximation for the housing production function \citep{Combes-Duranton-Gobillon2021}. To summarize, our setting may be viewed as a stylized model of public housing with presale when land is scarce.

If public housing were produced and allocated via VCG with the profits absorbed by the state, then our endogenous shareholding auctions would improve housing affordability for the winners while providing financial support to the losers, with these welfare benefits funded by the auction itself. That said, public housing is typically not allocated by auction, but rather the price is controlled at a low level below the competitive price and then the housing is allocated by lottery. Accordingly, theory work on public housing has investigated lotteries and wait lists (\citealp{Arnosti-Shi2020}; \citealp{Waldinger2021}). A simple rationale for price controls is that there are positive externalities; for example, in Singapore, where winners acquire transferable 99-year leases during which they effectively own their flats, policy documents cite the importance of giving Singaporeans a tangible stake in the country. Another rationale is that price controls can facilitate redistribution when wealth is not observable \citep{Dworczak-Kominers-Akbarpour2021}.

It is well-known that, in the absence of externalities, the usual practice is inefficient. In fact, because housing is not allocated to those who value it most, there is deadweight loss from misallocation that can in fact exceed the usual `Harberger's triangle' deadweight loss due to undersupply \citep{Glaeser-Luttmer2003}. Moreover, the usual practice also violates our no-envy requirement, because the lottery's losers envy the large windfalls enjoyed by its winners, and these large windfalls have received considerable recent attention: (i)~in Singapore, in 2023, the Minister for National Development acknowledged public grievances about excessive windfalls, particularly for those who win homes in highly demanded locations,\footnote{The minister stated, ``\elide the system must be fair: public housing should be primarily for owner occupation, not speculative investment. HDB flats will continue to be an important asset for Singaporeans, but many Singaporeans told us during the FSG conversations that it is not fair for a small minority to gain large windfalls when they sell their BTO flats in popular locations on the resale market." See ``Opening Remarks by Minister Desmond Lee at the Post-NDR Reach Engagement 2023," \url{https://www.mnd.gov.sg/newsroom/speeches/view/opening-remarks-by-minister-desmond-lee-at-the-post-ndr-reach-engagement-2023}.} and (ii)~in South Korea, in 2026, an assemblyman proposed a bill to allocate price-capped housing using bond purchases, effectively introducing an auction to transfer winner windfalls to the Housing and Urban Fund.\footnote{See ``Introduction of the Bidding System for National Housing Bonds, Representative Proposal by Assemblyman Ahn Tae-jun (Bill No. 2217684)." Available at \url{https://likms.assembly.go.kr/bill/bi/billDetailPage.do?billId=PRC_L2M6K0K2J2R3R0P9Q3O7P4N7O8W4W7}.} By contrast, our auctions provide windfalls to eligible participants whether or not they win, avoiding a guaranteed pool of eligible (often due to low income) participants who lose and receive no financial support.

\vspace{\baselineskip} \noindent \textsc{Example 2.} {\it Voluntary separation.} Downsizing is sometimes necessary in both the private sector and the public sector. For example, a private firm might downsize via mass layoffs \citep{Jacobsen-LaLonde-Sullivan1993}, a military might downsize at the end of a conflict \citep{Warner-Pleeter2001}, or a government might downsize when transitioning from a planned economy to a market economy \citep{Haltiwanger-Singh1999}. In this case, a voluntary separation can be ``produced" at a possibly negative cost, and this has been the primary instrument for downsizing at universities in the United States for decades because (i)~in 1994, mandatory retirement of tenured faculty was eliminated \citep{Ashenfelter-Card2002}, and (ii)~in 1998, higher education institutions were given an exemption from the Age Discrimination in Employment Act to allow for voluntary retirement incentive plans.\footnote{See 29 U.S. Code $\S$ 623(m), ``Voluntary retirement incentive plans," \url{https://www.law.cornell.edu/uscode/text/29/623}.} Our assumption that benefit depends only on headcount reduction is reasonable when the eligibility pool is sufficiently narrow that productivity is nearly homogeneous, though we caution that it is not enough for productivity to be simply unobserved because of adverse selection concerns.\footnote{The adverse selection concern is that the most productive employees are the ones with the best outside options (\citealp{Diwan1994}; \citealp{Rama1999}).} Unit demand captures the requirement that an employee can separate at most once, and in this context the employee's valuation is negative: the employee must be paid to separate. Convex costs are simply concave benefits. Altogether, our setting may be viewed as a stylized model of a hypothetical academic department that (i)~wishes to induce some senior tenured faculty to take early retirement in order to open up some tenure lines, and (ii)~happens to have senior tenured faculty that all have similar productivity.

If the department downsized via VCG with the profits absorbed by the university, then our endogenous shareholding auctions would redistribute some of the downsizing benefits from the university to both improve retirement benefits for those who retire as well as provide financial support to the senior faculty who stay. That said, downsizing is typically not achieved by auction, but rather by announcing take-it-or-leave-it financial incentives; this is true for universities \citep{Pencavel2001}, the military \citep{Warner-Pleeter2001}, and the public sector.\footnote{\cite{Haltiwanger-Singh1999} survey 41 public sector retrenchment programs across 37 countries, and categorize 19.5\% of the programs as neutral programs that ``offer a simple, common package to a wide class of workers with little or no attempt to target specific groups." A recent example is the 2025 U.S. federal deferred resignation program, where a mass email to about two million federal employees announced a uniform offer to voluntarily resign in exchange for about eight months of pay without work duties. Most of the surveyed programs targeted based on skills, age, or both, but none resembled auctions.} Accordingly, theory work on downsizing has investigated optimal screening mechanisms \citep{Jeon-Laffont1999}. A simple rationale for this practice is that it is easy to implement in a manner that is consistent with discrimination laws.

When the usual practice has a single round, then it is inefficient because it is difficult to forecast quit responses to take-it-or-leave-it offers. When there are multiple rounds, say because the university cannot credibly commit that it won't revise the initial offer if the initial response is poor, then those who take the initial offer may envy those who take a later offer. Indeed, this was the experience of the University of California with its Voluntary Early Retirement Program in the early 1990s, where there were three rounds and the third had a considerably larger severance incentive than the first two \citep{Pencavel2001}. By contrast, our auctions maximize the net benefit from downsizing, remain consistent with discrimination laws by generating a common offer for the targeted pool, and share some of the benefits with the faculty.

\vspace{\baselineskip} \noindent \textsc{Example 3.} {\it Monopoly regulation.} It is an ancient insight that monopolies are lucrative,\footnote{In {\it Politics}, Aristotle reports that Thales of Miletus, one of the Seven Sages, used astronomy to create a lucrative monopoly on olive presses during an unusually fruitful winter.} and legal arguments in favor of their regulation can be traced to the late sixteenth century, if not earlier.\footnote{In 1599, Coke argued successfully in {\it Davenant v. Hurdis} that monopolies are against the public good, and his opponent conceded this point. Over two hundred years earlier, during the reign of Edward~III, Parliament convicted one John Pecche for excessively exploiting his patent to exclusively sell sweet wines in London, referring in his sentencing to his ``extortionate prices." See \cite{Letwin1954}.} Modern rationales for monopoly regulation include efficiency, fairness, and consumer surplus \citep{Mas-Colell-Whinston-Green1995}. Of course, a regulator with limited information must design policies carefully, as regulations based on incorrect assumptions may be counterproductive, and this point has received considerable attention: beginning with \cite{Baron-Myerson1982}, there is now a substantial literature dedicated to regulatory design when the monopolist knows both the supply curve and the demand curve while the regulator knows less.\footnote{See for example \cite{Lewis-Sappington1988a}, \cite{Lewis-Sappington1988b}, \cite{Armstrong1999}, \cite{Guo-Shmaya2023}, and \cite{Mishra-Patil2025}.} We consider instead an alternative problem for a regulator with limited information: regulatory design when the monopolist knows only the supply curve while the regulator has the same information. Because there is no asymmetric information between the regulator and the monopolist, and because neither party knows the demand curve, our regulator is not concerned with screening the monopolist but rather with the choice of auction.

If the monopolist used VCG, then our endogenous shareholding auctions would redistribute some of the producer surplus to consumer surplus by paying each consumer an endogenous dividend of the auction's profit. That said, in the absence of regulation, the monopolist may find it optimal to deviate from VCG by using reserve prices and engaging in price discrimination \citep{Myerson1981}. A simple rationale is that the monopolist is sometimes willing to destroy surplus if he can keep a larger share of what remains.

In general, the optimal auction is neither efficient nor envy-free. By contrast, VCG is both efficient and envy-free, and at first glance it appears quite consumer-friendly: regulate the monopolist by requiring him to behave as though he is a competitive firm, and among the competitive prices select the one that is most consumer-friendly. That said, all of the producer surplus in this fictional competitive market accrues to one party in reality, and our results show that subject to our constraints, VCG is in fact the most producer-friendly auction.

\vspace{\baselineskip} For the rest of the paper, we favor the language of monopoly regulation: the monopolist's objective is expected producer surplus maximization, the well-studied benchmark is that the monopolist is unconstrained, and our axioms are hard constraints imposed by a regulator. Though this is our most stylized application, it is the most convenient for discussing the literature.

\hypertarget{Section2}{}
\section{Description of endogenous shareholding auctions}

In this section, we describe the class of endogenous shareholding auctions, as well as the larger class of endogenous subsidy auctions and the smaller class of valvular auctions. Some light notation will facilitate the discussion: (i)~let $n$ denote the number of consumers, each of whom has unit demand and quasi-linear preferences, and (ii)~for each quantity $q \in \{1, 2, ..., n\}$, let $\mathsf{S}(q)$ denote the associated marginal cost specified by the supply curve. We assume that costs are convex, so that $\mathsf{S}(1) \leq \mathsf{S}(2) \leq ... \leq \mathsf{S}(n)$.

\hypertarget{Section2.1}{}
\subsection{The myopic subsidy curve}

The formal definition of endogenous shareholding auction involves a {\it funded subsidy curve}: a function mapping certain prices to subsidies that satisfies several conditions. Rather than describe the conditions in detail at this point, we instead begin with a concrete example that satisfies all of the conditions. We refer to this example as the {\it myopic subsidy curve}.

To avoid distractions, let us suppose for the moment that all marginal costs are finite; then the myopic subsidy curve is drawn as follows.
\begin{itemize}
\item First, draw the {\it average profit curve} over $[\mathsf{S}(1), \mathsf{S}(n)]$: (i)~at $\mathsf{S}(1)$, the average profit is zero, and (ii)~for each $q \in \{1, 2, ..., n-1\}$, on the subinterval $[\mathsf{S}(q), \mathsf{S}(q+1)]$, the average profit increases with slope $\frac{q}{n}$. Intuitively, when a price-taking firm faces price $p$ and maximizes profit with the understanding that it will sell everything it produces, the average profit at $p$ is the maximum subsidy that could be distributed to each consumer from this profit.

\item Second, draw the {\it myopic subsidy curve}: (i)~at $\mathsf{S}(1)$, the subsidy is zero, and (ii)~for each $q \in \{1, 2, ..., n-1\}$, on the subinterval $[\mathsf{S}(q), \mathsf{S}(q+1)]$, the subsidy increases with slope $1$ until it reaches the average profit curve (which may or may not occur), after which it has slope $0$.
\end{itemize}
The myopic subsidy curve satisfies all of the requirements to be a funded subsidy curve, and therefore its associated auctions are endogenous shareholding auctions. In fact, the myopic subsidy curve satisfies additional requirements that guarantee its associated auctions are consumer-optimal among the endogenous shareholding auctions.

We illustrate the myopic subsidy curve for a concrete supply curve in \hyperlink{Figure1}{Figure~1}. In this figure, we also show an example reported demand curve, for which the auction's outcome is that (i)~each winner pays the minimum competitive price and receives the subsidy associated with that price, (ii)~each loser refuses the maximum competitive price and receives the subsidy associated with that price, and (iii)~the monopolist receives as revenue the negative of the transfer to the consumers and then pays the associated production costs. Equivalently, this same outcome is a competitive equilibrium with redistribution where (i)~each consumer receives the subsidy at the maximum competitive price, (ii)~this total is taken from the monopolist, and (iii)~the market price is the minimum competitive price plus the difference between the subsidy at the maximum competitive price and the subsidy at the minimum competitive price. With suitable qualifications to cover competitive prices outside the interval $[\mathsf{S}(1), \mathsf{S}(n)]$, this description is true of every outcome of every endogenous shareholding auction; see \hyperlink{Theorem1}{Theorem~1}.

\hypertarget{Figure1}{}
\begin{figure}[]
\centering
\includegraphics[width=140mm]{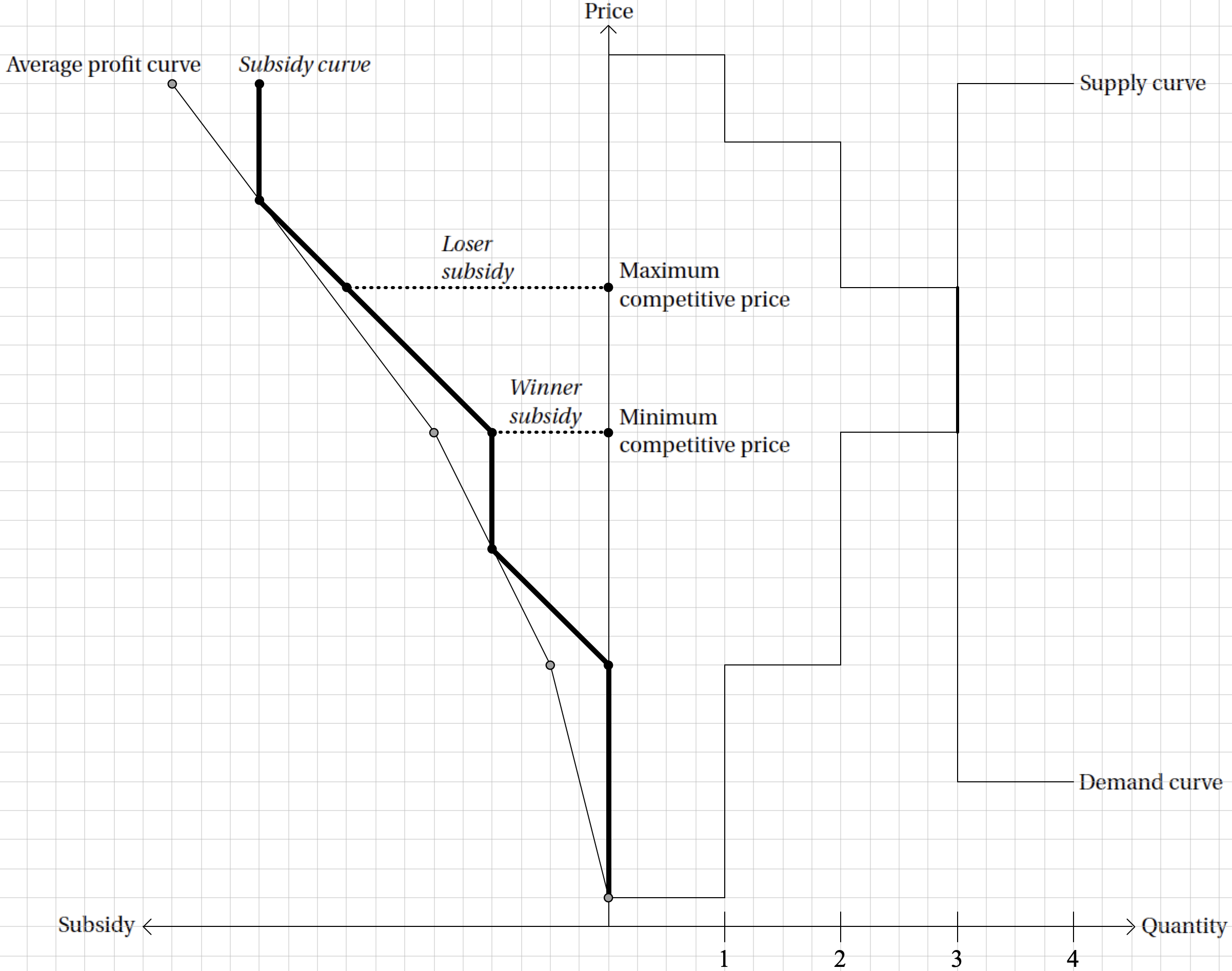}
\caption{{\it Myopic subsidy curve.} In this example, there are four consumers. Using the quantity and price axes: (i)~the known supply curve is such that $\mathsf{S}(1) = 1$, $\mathsf{S}(2) = 9$, $\mathsf{S}(3) = 17$, and $\mathsf{S}(4) = 29$, (ii)~the demand curve according to the reported valuation profile $v$ is such that $\mathsf{D}_v(1) = 30$, $\mathsf{D}_v(2) = 27$, $\mathsf{D}_v(3) = 22$, and $\mathsf{D}_v(4) = 5$, and (iii)~it follows that the minimum competitive price is $17$ and the maximum competitive price is $22$. Using the price and subsidy axes: (i)~the average profit curve is drawn using the known supply curve but without using the reported demand curve, and (ii)~the myopic subsidy curve iteratively increases with slope one until it hits the average profit curve and then is flat with slope zero until it reaches a new marginal cost. Each valvular auction associated with this subsidy curve is such that at any valuation profile, (i)~each winner pays the minimum competitive price and receives the subsidy associated with that price, (ii)~each loser refuses the maximum competitive price and receives the subsidy associated with that price, and (iii)~the monopolist receives as revenue the negative of the transfer to the consumers and then pays the associated production costs. For the reported demand curve, (i)~each consumer with one of the three highest valuations wins an object, receives a subsidy of $4$, and pays $17$, altogether consuming transfer $-13$, (ii)~the consumer with the lowest valuation loses and receives a subsidy of $9$, and (iii)~the monopolist receives as revenue $30 = -(-13 \cdot 3 + 9)$ and pays as cost $27=1+9+17$, altogether consuming a profit of $3$. This is a competitive equilibrium with redistribution where (i)~each consumer receives the subsidy at the maximum competitive price, $9$, (ii)~this total $4 \cdot 9 = 36$ is taken from the monopolist, and (iii)~the market price is the minimum competitive price plus the difference between the subsidy at the maximum competitive price and the subsidy at the minimum competitive price, $17 + (9-4) = 22$.}
\label{overflow}
\end{figure}

\hypertarget{Section2.2}{}
\subsection{Clock auctions}

Though we do not formally state any results about clock auctions, they are useful for an informal discussion about incentive compatibility. In particular, in order to focus on the intuition for this discussion, we ignore all details about tie-breaking, extreme valuations, and whether time is continuous or discrete.

To begin, let us say an {\it ordinary clock auction} is an extensive form mechanism and associated intended strategy profile with the following features.
\begin{itemize}
\item There is a public {\it clock price} that begins at $\mathsf{S}(1)$, and that steadily increases toward a price ceiling of $\mathsf{S}(n)$ over the course of the auction.

\item At each clock price, each active consumer can either (i)~{\it stay} and remain tentatively attached to the {\it winning bundle}, or (ii)~{\it exit} and become permanently attached to the losing bundle. The winning bundle is paying the clock price to receive an object, the losing bundle is paying nothing and receiving no object, and to facilitate comparison with other clock auctions, we take the perspective that both bundles are continuously updated alongside the clock price even though the losing bundle is constant. For each consumer, the intended strategy is that he stays if and only if the clock price is below his private valuation. Exits are not observed by other consumers.

\item If the clock price is below the price ceiling, then it increases so long as the number of consumers who have not yet exited is greater than the number of marginal costs at or below the clock price. In other words, the clock price increases so long as demand exceeds supply. Whenever the clock price stops, each active consumer receives the associated winning bundle, and each consumer who has exited receives the associated losing bundle.
\end{itemize}
Following the logic in \cite{Mishra-Talman2010}, an ordinary clock auction terminates at the minimum competitive price.\footnote{To be precise, this is true so long as valuations are not extreme in the sense that they fall between $\mathsf{S}(1)$ and $\mathsf{S}(n)$. In general, an ordinary clock auction terminates at the {\it rounded minimum competitive price}, or the minimum competitive price rounded up to $\mathsf{S}(1)$ if necessary, which provides intuition for why the latter appears in the statement of \hyperlink{Theorem1}{Theorem~1}.} Moreover, the intended strategies are dominant for the following reasons: (i)~because the winning bundle never gets better, the losing bundle never gets worse, and they are equivalent at the intended exit, there is no benefit to staying late and winning or to exiting early, and (ii)~because moreover the losing bundle never changes, there is no benefit to exiting late.

Consider now an {\it extended clock auction}, which differs in that (i)~the clock price continues to increase even if demand does not exceed supply, so long as demand is at least supply, and (ii)~as soon as supply is at least demand, the set of winners and the winning bundle both freeze, while the losing bundle continues to be updated alongside the clock price even though in this auction it is again constant. Following the logic in \cite{Mishra-Talman2010}, an extended clock auction terminates at the maximum competitive price.\footnote{A remark analogous to that in the previous footnote applies for the {\it rounded maximum competitive price}, or the maximum competitive price rounded down to $\mathsf{S}(n)$ if necessary.} Moreover, the natural extensions of the previous intended strategies remain dominant, and for all the same reasons.

Finally, consider a {\it modified extended clock auction} for an associated subsidy curve. The subsidy curve must be chosen such that as the price increases, the subsidy never falls, and though it may increase, it never increases more rapidly than the price. The extended clock auction is modified as follows: (i)~until the winning bundle is frozen, the winning bundle has the subsidy associated with the clock price added to it, and (ii)~throughout the auction, the losing bundle has the subsidy associated with the clock price added to it. For the same reasons as before, there is no benefit to staying late and winning, and there is no benefit to exiting early. Moreover, {\it because the losing bundle is never changed if a loser exits earlier}, there is no benefit to exiting late. Indeed, the losing bundle is determined by the terminal price at which demand falls below supply, and if the terminal price changes to $p$ when $i$ exits earlier, then demand must be at least supply at $p$ when $i$ exits later, so that $i$ is a winner when he exits later.

This informal discussion of clock auctions provides some rough intuition for why the endogenous shareholding auctions---and more generally the endogenous subsidy auctions, which differ only in that they need not be voluntary---have dominant strategies.

\hypertarget{Section2.3}{}
\subsection{Allocation versus production}

The problem of allocating one object is modeled by setting $\mathsf{S}(1) = 0$ and $\mathsf{S}(2) = \infty$. In this case, the only endogenous shareholding auctions are the VCG mechanisms, also known for this problem as the Vickrey auctions. To avoid distractions, let us again set aside extreme valuations for this discussion by assuming all valuations are non-negative; then in a Vickrey auction, one consumer whose bid is highest wins and pays the second-highest bid, and the other consumers lose and pay nothing. We highlight four observations about VCG mechanisms for this problem.

First, the VCG mechanisms select minimum competitive price equilibria (\citealp{Demange1982}; \citealp{Leonard1983}). Indeed, the set of competitive prices is the interval between the second-highest valuation and the highest valuation, and thus the minimum competitive price is accepted by the winner and refused by the losers. For this reason, the VCG mechanisms are efficient, envy-free, and voluntary.

Second, recall that for each consumer $i$ and each valuation profile reported by his peers $v_{-i}$, the associated {\it option set} is the set of feasible consumption bundles that $i$ can obtain by varying his own report $v_i$. In VCG mechanisms, each option set is simply a budget set with no subsidy from which $i$ selects optimally by reporting his true valuation. In particular, (i)~the winner faces the second-highest valuation, or the minimum competitive price, and (ii)~each loser faces the highest valuation, or the maximum competitive price. For this reason, the VCG mechanisms are strategy-proof.

Third, the outcome of a VCG mechanism is determined by a surplus-maximizing group of winners and the minimum competitive price, which informally are outputs of an ordinary clock auction.

Finally, all option sets for VCG mechanisms are determined by the minimum competitive price and the maximum competitive price, which informally are outputs of an extended clock auction.

These four observations generalize to the problem of allocating $q$ identical objects without production. As in the one object case, each VCG mechanism selects minimum competitive price equilibria, delivers each winner's option set as the budget set given by no subsidy and the minimum competitive price, and delivers each loser's option set as the budget set given by no subsidy and the maximum competitive price. Moreover, each VCG mechanism has outcomes determined by ordinary clock auctions and option sets determined by extended clock auctions. Finally, it remains the case that these are the only endogenous shareholding auctions.

For the general problem in which identical objects can be produced, all four of our observations about VCG mechanisms remain valid,\footnote{See the \hyperlink{PriceLemma}{Price~Lemma} and the \hyperlink{VCGLemma}{VCG~Lemma}.} but the VCG mechanisms are no longer the only endogenous shareholding auctions. Endogenous shareholding auctions differ from VCG mechanisms in a variety of ways: (i)~while they select competitive equilibria, these equilibria may involve redistribution and they need not be associated with the minimum competitive price, (ii)~while they deliver option sets as budget sets whose prices always match those of VCG mechanisms, the subsidies of these budget sets may be positive, and (iii)~their outcomes need not be determined by ordinary clock auctions, because the outcome can depend on both the minimum competitive price and the maximum competitive price. That said, the fourth observation remains intact: as suggested by our earlier clock auction discussion, outcomes and option sets are determined by extended clock auctions (along with the associated subsidy curve).

\hypertarget{Section2.4}{}
\subsection{Valvular auctions}

We conclude this discussion with a description of the valvular auctions, which are consumer-optimal among the endogenous shareholding auctions in a suitable sense (\hyperlink{Theorem5}{Theorem~5}). The auctions associated with the myopic subsidy curve are valvular, and more generally, all valvular auctions have subsidy curves that are piecewise linear with each segment having either slope one or slope zero.

The term {\it valvular} is intended to suggest a water system that is run alongside an extended clock auction, where the water represents the auction's profit. In this water system, (i)~all water is initially in the {\it profit reservoir}, (ii)~there are $n-1$ {\it upper pipes} from the profit reservoir to the {\it producer tank}, and (iii)~there are $n$ {\it lower pipes} from the producer tank to the {\it consumer tank}. Each upper pipe has its own {\it profit valve} that is initially closed, and profit valve $q$ is opened when the clock price reaches the marginal cost $\mathsf{S}(q)$. The $n$ lower pipes are jointly controlled by the {\it regulator's valve}, which is closed while the subsidy curve is flat with slope zero and open while the subsidy curve climbs with slope one. All $2n-1$ pipes have the same size. The process ends when all valves are closed, which occurs either (i)~at the end of the extended clock auction, at the rounded maximum competitive price, or (ii)~before the end of the extended clock auction, at the first price after the rounded minimum competitive price where the regulator's valve is closed. At this point, the auction's profit is distributed across the sides of the market in proportion to the water in their tanks, with the consumer portion divided evenly across the consumers. In this way, the ownership of the auction's profit is determined endogenously. We illustrate the process with an example in \hyperlink{Figure2}{Figure~2}.

The myopic subsidy curve corresponds to using the regulator's valve as follows: (i)~at each marginal cost, the valve is opened, and (ii)~whenever the producer tank is empty, the valve is closed. At the start of the auction, the valve is opened and then immediately closed before any water reaches the consumer tank, so water only begins flowing into the consumer tank at the second distinct marginal cost. In order for any water to actually reach the consumer tank, then, there must be at least three distinct marginal costs. When there is no production, the only distinct marginal costs are zero and infinity, so the myopic subsidy curve is the zero subsidy curve for the VCG mechanisms, the regulator's valve is never opened, and the process ends at the rounded minimum competitive price.

\hypertarget{Figure2}{}
\begin{figure}[]
\centering
\includegraphics[width=\linewidth]{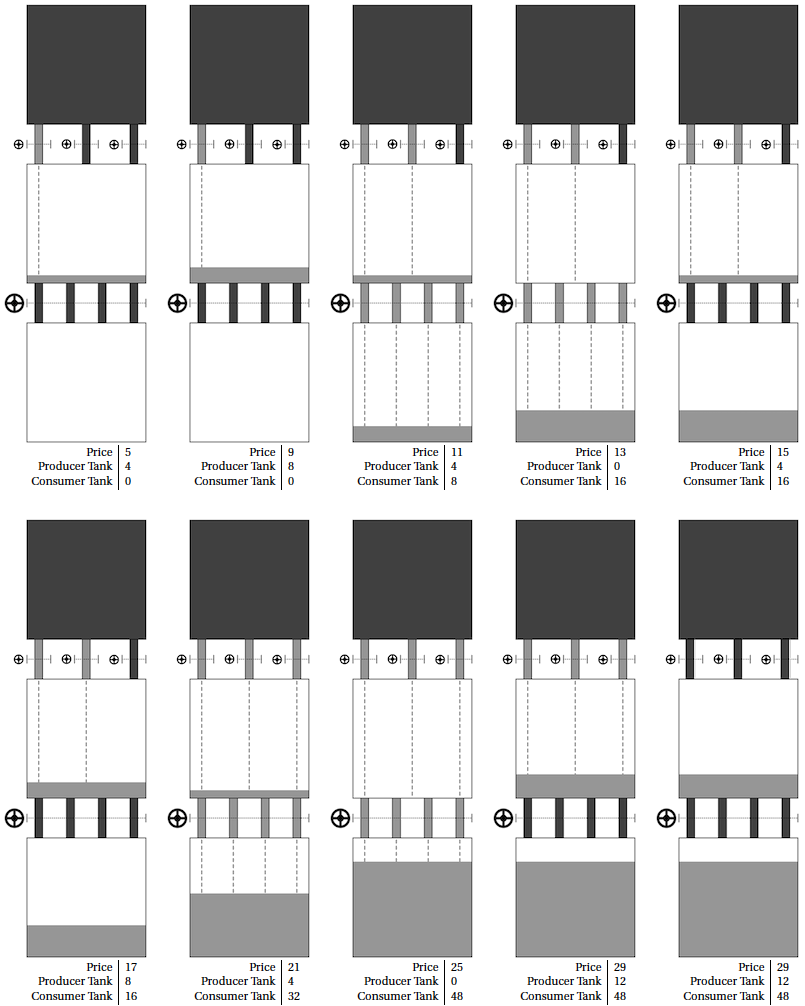}
\caption{{\it Valvular auctions and water systems.} The top tank is the profit reservoir (drawn as opaque), there are $n-1$ upper pipes from the profit reservoir to the producer tank, and there are $n$ lower pipes from the producer tank to the consumer tank. Each upper pipe has its own profit valve and the $n$ lower pipes are jointly controlled by the regulator's valve. We illustrate the full process for (i)~the example subsidy curve from Figure~1, and (ii)~any demand curve where all four agents have valuations above $\mathsf{S}(n)$. For the demand curve in Figure~1, this process simply terminates earlier, at Price $22$, Producer Tank $3$, Consumer Tank $36$.}
\label{overflow}
\end{figure}

\hypertarget{Section3}{}
\section{Model}

In order to facilitate discussion of the literature, we favor the language of monopoly regulation to present the formal model and results, with just a few brief remarks about interpretation for public housing and voluntary separation.

\hypertarget{Section3.1}{}
\subsection{Monopolist environments}

We consider a monopolist who can produce identical indivisible objects at a monetary cost. The cost function is weakly convex, and therefore specified by a non-decreasing sequence of marginal costs that (in a slight abuse of terminology\footnote{In standard terminology, {\it supply curves} map prices to quantities and {\it inverse supply curves} map quantities to prices, but for brevity we use the former term for the latter concept.}) we refer to as the {\it supply curve}. We allow for both negative marginal costs to allow for objects whose consumption benefits the monopolist and infinite marginal costs to capture infeasible production, and in order to simplify certain edge case arguments, we adopt the convention that the marginal cost at zero is negative infinity.

\vspace{\baselineskip} \noindent \textsc{Definition:} The set of {\it (object) quantities} is $Q \equiv \{0, 1, ...\}$. We say that a function $\mathsf{S}:Q \to \mathbb{R} \cup \{-\infty, \infty\}$ is a {\it supply curve} if (i)~$\mathsf{S}(0) = -\infty$, (ii)~for each $q \in \{1, 2, ...\}$ we have $\mathsf{S}(q) \in \mathbb{R} \cup \{\infty\}$, and (iii)~for each $q \in Q$ we have $\mathsf{S}(q) \leq \mathsf{S}(q+1)$. For each supply curve $\mathsf{S}$, the associated {\it cost function} is the function $\mathsf{C}: Q \to \mathbb{R} \cup \{\infty\}$ such that for each $q \in Q$, we have
\begin{align*}
\mathsf{C}(q) \equiv \left\{
     \begin{array}{lr}
       \sum_{q' \in \llbracket 1, q\rrbracket} \mathsf{S}(q'),\footnotemark & \mathsf{S}(q) < \infty, \\
       \infty, & \mathsf{S}(q) = \infty,
     \end{array}
   \right.
\end{align*}
\footnotetext{Throughout the paper, we often use integer interval notation: for each pair $a, b \in \mathbb{Z}$, $\llbracket a, b \rrbracket$ denotes $[a, b] \cap \mathbb{Z} = \{a, a+1, ..., b\}$.}
using the convention that $\mathsf{C}(0) = 0$.

\vspace{\baselineskip} The monopolist cannot consume objects, but there are also consumers who can each consume at most one object, and everybody can consume money. We assume the supply curve and the finite set of consumers are common knowledge, while each consumer's willingness to pay for an object is his private information.

\vspace{\baselineskip} \noindent \textsc{Definition:} A {\it (monopolist) environment} is a pair $(\mathsf{S}, N)$ defined as follows.
\begin{itemize}
\item $\mathsf{S}$ is the supply curve and $\mathsf{C}$ is the associated cost function.

\item $N$ is the finite and nonempty set of {\it consumers}. We define $n \equiv |N|$. There is one producer, the {\it monopolist}, denoted by $i_0 \not \in N$. Each $i \in N \cup \{i_0\}$ is an {\it agent}.

\item For each $i \in N$, the set of {\it (consumption) bundles for $i$}, denoted $X_i$, is a personal copy of $\mathbb{R} \times \{0, 1\}$ for $i$,\footnote{Formally, whenever we say that $A_i$ is a personal copy of space $B$ for agent $i$, we mean that $A_i = \{i\} \times B$, so that the space's owner $i$ is not only suggested by our usual notation but moreover determined by the space. We then suppress $i$ in an abuse of notation: for each $(i, b) \in \{i\} \times B = A_i$, we write $b \in A_i$.} where each bundle $x_i = (t_i, a_i)$ specifies {\it (monetary) transfer}~$t_i$ and {\it (object) assignment} $a_i$. We refer to a consumer who receives an object as a {\it winner} and a consumer who does not as a {\it loser}. As the monopolist cannot consume objects, the set of consumption bundles for $i_0$, denoted $X_{i_0}$, is a personal copy of $\mathbb{R} \times \{0\}$ for $i_0$.

\item Each $i \in N$ has a {\it valuation} $v_i \in \mathbb{R}$, which represents the quasi-linear preference relation $\succsim_{i|v_i}$ over bundles such that for each pair $(t_i, a_i), (t_i', a_i') \in X_i$,
\begin{align*}
(t_i, a_i) \mathrel{\succsim_{i|v_i}} (t_i', a_i') \text{ if and only if } v_i \cdot a_i + t_i \geq v_i \cdot a_i' + t_i'.
\end{align*}
This is private information, and the set of possible {\it valuations for $i$}, denoted $V_i$, is a personal copy of $\mathbb{R}$ for $i$. It is common knowledge that the monopolist's preference relation over $X_{i_0}$, $\succsim_{i_0}$, is the strictly monotonic one. Altogether, the set of possible {\it valuation profiles} is $V \equiv \times_{i \in N} V_i$, and to unify notation across consumers and the monopolist, we sometimes write $(\succsim_{i|v})_{i \in N \cup \{i_0\}}$ for the preference profile determined by $v$. As usual, we use $\succ$ instead of $\succsim$ to denote strict preference.

\item We define the {\it production capacity} $\qcap \equiv \max \{q \in \llbracket 0, n \rrbracket | \mathsf{C}(q) < \infty \}$ to be the maximum quantity that can feasibly be produced and consumed. The collection of {\it feasible winner sets} is $\mathbb{W} \equiv \{W \subseteq N | |W| \leq \qcap\}$, and an {\it outcome} is a pair $(t, W) \in \mathbb{R}^N \times \mathbb{W}$. We let~$X$ denote the set of outcomes, and we associate each $(t, W) \in X$ with the allocation in $\times_{i \in N \cup \{i_0\}} X_i$ such that (i)~each consumer $i$ receives transfer $t_i$, (ii)~$W$ is the set of winners, and (iii)~the monopolist receives the net payment from the consumers and then pays the production costs. Formally, (i)~for each $i \in N$, $x_i(t, W) \equiv (t_i, |W \cap \{i\}|)$, and (ii)~$x_{i_0}(t, W) \equiv ([-\sum_{i \in N} t_i] - \mathsf{C}(|W|), 0)$.
\end{itemize}
Whenever we refer to an arbitrary environment, we implicitly assume all of this notation.

\vspace{\baselineskip} In the context of public housing, the interpretation is that the designer is the housing authority, the objects are houses and valuations are presumably positive, $\mathsf{S}$ gathers the marginal costs which are presumably positive, and $\mathsf{S}$ is known to the housing authority (say because of an existing contract with a contractor). In the context of voluntary separation, the interpretation is that the designer is the university, the objects are retirements and valuations are presumably negative, $\mathsf{S}$ gathers the negative marginal benefits which are presumably negative, and $\mathsf{S}$ is known to the university.

\hypertarget{Section3.2}{}
\subsection{Surplus and efficiency}

In the context of monopoly regulation, if the valuation profile were known, then the regulator would be able to evaluate each outcome on the basis of its consumer surplus, producer surplus, and total surplus.

\vspace{\baselineskip} \noindent \textsc{Definition:} Fix an environment. For each $v \in V$, each $t \in \mathbb{R}^N$, and each $W \in \mathbb{W}$, we define the associated consumer surplus $\mathsf{CS}_v(t, W)$, producer surplus $\mathsf{PS}(t, W)$, and total surplus $\mathsf{TS}_v(W)$ by:
\begin{align*}
\mathsf{CS}_v(t, W) &\equiv \sum_{i \in W} v_i + \sum_{i \in N} t_i,
\\ \mathsf{PS}(t, W) &\equiv [-\sum_{i \in N} t_i] - \mathsf{C}(|W|), \text{ and}
\\ \mathsf{TS}_v(W) &\equiv \sum_{i \in W} v_i - \mathsf{C}(|W|).
\end{align*}
Moreover, for each $W \in 2^N \backslash \mathbb{W}$, we define $\mathsf{TS}_v(W) \equiv -\infty$. Observe that because there are no fixed costs, producer surplus is equal to profit: this is precisely the monetary transfer consumed by the monopolist.

\vspace{\baselineskip} Because of our assumption that preferences are quasi-linear, an outcome maximizes total surplus if and only if it is (Pareto) efficient.

\vspace{\baselineskip} \noindent \textsc{Definition:} Fix an environment. For each $v \in V$ and each $(t, W) \in X$, we say that $(t, W)$ is {\it $v$-efficient} if and only if for each $W' \in \mathbb{W}$, $\mathsf{TS}_v(W) \geq \mathsf{TS}_v(W')$. Equivalently,\footnote{We omit the simple proof; see Chapter 10.E of \cite{Mas-Colell-Whinston-Green1995}.} $(t, W)$ is {\it $v$-efficient} if and only if there is no $(t', W') \in X$ such that (i)~for each $i \in N \cup \{i_0\}$, $x_i(t', W') \mathrel{\succsim_{i|v}} x_i(t, W)$, and (ii)~for some $i \in N \cup \{i_0\}$, $x_i(t', W') \mathrel{\succ_{i|v}} x_i(t, W)$.

\hypertarget{Section3.3}{}
\subsection{Mechanisms and axioms}

The monopolist uses a mechanism to elicit consumer demand and then determine the outcome, and we assume that the regulator is able to constrain which mechanisms are available to the monopolist. As we focus on dominant strategy implementation, we can safely restrict attention to {\it (direct) mechanisms} where the consumers simultaneously report their valuations by the classic revelation principle (\citealp{Gibbard1973}; \citealp{Myerson1981}):

\vspace{\baselineskip} \noindent \textsc{Definition:} Fix an environment. A {\it transfer policy} is a mapping $\tau: V \to \mathbb{R}^N$, and an {\it assignment policy} is a mapping $\alpha: V \to \mathbb{W}$. For each assignment policy $\alpha$ and each $i \in N$, we let $\alpha_i$ denote the associated policy for determining whether or not $i$ is a winner; this is the mapping $\alpha_i: V \to \{0, 1\}$ such that for each $v \in V$, $\alpha_i(v) = |\alpha(v) \cap \{i\}|$. A {\it mechanism} $(\tau, \alpha)$ consists of a transfer policy $\tau$ and an assignment policy $\alpha$.

\vspace{\baselineskip} The regulator may require the mechanism to satisfy certain properties that we refer to as {\it axioms}, and our analysis involves the following ones.

\vspace{\baselineskip} \noindent \textsc{Definition:} Fix an environment. A mechanism $(\tau, \alpha)$ satisfies
\begin{itemize}
\item {\it efficiency} if for each $v \in V$, $(\tau(v), \alpha(v))$ is $v$-efficient;

\item {\it strategy-proofness} if for each $i \in N$, each $v_{-i} \in V_{-i}$, and each pair $v_i, v_i' \in V_i$, we have $v_i \cdot \alpha_i(v) + \tau_i(v) \geq v_i \cdot \alpha_i(v_i', v_{-i}) + \tau_i(v_i', v_{-i})$;

\item {\it no-envy} if for each $v \in V$ and each pair $i, j \in N$, $v_i \cdot \alpha_i(v) + \tau_i(v) \geq v_i \cdot \alpha_j(v) + \tau_j(v)$;

\item {\it consumer voluntary participation} if for each $v \in V$ and each $i \in N$, we have $v_i \cdot \alpha_i(v) + \tau_i(v) \geq 0$;

\item {\it producer voluntary participation} if for each $v \in V$, $[-\sum_{i \in N} \tau_i(v)] - \mathsf{C}(|\alpha(v)|) \geq 0$; and

\item {\it voluntary participation} if it satisfies both consumer voluntary participation and producer voluntary participation.
\end{itemize}
We say that an assignment policy $\alpha$ is {\it surplus-maximizing} if and only if for each $v \in V$ and each $W \in \mathbb{W}$, $\mathsf{TS}_v(\alpha(v)) \geq \mathsf{TS}_v(W)$. Equivalently, $\alpha$ is surplus-maximizing if and only if for each transfer policy $\tau$, the mechanism $(\tau, \alpha)$ satisfies {\it efficiency}.

\vspace{\baselineskip} The first three axioms represent constraints imposed by the regulator that we fix throughout our analysis. The first implies that the monopolist cannot reduce production from the efficient level in order to increase the price, the second requires that it is always a dominant strategy for each customer to honestly reveal his valuation, and the third captures the idea that the auction should output equal opportunities. The remaining axioms are ex-post voluntary participation constraints that prevent arbitrary transfers from one side of the market to the other; these are also known as individual rationality constraints in the literature.

In the context of public housing, the interpretation is that the axioms are self-imposed by the housing authority. In the context of voluntary separation, the interpretation is that the axioms are self-imposed by the university, but this may be influenced by other parties: the restriction to efficient auctions may reflect an agreement with a faculty labor union, and the restriction to envy-free auctions may reflect concerns about discrimination lawsuits due to regulations outside the model.

\hypertarget{Section4}{}
\section{Preliminary lemmas}

We study a particularly simple auction model: there is unit demand, objects are identical, and preferences are quasi-linear. As a result, many known results apply to our model as a special case, and our main results involve many familiar ideas. The purpose of this section is to gather the familiar ideas that are necessary for stating our main results and discussing the underlying intuition. For the proofs of all lemmas in this section, see \hyperlink{AppendixA}{Appendix~A}.

\hypertarget{Section4.1}{}
\subsection{Complete information}

When we pair an environment with a valuation profile, we have a complete information economy that is ripe for classic supply-demand analysis. We begin by introducing the reported demand curve and using it to provide explicit formulas for the endpoints of the interval of efficient quantities. Moreover, in anticipation of their later usefulness, we also introduce these ideas for economies with a single consumer removed.

\vspace{\baselineskip} \noindent \textsc{Definition:} Fix an environment. For each $v \in V$, we define (i)~the {\it demand curve given~$v$}, $\mathsf{D}_v:Q \to \mathbb{R} \cup \{-\infty, \infty\}$, (ii)~the {\it minimum efficient quantity given $v$}, $\qmin_v \in Q$, and (iii)~the {\it maximum efficient quantity given $v$}, $\qmax_v \in Q$, as follows.
\begin{itemize}

\item $\mathsf{D}_v(0) \equiv \infty$, $(\mathsf{D}_v(q))_{q \in \llbracket 1, n\rrbracket}$ re-orders $(v_j)_{j \in N}$ such that $\mathsf{D}_v(1) \geq \mathsf{D}_v(2) \geq ... \geq \mathsf{D}_v(n)$, and $q > n$ implies $\mathsf{D}_v(q) \equiv -\infty$;

\item $\qmin_v \equiv \max \{q \in Q | \mathsf{D}_v(q) > \mathsf{S}(q) \} $ and $\qmax_v \equiv \max \{q \in Q | \mathsf{D}_v(q) \geq \mathsf{S}(q) \}$.
\end{itemize}
Similarly, for each $i \in N$ and each $v_{-i} \in V_{-i}$, we define $\mathsf{D}_{v_{-i}}:Q \to \mathbb{R} \cup \{-\infty, \infty\}$, $\qmin_{v_{-i}} \in Q$, and $\qmax_{v_{-i}} \in Q$ as follows.
\begin{itemize}
\item$\mathsf{D}_{v_{-i}}(0) \equiv \infty$, $(\mathsf{D}_{v_{-i}}(q))_{q \in \llbracket 1, n-1 \rrbracket}$ re-orders $(v_j)_{j \in N \backslash \{i\}}$ such that $\mathsf{D}_{v_{-i}}(1) \geq \mathsf{D}_{v_{-i}}(2) \geq ... \geq \mathsf{D}_{v_{-i}}(n-1)$, and $q > n-1$ implies $\mathsf{D}_{v_{-i}}(q) \equiv -\infty$; 

\item $\qmin_{v_{-i}} \equiv \max \{q \in Q | \mathsf{D}_{v_{-i}}(q) > \mathsf{S}(q) \}$ and $\qmax_{v_{-i}} \equiv \max \{q \in Q | \mathsf{D}_{v_{-i}}(q) \geq \mathsf{S}(q) \}$.
\end{itemize}

\vspace{\baselineskip} Our terminology for these quantities is justified by the following lemma.

\hypertarget{QuantityLemma}{}
\vspace{\baselineskip} \noindent \textsc{Quantity~Lemma:} Fix an environment and let $v \in V$. For each $(t, W) \in X$, $(t, W)$ is $v$-efficient if and only if (i)~$|W| \in \llbracket \qmin_v, \qmax_v \rrbracket$, and (ii)~there is no loser with a higher valuation than a winner: for each $i \in W$ and each $j \in N \backslash W$, we have $v_i \geq v_j$.

\vspace{\baselineskip} Similarly, we can use the reported demand curve and the known supply curve to determine an interval on the price axis. In particular, recall that in the sense of the second welfare theorem, a competitive equilibrium is an efficient outcome supported by some redistribution and market prices. In general, redistribution might involve a combination of commodities and ownership shares, but in partial equilibrium models such as this one, redistribution of the numeraire suffices to support all efficient outcomes; see Chapter 10.D of \cite{Mas-Colell-Whinston-Green1995}. We therefore consider competitive equilibria that are supported by (i)~a profile of {\it subsidies}, or monetary transfers, with a negative subsidy understood as a tax, and (ii)~a single market {\it price} for objects. To define them, we first introduce the relevant budget sets and demand sets for consumers.

\vspace{\baselineskip} \noindent \textsc{Definition:} Fix an environment. For each $i \in N$, each subsidy $s_i \in \mathbb{R}$, each price $p \in \mathbb{R} \cup \{\infty\}$, and each valuation $v_i \in V_i$, the associated {\it budget set boundary} $B_i(s_i, p) \subseteq X_i$ and {\it demand set} $B^\delta_i(s_i, p | v_i) \subseteq B_i(s_i, p)$ are
\begin{align*}
B_i(s_i, p) &\equiv \left\{
     \begin{array}{lr}
      \{ (s_i, 0), (s_i - p, 1)\}, & p \in \mathbb{R}, \\     
      \{ (s_i, 0) \}, & p = \infty; \text{ and}
     \end{array}
   \right.
\\ B^\delta_i(s_i, p | v_i) & \equiv \argmax_{(t_i, a_i) \in B_i(s_i, p)} [v_i \cdot a_i + t_i].
\end{align*}

\vspace{\baselineskip} In our setting, a simple interpretation of competitive equilibrium is that only the numeraire is redistributed and then ``the market works," with the monopolist behaving as a price taker. That said, certain subsidy profiles allow for the interpretation that only {\it ownership shares} are redistributed, from the monopolist to the consumers, with all consumers receiving the same shareholding.

\vspace{\baselineskip} \noindent \textsc{Definition:} Fix an environment, let $v \in V$, let $(t^*, W^*) \in X$, let $(a^*_i)_{i \in N} \in \{0, 1\}^N$ denote the associated assignment profile, and define $q^* \equiv |W^*|$. For each subsidy profile $s \in \mathbb{R}^N$ and each price $p \in \mathbb{R} \cup \{\infty\}$, we say that $(t^*, W^*)$ is a {\it competitive equilibrium supported by $(s, p)$} if and only if
\begin{itemize}
\item for each $i \in N$, we have $(t^*_i, a^*_i) \in B^\delta_i(s_i, p | v_i)$, and

\item the monopolist selects $q^*$ to maximize profit in response to $p$, or equivalently,

\hspace{5mm} (i)~$p \in \mathbb{R}$ implies $q^* \in \argmax_{q \in \{ q' \in Q | \mathsf{C}(q') < \infty \}} [p \cdot q - \mathsf{C}(q)]$, and

\hspace{5mm} (ii)~$p = \infty$ implies $q^* = \max \{ q' \in Q | \mathsf{C}(q') < \infty \}$.
\end{itemize}
If moreover there is a consumer subsidy $s_N \in \mathbb{R}$ such that for each $i \in N$ we have $s_i = s_N$, then we say that $(t^*, W^*)$ is an {\it equal subsidy equilibrium supported by $(s_N, p)$}. Finally, if moreover there is a consumer shareholding $\textfrak{s:}_N \in [0, \frac{1}{n}]$ such that either (i)~$|W^*| = 0$ and $s_N = 0$, or (ii)~$|W^*| > 0$ and $s_N = \textfrak{s:}_N \cdot (p \cdot |W^*| - \mathsf{C}(|W^*|))$, so that the consumer subsidy is the associated dividend of the profit, then we say that $(t^*, W^*)$ is an {\it equal shareholding equilibrium supported by $(\textfrak{s:}_N, p)$}. We emphasize that if an equal shareholding equilibrium has zero profit, which includes the case of an infinite price with no production, then the equilibrium can be supported by any consumer shareholding.

\vspace{\baselineskip} We can establish the existence of competitive equilibria in our setting by appealing to known results about the {\it assignment game}, in which unit demand buyers match with unit supply sellers and money can be transferred. Indeed, we can analyze each environment and valuation profile using an associated assignment game with $n$ buyers and $\qcap$ sellers whose valuations are respectively summarized by $\mathsf{D}_v$ and $\mathsf{S}$, as the competitive equilibria for the two economies coincide. In the assignment game, the set of competitive prices is nonempty \citep{Koopmans-Beckmann1957} and moreover a lattice \citep{Shapley-Shubik1972}, and because we consider the special case where the objects are identical \citep{Bohm-Bawerk1888}, this lattice is simply an interval.

As with the interval of efficient quantities, we can provide explicit formulas for the endpoints of the interval of competitive prices. In general, a competitive price may lie outside the interval of relevant marginal costs, but these extreme prices are never necessary: (i)~$p<\mathsf{S}(1)$ implies no production, which can also be supported with $p'=\mathsf{S}(1)$, and (ii)~$p>\mathsf{S}(n)$ implies full production, which can also be supported with $p'=\mathsf{S}(n)$. We therefore also introduce {\it rounded} competitive prices for situations where it is useful to avoid these unnecessary extreme prices.

\vspace{\baselineskip} \noindent \textsc{Definition:} Fix an environment. For each $v \in V$, we define (i)~the {\it minimum competitive price given $v$}, $\pmin_v \in \mathbb{R}$, (ii)~the {\it maximum competitive price given $v$}, $\pmax_v \in \mathbb{R} \cup \{\infty\}$, (iii)~the {\it rounded minimum competitive price given $v$}, $\pminup_v \in \mathbb{R} \cup \{\infty\}$, and (iv)~the {\it rounded maximum competitive price given $v$}, $\pmaxdown_v \in \mathbb{R} \cup \{\infty\}$, as follows.
\begin{itemize}
\item $\pmin_v \equiv \max \{ \mathsf{D}_v(\qmax_v+1), \mathsf{S}(\qmax_v)\}$ and $\pmax_v \equiv \min \{ \mathsf{D}_v(\qmax_v), \mathsf{S}(\qmax_v+1)\}$; and

\item $\pminup_v \equiv \max\{\pmin_v, \mathsf{S}(1)\}$ and $\pmaxdown_v \equiv \min \{\pmax_v, \mathsf{S}(n)\}$.
\end{itemize}

\vspace{\baselineskip} Our terminology for these competitive prices will be justified by our next lemma, which also relates them to the following {\it individualized} prices.

\vspace{\baselineskip} \noindent \textsc{Definition:} Fix an environment. For each $i \in N$ and each $v_{-i} \in V_{-i}$, we define the {\it Groves price given $v_{-i}$}, $\pgro_{v_{-i}} \in \mathbb{R} \cup \{\infty\}$, by
\begin{align*}
\pgro_{v_{-i}} \equiv \min \{ \mathsf{D}_{v_{-i}}(\qmax_{v_{-i}}), \mathsf{S}(\qmax_{v_{-i}}+1)\}.
\end{align*}

\vspace{\baselineskip} When the Groves price $\pgro_{v_{-i}}$ is finite, it has the property that $v_i - \pgro_{v_{-i}}$ is the marginal contribution of $i$ to total surplus: it is the difference between the optimal total surplus of a group that includes $i$ and the optimal total surplus of a group that excludes $i$. Intuitively, it has this particular structure in our model because we can always optimally exclude $i$ by selecting the $\qmax_{v_{-i}}$ highest-valuation peers of $i$, but sometimes it is optimal for $i$ to replace one of these peers and other times it is optimal for $i$ to join them. In the general context of Groves mechanisms, which are characterized by efficiency and strategy-proofness \citep{Holmstrom1979}, $\pgro_{v_{-i}}$ is sometimes described as the externality that $i$ imposes on his peers, or the VCG payment, and our terminology reflects its central role in the full class of Groves mechanisms. That said, we emphasize that our next lemma involves only efficiency.

In the specific context of auctions with unit demand and quasi-linear preferences, it is well-known that a VCG mechanism is a {\it minimum competitive price mechanism} (\citealp{Demange1982}; \citealp{Leonard1983}), in which winners {\it pay} and the losers {\it refuse} the {\it minimum} competitive prices. For example, in an auction for one object, the winner pays the second-highest valuation, the losers refuse the second-highest valuation, and a price is competitive if and only if it is between the second-highest valuation and the highest valuation. Rarely emphasized, but important to our analysis, is the observation that the losers also refuse the {\it maximum} competitive prices. For example, in an auction for one object, each loser refuses the highest valuation. In the following lemma, we gather this observation and other useful relationships between competitive prices, rounded competitive prices, and Groves prices.

\hypertarget{PriceLemma}{}
\vspace{\baselineskip} \noindent \textsc{Price~Lemma:} Fix an environment and let $v \in V$. First, for each $p \in \mathbb{R} \cup \{\infty\}$, there are a subsidy profile $s \in \mathbb{R}^N$ and a competitive equilibrium supported by $(s, p)$ if and only if $p \in [\pmin_v, \pmax_v]$. Second, for each $i \in N$, we have either
\begin{itemize}
\item $v_i \geq \mathsf{D}_v(\qmax_v) \geq \pmax_v \geq \pmaxdown_v \geq \pminup_v = \pmin_v = \pgro_{v_{-i}} \geq \mathsf{D}_v(\qmax_v+1)$, or

\item $\mathsf{D}_v(\qmax_v) \geq \pgro_{v_{-i}} = \pmax_v = \pmaxdown_v \geq \pminup_v \geq \pmin_v \geq \mathsf{D}_v(\qmax_v+1) \geq v_i$.
\end{itemize}
Third, for each $(t, W) \in X$ that is $v$-efficient and each $i \in N$, $v_i > \pgro_{v_{-i}}$ implies $i \in W$ and $\pgro_{v_{-i}} > v_i$ implies $i \not \in W$. Finally, if $\mathsf{S}(1) \neq \infty$, then for each $i \in N$,
\begin{align*}
\pgro_{v_{-i}} = v_i - [\max_{W \in \mathbb{W}, i \in W} \mathsf{TS}_v(W) - \max_{W' \in \mathbb{W}, i \not \in W'} \mathsf{TS}_v(W')].
\end{align*}

\vspace{\baselineskip} Our \hyperlink{PriceLemma}{Price~Lemma} reinforces insights from a link between extreme competitive prices and VCG payments in the context of two-sided environments with single-object traders \citep{Delacretaz-Loertscher-Mezzetti2022}: ``\elide the--extremal--Walrasian prices provide the traders with precisely the right incentives to reveal their valuations. The subtle but important twist is that incentive compatible information revelation requires the use of two different Walrasian prices for every object that is traded \elide" 

Finally, it will prove useful to distinguish the interval of Groves prices from the interval of competitive prices, to divide the former into subintervals using the marginal costs, and to observe that whenever a possible Groves price is the common valuation of all consumers, it is also the common Groves price of all consumers.

\vspace{\baselineskip} \noindent \textsc{Definition:} Fix an environment. We define the {\it Groves price interval}, $\mathbb{P} \subseteq \mathbb{R} \cup \{\infty\}$, by
\begin{align*}
\mathbb{P} &= \left\{
     \begin{array}{lr}
       [\mathsf{S}(1), \mathsf{S}(n)] \backslash \{\infty\}, & \mathsf{S}(1) \neq \infty, \\ \relax
       \{ \infty \}, & \mathsf{S}(1) = \infty.
     \end{array}
   \right.
\end{align*}
We define the set of {\it Groves price subintervals}, $\{\mathbb{P}_q \}_{q \in \llbracket 1, \qcap \rrbracket}$, as follows: (i)~for each $q \in \llbracket 1, \qcap -1 \rrbracket$, $\mathbb{P}_q \equiv [\mathsf{S}(q), \mathsf{S}(q+1)]$, (ii)~if $\qcap = n$, then $\mathbb{P}_{\qcap} = \{\mathsf{S}(\qcap)\}$, and (ii)~if $\qcap < n$, then $\mathbb{P}_{\qcap} \equiv [\mathsf{S}(\qcap), \infty)$. Observe that if $\mathsf{S}(1) = \infty$, then there are no Groves price subintervals.

\hypertarget{IntervalLemma}{}
\vspace{\baselineskip} \noindent \textsc{Interval~Lemma:} Fix an environment. First, for each $p \in \mathbb{P}$, if $v \in V$ is such that each consumer has valuation $p$, then for each $i \in N$ we have $\pgro_{v_{-i}} = p$. Second, for each $i \in N$, $\mathbb{P} = \{\pgro_{v_{-i}} | v_{-i} \in V_{-i}\}$. Finally, $\mathsf{S}(1) \neq \infty$ implies $\mathbb{P} = \cup_{q \in \llbracket 1, \qcap \rrbracket} \mathbb{P}_q$.

\hypertarget{Section4.2}{}
\subsection{Incomplete information}

We now turn to our preliminary lemmas about mechanisms, and begin by observing that efficiency, strategy-proofness, and no-envy are all closely related to budget sets. First, by the second welfare theorem, efficiency is equivalent to the requirement that at each valuation profile, the consumers select from personalized budget sets that share a market price, and moreover the monopolist selects a quantity to maximize profit given this market price. Second, strategy-proofness is equivalent to the requirement that at each valuation profile, each consumer selects from a personalized budget set determined by his peers. Finally, no-envy is equivalent to the requirement that at each valuation profile, the consumers select from a common consumer budget set.

\hypertarget{BudgetSetLemma}{}
\vspace{\baselineskip} \noindent \textsc{Budget Set Lemma:} Fix an environment. A mechanism $(\tau, \alpha)$ satisfies
\begin{itemize}
\item[(i)] {\it efficiency} if and only if for each $v \in V$, there is personalized subsidy profile $(s_i(v))_{i \in N} \in \mathbb{R}^N$ and market price $p_{N_0}(v) \in [\pmin_v, \pmax_v]$ such that $(\tau(v), \alpha(v))$ is a competitive equilibrium supported by $((s_i(v))_{i \in N}, p_{N_0}(v))$;

\item[(ii)] {\it strategy-proofness} if and only if for each $i \in N$ and each $v_{-i} \in V_{-i}$, either
\begin{itemize}
\item[$\bullet$] there is personalized bundle $x_i(v_{-i}) \in X_i$ such that for each $v_i \in V_i$, we have $(\tau_i(v), \alpha_i(v)) = x_i(v_{-i})$,~or

\item[$\bullet$] there are personalized subsidy $s_i(v_{-i}) \in \mathbb{R}$ and personalized price $p_i(v_{-i}) \in \mathbb{R}$ such that for each $v_i \in V_i$, we have $(\tau_i(v), \alpha_i(v)) \in B^\delta_i(s_i(v_{-i}), p_i(v_{-i}) | v_i)$; and
\end{itemize}

\item[(iii)] {\it no-envy} if and only if for each $v \in V$, there are consumer subsidy $s_N(v) \in \mathbb{R}$ and consumer price $p_N(v) \in \mathbb{R}$ such that for each $i \in N$, we have $(\tau_i(v), \alpha_i(v)) \in B^\delta_i(s_N(v), p_N(v) | v_i)$.
\end{itemize}

\vspace{\baselineskip} Though it is not obvious from inspection that any mechanism simultaneously satisfies all three requirements, it is well-known that in fact the VCG mechanisms do, and that they moreover satisfy voluntary participation. Our primary contribution is to show that they are not alone.

The VCG mechanisms are certainly not alone if we only require the first two axioms: by a superficial modification of \cite{Holmstrom1979} to allow for cost functions, the class of mechanisms satisfying efficiency and strategy-proofness is precisely the class of Groves mechanisms. In the usual definition, each consumer's transfer is the sum of (i)~the reported peer surplus at the selected decision, and (ii)~a bonus determined entirely by his peers. For our purposes, it is useful to work with an equivalent definition: each consumer selects his favorite bundle from his {\it option set}, or collection of attainable bundles given his peers' reports, and this is a budget set whose subsidy is determined entirely by his peers and whose price is the Groves price.

\vspace{\baselineskip} \noindent \textsc{Definition:} Fix an environment. A mechanism $(\tau, \alpha)$ is a {\it Groves mechanism} if and only if (i)~$\alpha$ is surplus-maximizing, and (ii)~for each $i \in N$, there is {\it (peer) bonus function} $\beta_i: V_{-i} \to \mathbb{R}$ such that for each $v \in V$, we have
\begin{align*}
\tau_i(v) = \big( [\max_{W \subseteq N} \mathsf{TS}_v(W)] - \alpha_i(v) v_i \big) + \beta_i(v_{-i}).
\end{align*}

\hypertarget{GrovesLemma}{}
\vspace{\baselineskip} \noindent \textsc{Groves Lemma:} Fix an environment and let $(\tau, \alpha)$ be an {\it efficient} mechanism. The following are equivalent:
\begin{itemize}
\item[(i)] $(\tau, \alpha)$ is {\it strategy-proof};

\item[(ii)] $(\tau, \alpha)$ is a Groves mechanism; and

\item[(iii)] for each $i \in N$, there is a subsidy function $\sigma_i:V_{-i} \to \mathbb{R}$ such that for each $v \in V$, we have $(\tau_i(v), \alpha_i(v)) \in B^\delta_i(\sigma_i(v_{-i}), \pgro_{v_{-i}} | v_i)$.
\end{itemize}

\vspace{\baselineskip} We define the VCG mechanisms as Groves mechanisms for which each consumer's transfer is precisely the difference between (i)~the reported peer surplus at the selected decision, and (ii)~the optimal reported peer surplus {\it provided that $i$ must be excluded}. This exclusion provision is a twist on the usual definition that protects a loser from paying when there are negative costs, but has no bite in the usual scenario where costs are non-negative; see the comparison of standard pivot mechanisms and exclusion pivot mechanisms in \cite{Mackenzie-Trudeau2023}. Again, there is a useful equivalent definition involving option sets: the VCG mechanisms are the Groves mechanisms for which each option set is a budget whose subsidy is {\it zero} and whose price is the Groves price. Finally, there is a third equivalent definition: a VCG mechanism is a minimum competitive price mechanism (\citealp{Demange1982}; \citealp{Leonard1983}).

\vspace{\baselineskip} \noindent \textsc{Definition:} Fix an environment. A Groves mechanism $(\tau, \alpha)$ is moreover a {\it VCG mechanism} if and only if for each $i \in N$ and each $v \in V$, we have
\begin{align*}
\tau_i(v) = \big( [\max_{W \subseteq N} \mathsf{TS}_v(W)] - \alpha_i(v) v_i \big) - [ \max_{W \subseteq N \backslash \{i\}} \mathsf{TS}_{(0, v_{-i})}(W) ].
\end{align*}

\hypertarget{VCGLemma}{}
\vspace{\baselineskip} \noindent \textsc{VCG Lemma:}  Fix an environment and let $(\tau, \alpha)$ be a Groves mechanism. The following are equivalent:
\begin{itemize}
\item[(i)] $(\tau, \alpha)$ is a VCG mechanism;

\item[(ii)] for each $i \in N$ and each $v \in V$, $(\tau_i(v), \alpha_i(v)) \in B^\delta_i(0, \pgro_{v_{-i}}| v_i)$; and

\item[(iii)] for each $i \in N$ and each $v \in V$, $(\tau_i(v), \alpha_i(v)) \in B^\delta_i(0, \pmin_v | v_i)$.
\end{itemize}

\hypertarget{Section5}{}
\section{Main results}

Before we proceed, we remark that the proofs of results in \hyperlink{Section5.1}{Section~5.1}, \hyperlink{Section5.2}{Section~5.2}, \hyperlink{Section5.3}{Section~5.3}, and \hyperlink{Section5.4}{Section~5.4} can be found, respectively, in \hyperlink{AppendixB}{Appendix~B}, \hyperlink{AppendixC}{Appendix~C}, \hyperlink{AppendixD}{Appendix~D}, and \hyperlink{AppendixE}{Appendix~E}.

\hypertarget{Section5.1}{}
\subsection{Endogenous subsidy auctions}

It is easy to see that the VCG mechanisms are not the only envy-free Groves mechanisms. Indeed, we can modify any VCG mechanism by always transferring an exogenous subsidy to each consumer from the monopolist, and the quasilinearity of preferences ensures that all axioms are preserved. This is not particularly appealing, however, because the only exogenous subsidy that guarantees ex-post voluntary participation for all parties is zero.

A more appealing direction would be to design subsidies that preserve all axioms and yet vary with the auction's outcome, and the simplest approach would be to assign an exogenous share of the auction's profit to each consumer. Unfortunately, however, this simple approach does not preserve strategy-proofness. For example, in an auction for one object, if we modify the Vickrey auction by assigning a positive share of the profit to each consumer, then the profit is the second highest bid; thus for any consumer who is convinced that the highest bid will be above his own valuation, there is an incentive to misreport.

That said, it turns out that it is possible to design an endogenous subsidy that preserves all axioms. The construction is rather delicate: the subsidy to each consumer must be determined by his peers, it must moreover vary only with the Groves price, and it must moreover vary continuously with the Groves price in a particular way. All of this structure is driven by our key lemma.

\hypertarget{InvarianceLemma}{}
\vspace{\baselineskip} \noindent \textsc{Invariance~Lemma:} Fix an environment. If a mechanism $(\tau, \alpha)$ satisfies {\it efficiency}, {\it strategy-proofness}, and {\it no-envy}, then for each $i \in N$, there is a subsidy function $\sigma_i: V_{-i} \to \mathbb{R}$ such that

(i)~for each $v \in V$, $(\tau_i(v), \alpha_i(v)) \in B^\delta_i(\sigma_i(v_{-i}), \pgro_{v_{-i}} | v_i)$, and

(ii)~for each pair $v_{-i}, v_{-i}' \in V_{-i}$, $\pgro_{v_{-i}} = \pgro_{v_{-i}'}$ implies $\sigma_i(v_{-i}) = \sigma_i(v_{-i}')$.

\vspace{\baselineskip} For this particular proof, we provide a proof sketch in \hyperlink{Section6}{Section~6}. As a result of our key lemma, under these three axioms, the subsidy {\it function} of a consumer $i$, $\sigma_i: V_{-i} \to \mathbb{R}$, is determined entirely by an associated function that maps Groves prices to subsidies. In fact, there is one such function shared by all consumers, and it has some structure.

\vspace{\baselineskip} \noindent \textsc{Definition:} Fix an environment. A {\it subsidy curve} is a function $\varsigma: \mathbb{P} \to \mathbb{R}$ such that for each pair $p, p' \in \mathbb{P}$ such that $p' > p$, we have $\frac{\varsigma(p') - \varsigma(p)}{p'-p} \in [0,1]$. The associated {\it extended subsidy curve}, $\varsigma^\leftrightarrow$, is defined in two cases.
\begin{itemize}
\item If $\mathsf{S}(1) \neq \infty$, then $\varsigma^\leftrightarrow:\mathbb{R} \to \mathbb{R}$ is the function such that for each $p \in \mathbb{R}$, (i)~$p < \mathsf{S}(1)$ implies $\varsigma^\leftrightarrow(p) = \varsigma(\mathsf{S}(1))$, (ii)~$p > \mathsf{S}(n)$ implies $\varsigma^\leftrightarrow(p) = \varsigma(\mathsf{S}(n))$, and (iii)~otherwise $\varsigma^\leftrightarrow(p) = \varsigma(p)$. In this case, $\varsigma^\leftrightarrow$ extends $\varsigma$ from $\mathbb{P} \subseteq \mathbb{R}$ to $\mathbb{R}$.

\item If $\mathsf{S}(1) = \infty$, then $\varsigma^{\leftrightarrow}: \mathbb{R} \cup \{\infty\} \to \mathbb{R}$ is the function such that for each $p \in \mathbb{R} \cup \{\infty\}$, $\varsigma^\leftrightarrow(p) = \varsigma(\infty)$. In this case, $\varsigma^\leftrightarrow$ extends $\varsigma$ from $\mathbb{P} = \{\infty\}$ to $\mathbb{R} \cup \{\infty\}$.
\end{itemize}

\vspace{\baselineskip} In an {\it endogenous subsidy auction}, a single subsidy curve is used to determine the subsidies for all consumers. More precisely, for each surplus-maximizing assignment policy~$\alpha$ and each subsidy curve $\varsigma$, the associated endogenous subsidy auction can be described as follows: (i)~take the outcome of the VCG mechanism associated with $\alpha$, and then (ii)~transfer to each consumer the subsidy that $\varsigma$ assigns to his personal Groves price. In the special case that the subsidy curve is flat at zero, this is simply a VCG mechanism.

\vspace{\baselineskip} \noindent \textsc{Definition:} Fix an environment. A mechanism $(\tau, \alpha)$ is an {\it endogenous subsidy auction} if and only if (i)~$\alpha$ is surplus-maximizing, and (ii)~there is a subsidy curve $\varsigma$ such that for each $i \in N$ and each $v \in V$,
\begin{align*}
\tau_i(v) = \big( [\max_{W \subseteq N} \mathsf{TS}_v(W)] - \alpha_i(v) v_i \big) - [ \max_{W \subseteq N \backslash \{i\}} \mathsf{TS}_{(0, v_{-i})}(W) ] + \varsigma(\pgro_{v_{-i}}).
\end{align*}
In this case, we also say that $(\tau, \alpha)$ is {\it supported by $\varsigma$}.

\vspace{\baselineskip} Endogenous subsidy auctions, just like VCG mechanisms, have two alternative definitions. First, each endogenous subsidy auction has the property that each consumer $i$ always selects from the personalized budget set whose subsidy is $\varsigma(\pgro_{v_{-i}})$ and whose price is $\pgro_{v_{-i}}$. Second, each endogenous subsidy auction has the property that each consumer always selects from the common budget set whose subsidy is $\varsigma(\pmaxdown_v)$ and whose price is $\pmin_v + [\varsigma(\pmaxdown_v) - \varsigma(\pminup_v)]$, which can alternatively be written without the rounded prices by using the extended subsidy curve. In the special case that the subsidy curve is flat at zero, this is simply the \hyperlink{VCGLemma}{VCG~Lemma}.

\hypertarget{Theorem1}{}
\vspace{\baselineskip} \noindent \textsc{Theorem 1:} Fix an environment, let $(\tau, \alpha)$ be a Groves mechanism, and let $\varsigma$ be a subsidy curve. The following are equivalent:
\begin{itemize}
\item[(i)] $(\tau, \alpha)$ is an endogenous subsidy auction supported by $\varsigma$;

\item[(ii)] for each $i \in N$ and each $v \in V$, $(\tau_i(v), \alpha_i(v)) \in B^\delta_i(\varsigma(\pgro_{v_{-i}}), \pgro_{v_{-i}}| v_i)$; and

\item[(iii)] for each $i \in N$ and each $v \in V$,
\begin{align*}
(\tau_i(v), \alpha_i(v)) &\in B^\delta_i(\varsigma(\pmaxdown_v), \pmin_v + [\varsigma(\pmaxdown_v) - \varsigma(\pminup_v)] | v_i)
\\ &=B^\delta_i(\varsigma^\leftrightarrow(\pmax_v), \pmin_v + [\varsigma^\leftrightarrow(\pmax_v) - \varsigma^\leftrightarrow(\pmin_v)] | v_i).
\end{align*}
\end{itemize}

\vspace{\baselineskip} It is clear from the third definition that endogenous subsidy auctions are {\it envy-free}. In fact, these are the {\it only} envy-free Groves mechanisms, and moreover the price of the common budget set for the consumers can also serve as a market price for the monopolist.

\hypertarget{Theorem2}{}
\vspace{\baselineskip} \noindent \textsc{Theorem 2:} Fix an environment. A mechanism satisfies {\it efficiency}, {\it strategy-proofness}, and {\it no-envy} if and only if it is an endogenous subsidy auction. In this case, the outcome selected at each valuation profile is an equal subsidy equilibrium.

\vspace{\baselineskip} The theorem is tight, in the sense that it is false if any of the axioms is dropped. Indeed, this is true even when selling one object without production: (i)~the constant auction with no winners and no transfers violates only efficiency, (ii)~an `exogenous shareholding auction,' such as VCG modified to exogenously divide the auction's profit equally among the consumers, violates only strategy-proofness, and (iii)~the Bailey-Cavallo mechanism violates only no-envy.

We remark that in the more general model of costly inclusion, which allows for any cost function $\mathsf{C}: 2^N \to \mathbb{R} \cup \{\infty\}$ with $\mathsf{C}(\emptyset) = 0$ and therefore does not require cost to only depend on the number of winners, it was previously shown that the three axioms in \hyperlink{Theorem2}{Theorem~2} are compatible if and only if the cost function is symmetric and convex, which is this paper's model \citep{Mackenzie-Trudeau2023}. By combining these results, we therefore have a complete characterization of the envy-free Groves mechanisms in the more general model.

\hypertarget{Section5.2}{}
\subsection{Endogenous shareholding auctions}

In general, an endogenous subsidy auction need not be voluntary: negative subsidies tax consumers, while excessively large subsidies tax the monopolist. In order to be voluntary, an endogenous subsidy auction must only select equal subsidy equilibria that are moreover equal shareholding equilibria: the consumer subsidies must be individually non-negative and collectively funded out of the auction's profit. This is guaranteed when the subsidy curve satisfies some additional conditions involving the {\it average profit curve}.

\vspace{\baselineskip} \noindent \textsc{Definition:} Fix an environment. The {\it average profit curve} is the mapping $\piavg: \mathbb{P} \to \mathbb{R}$ such that for each $p \in \mathbb{P}$,
\begin{align*}
\piavg(p) &\equiv \left\{
     \begin{array}{lr}
       \tfrac{1}{n} \cdot \max_{q \in \{q' \in Q | \mathsf{C}(q') < \infty \}} [q \cdot p - \mathsf{C}(q) ], & \mathsf{S}(1) \neq \infty \text{ and } p \in \mathbb{R}, \\ \relax
       0, & \mathsf{S}(1) = \infty \text{ and } p = \infty.
     \end{array}
   \right.
\end{align*}
By definition of $\mathbb{P}$, the two cases above are exhaustive. Equivalently,\footnote{We omit the simple proof, which is straightforward using the monotonicity of $\mathsf{S}$.} (i)~$\mathsf{S}(1) \neq \infty$ implies for each $q \in \llbracket 1, \qcap \rrbracket$ and each $p \in \mathbb{P}_q$, we have $\piavg(p) = \frac{q \cdot p - \mathsf{C}(q)}{n}$, and (ii)~$\mathsf{S}(1) = \infty$ implies $\piavg$ only assigns zero.

\vspace{\baselineskip} When a price-taking firm faces price $p$ and maximizes profit with the understanding that it will sell everything it produces, $\piavg(p)$ specifies the maximum subsidy that could be distributed to each consumer without leaving the firm worse off than shutting down.

In order for an endogenous subsidy auction to be voluntary, it is necessary for the subsidy curve to always fall between the horizontal axis and the average profit curve. This is not sufficient, however, because an outcome may involve both a small Groves price for winners and a high Groves price---and thus a high subsidy---for losers. To illustrate this point, suppose there are two consumers, suppose the supply curve is given by $(\mathsf{S}(1), \mathsf{S}(2)) = (1, 2)$, and consider an endogenous subsidy auction whose supporting subsidy curve is the average profit curve itself. At the profile $(v_1, v_2) = (3, 0)$, the first consumer wins with Groves price $1$ and thus receives transfer $-1$, the second consumer loses with Groves price $2$ and thus receives transfer~$\frac{1}{2}$, and thus the net payment from the consumers to the monopolist is $\frac{1}{2}$, but the monopolist is responsible for paying the production cost of $1$, so we have a violation of producer voluntary participation.

It is easy to see that the only constraint for consumer voluntary participation is that the subsidy curve is never negative, and as the previous example suggests, the tightest constraints for producer voluntary participation concern price pairs with a large gap that can simultaneously occur as the winner Groves price and loser Groves price.

\vspace{\baselineskip} \noindent \textsc{Definition:} Fix an environment. A subsidy curve $\varsigma$ is {\it funded} if and only if
\begin{itemize}
\item[$\text{[}\mathcal{F}_1\text{]}$] $\varsigma(\mathsf{S}(1)) = 0$;

\item[$\text{[}\mathcal{F}_2\text{]}$] for each $q \in \llbracket 1, \qcap-1 \rrbracket$,
\begin{align*}
\big[ \tfrac{q}{n} \big] \cdot \varsigma(\mathsf{S}(q)) + \big[ 1 - \tfrac{q}{n} \big] \cdot \varsigma(\mathsf{S}(q+1)) \leq \piavg(\mathsf{S}(q)); \text{ and}
\end{align*}

\item[$\text{[}\mathcal{F}_3\text{]}$] if $\qcap \in \llbracket 1, n-1 \rrbracket$, then for each $p \in (\mathsf{S}(\qcap), \infty)$,
\begin{align*}
\big[ \tfrac{\qcap}{n} \big] \cdot \varsigma(\mathsf{S}(\qcap)) + \big[ 1 - \tfrac{\qcap}{n} \big] \cdot \varsigma(p) \leq \piavg(\mathsf{S}(\qcap)).
\end{align*}
\end{itemize}
For each subsidy curve $\varsigma$, we say that an endogenous subsidy auction supported by $\varsigma$ is moreover an {\it endogenous shareholding auction} if and only if $\varsigma$ is funded.

\vspace{\baselineskip} Clearly, a funded subsidy curve never falls below the horizontal axis. By the following lemma, a funded subsidy curve never falls above the average profit curve, and therefore satisfies the necessary condition from our discussion.

\hypertarget{CeilingLemma}{}
\vspace{\baselineskip} \noindent \textsc{Ceiling~Lemma:} Fix an environment. If $\varsigma$ is a funded subsidy curve, then for each $p \in \mathbb{P}$ we have $\varsigma(p) \leq \piavg(p)$.

\vspace{\baselineskip} By the following theorem, the funded subsidy curve conditions are stronger than the necessary condition from our discussion: they are both necessary and sufficient for voluntary participation. In this case, the auction selects equal shareholding equilibria, and thus it endogenously determines ownership shares for its own profit.

\hypertarget{Theorem3}{}
\vspace{\baselineskip} \noindent \textsc{Theorem 3:} Fix an environment. A mechanism satisfies {\it efficiency}, {\it strategy-proofness}, {\it no-envy}, and {\it voluntary participation} if and only if it is an endogenous shareholding auction. In this case, the outcome selected at each valuation profile is an equal shareholding equilibrium.

\vspace{\baselineskip} Many characterizations of the VCG mechanisms are available as corollaries, for example on the basis of strategic simplicity, privacy, and preserving all properties on larger preference domains; we omit the details. Instead, the rest of our results concern endogenous shareholding auctions that are in some sense optimal. From this perspective, we do provide one characterization of the VCG mechanisms: they are unambiguously the monopolist's favorites. For this reason, a monopolist has good reason to advocate for the importance of strategic simplicity, privacy, and preserving all properties on larger preference domains.

\hypertarget{Section5.3}{}
\subsection{Prior-free optimization}

There are many endogenous shareholding auctions, and there are a variety of regulatory objectives that might guide the selection of an auction from this class. In this section, we take a prior-free approach to describe those that are unambiguously the most producer-friendly and the most consumer-friendly.

To begin, as discussed in the introduction, the literature has shown that the VCG mechanisms---or more generally when preferences need not be quasi-linear, the minimum price Walrasian mechanisms---are optimal for the seller in other settings (\citealp{Krishna-Perry2000}; \citealp{Kazumura-Mishra-Serizawa2020}; \citealp{Sakai-Serizawa2023}). The following result reinforces this message.

\vspace{\baselineskip} \noindent \textsc{Definition:}\footnote{The meanings of weak dominance, dominance, and strict dominance are not consistent across the literature. In this article, whenever we compare functions $f$ and $g$ on some basis, $f$ {\it weakly dominates} $g$ if $f$ is always at least as good as $g$, $f$ {\it dominates} $g$ if $f$ weakly dominates $g$ and moreover $f$ is sometimes better than $g$, and we require no terminology for when $f$ is always better than $g$.} Fix an environment. Let $(\tau^*, \alpha^*)$ and $(\tau, \alpha)$ be mechanisms. We say that $(\tau^*, \alpha^*)$ {\it weakly producer-dominates} $(\tau, \alpha)$ if and only if for each $v \in V$, $\mathsf{PS}(\tau^*(v), \alpha^*(v)) \geq \mathsf{PS}(\tau(v), \alpha(v))$. We say that an endogenous shareholding auction is {\it weakly producer-dominant} if and only if it weakly producer-dominates all endogenous shareholding auctions.

\hypertarget{Theorem4}{}
\vspace{\baselineskip} \noindent \textsc{Theorem 4:} Fix an environment. An endogenous shareholding auction is weakly producer-dominant if and only if it is a VCG mechanism.

\vspace{\baselineskip} The intuition is simple: within the class of endogenous shareholding auctions, the consumer shareholding is always non-negative, and thus the VCG mechanisms are optimal for the monopolist because they always set the consumer shareholding to zero. By contrast, in general no endogenous shareholding auction is dominant for consumers, so we instead describe those that are undominated for consumers. In doing so, we follow several previous contributions (\citealp{Guo-Markakis-Apt-Conitzer2013}; \citealp{Athanasiou-Valletta2021}; \citealp{Borgers-Li-Wang2025}; \citealp{Mishra-Patil2025}); see Footnote~9 for details.

\vspace{\baselineskip} \noindent \textsc{Definition:} Fix an environment. Let $(\tau^*, \alpha^*)$ and $(\tau, \alpha)$ be mechanisms. We say that $(\tau^*, \alpha^*)$ {\it consumer-dominates} $(\tau, \alpha)$ if and only if
\begin{itemize}
\item for each $v \in V$ and each $i \in N$, $v_i \cdot \alpha^*_i(v) + \tau^*_i(v) \geq v_i \cdot \alpha_i(v) + \tau_i(v)$, and

\item there are $v \in V$ and $i \in N$ such that $v_i \cdot \alpha^*_i(v) + \tau^*_i(v) > v_i \cdot \alpha_i(v) + \tau_i(v)$.
\end{itemize}
We say that an endogenous shareholding auction is {\it consumer-optimal} if and only if it is not consumer-dominated by another endogenous shareholding auction.

\vspace{\baselineskip} The description of the consumer-optimal mechanisms is greatly facilitated by the simple observation that comparing two endogenous subsidy auctions is equivalent to comparing their subsidy curves.

\vspace{\baselineskip} \noindent \textsc{Definition:} Fix an environment. Let $\varsigma^*: \mathbb{P} \to \mathbb{R}$ and $\varsigma: \mathbb{P} \to \mathbb{R}$ be subsidy curves. We say that $\varsigma^*$ {\it dominates} $\varsigma$ if and only if (i)~for each $p \in \mathbb{P}$, $\varsigma^*(p) \geq \varsigma(p)$, and (ii)~there is $p \in \mathbb{P}$ such that $\varsigma^*(p) > \varsigma(p)$.

\hypertarget{DominationLemma}{}
\vspace{\baselineskip} \noindent \textsc{Domination~Lemma:} Fix an environment. If $\varsigma^*$ and $\varsigma$ are subsidy curves, then for each pair of endogenous subsidy auctions $(\tau^*, \alpha^*)$ and $(\tau, \alpha)$ such that $(\tau^*, \alpha^*)$ is supported by $\varsigma^*$ and $(\tau, \alpha)$ is supported by $\varsigma$, we have that $(\tau^*, \alpha^*)$ consumer-dominates $(\tau, \alpha)$ if and only if $\varsigma^*$ dominates $\varsigma$.

\vspace{\baselineskip} The consumer-optimal endogenous shareholding auctions satisfy four constraints. The first three require that we draw the graph of the subsidy curve by (i)~starting with zero subsidy at the first marginal cost, (ii)~within each Groves sub-interval, increasing with slope one until a cutoff price and then remaining flat thereafter, and (iii)~never crossing above the average profit curve. The final constraint states that an unnecessarily early cutoff in one Groves subinterval can only be justified by hitting the ceiling of the average profit curve in a future Groves subinterval, with the subsidy curve flat until that future Groves subinterval.

\vspace{\baselineskip} \noindent \textsc{Definition:} Fix an environment. A subsidy curve $\varsigma$ is {\it valvular} if and only if there is a vector of cutoff prices $(\kappa_q)_{q \in \llbracket 1, \qcap \rrbracket}$ such that
\begin{itemize}
\item[$\text{[}\mathcal{V}_1\text{]}$] $\varsigma(\mathsf{S}(1)) = 0$;

\item[$\text{[}\mathcal{V}_2\text{]}$] for each $q \in \llbracket 1, \qcap \rrbracket$, we have $\kappa_q \in \mathbb{P}_q$, and for each $p \in \mathbb{P}_q$,

\hspace{5mm} (i)~$p \leq \kappa_q$ implies $\varsigma(p) = \varsigma(\mathsf{S}(q)) + (p - \mathsf{S}(q))$, and

\hspace{5mm} (ii)~$p \geq \kappa_q$ implies  $\varsigma(p) = \varsigma(\kappa_q)$;

\item[$\text{[}\mathcal{V}_3\text{]}$] for each $q \in \llbracket 1, \qcap \rrbracket$, $\varsigma(\kappa_q) \leq \piavg(\kappa_q)$; and

\item[$\text{[}\mathcal{V}_4\text{]}$] for each $q \in \llbracket 1, \qcap \rrbracket$ such that $\kappa_q < \sup \mathbb{P}_q$ and $\varsigma(\kappa_q) < \piavg(\kappa_q)$, there is $q^* \in \llbracket 1, \qcap \rrbracket$ such that (i)~$q^* > q$, (ii)~$\varsigma(\kappa_{q^*}) = \piavg(\kappa_{q^*})$, and (iii)~$\varsigma(\mathsf{S}(q^*)) = \varsigma(\kappa_q)$.
\end{itemize}
For each subsidy curve $\varsigma$, we say that an endogenous subsidy auction supported by $\varsigma$ is moreover a {\it valvular auction} if and only if $\varsigma$ is valvular.

\vspace{\baselineskip} The term {\it valvular auction} is meant to suggest a valve-based water system whose water represents the auction's profit, as discussed in \hyperlink{Section2}{Section~2}. The next theorem states that the valvular auctions are precisely the consumer-optimal endogenous shareholding auctions.

\hypertarget{Theorem5}{}
\vspace{\baselineskip} \noindent \textsc{Theorem 5:} Fix an environment. An endogenous shareholding auction is consumer-optimal if and only if it is a valvular auction.

\vspace{\baselineskip} It follows that when all marginal costs are finite, all endogenous shareholding auctions supported by the myopic subsidy curve, described in \hyperlink{Section2}{Section~2} and illustrated in \hyperlink{Figure1}{Figure~1}, are consumer-optimal. It is straightforward to extend that section's description to the case that some marginal costs are infinite so that the preceding statements still holds; we omit the details.

\hypertarget{Section5.4}{}
\subsection{Subjective expected welfare}

We conclude by modeling a regulator selecting an endogenous shareholding auction as an agent facing a decision under uncertainty who is rational in the sense of \cite{Savage1954}. In this model, (i)~there is a set of uncertain {\it states} with an associated $\sigma$-algebra of {\it events}, (ii)~there is a set of {\it consequences} with an associated $\sigma$-algebra, and (iii)~the agent ranks {\it acts}, or actions with uncertain consequences, modeled as measurable functions from states to consequences. For the first piece, we declare the set of uncertain states to be $V$ and declare the associated $\sigma$-algebra to be the usual Borel $\sigma$-algebra.

\vspace{\baselineskip} \noindent \textsc{Definition:} Fix an environment. Let $\mathcal{B}(V) \subseteq 2^V$ denote the usual Borel $\sigma$-algebra of $V = \mathbb{R}^N$. A {\it prior} is a countably additive probability measure $\mu: \mathcal{B}(V) \to [0, 1]$. We say that $\mu$ has {\it compact support} if when $V$ is endowed with its Euclidean topology, there is compact $V^\mathsf{c} \subseteq V$ such that $\mu(V \backslash V^\mathsf{c}) = 0$.

\vspace{\baselineskip} For the second piece, observe that an outcome in $X$ may be efficient, envy-free, and voluntary at one valuation profile, and yet none of these at another valuation profile. For example, consider the problem of selling one object to two consumers, and consider the outcome where consumer $1$ wins and pays $10$ while consumer $2$ loses and pays nothing: this satisfies all desiderata at $(v_1, v_2) = (20, 5)$, but satisfies none at $(v'_1, v'_2) = (5, 20)$.

For this reason, it is tempting to declare the set of consequences to be $V \times X$, then associate each mechanism with the act $v \mapsto (v, (\tau(v), \alpha(v))$. Unfortunately, with this specification, mechanisms need not be associated with measurable functions, because tie-breaking may be done arbitrarily. Fortunately, however, due to the rich mathematical structure of the endogenous shareholding auctions, we are able to avoid this technical issue by asserting that the regulator is {\it welfarist}, in the sense that the set of consequences is the set of {\it utility profiles} with its Borel $\sigma$-algebra.

\vspace{\baselineskip} \noindent \textsc{Definition:} Fix an environment. The set of {\it utility profiles} is $\mathbb{U} \equiv \mathbb{R}^{N \cup \{i_0\}}$. A {\it social welfare function} is a function $\mathcal{W}: \mathbb{U} \to \mathbb{R}$, and is {\it continuous} if it is a continuous function when both $\mathbb{U}$ and $\mathbb{R}$ are endowed with their Euclidean topologies.

\vspace{\baselineskip} If the regulator ranks all acts including those that are not associated with feasible mechanisms, and if this ranking satisfies suitable variants of the axioms of \cite{Savage1954}, then the regulator's preferences over acts can be represented by {\it subjective expected welfare} with respect to (i)~a (countably additive) prior with compact support representing the regulator's subjective beliefs about consumer preferences, and (ii)~a continuous social welfare function.\footnote{In particular, we differ from Savage in five ways: (i)~we do not require the $\sigma$-algebra of events to be a power set, (ii)~we require the prior to be countably additive, (iii)~we do not require the social welfare function to be bounded, (iv)~we require the prior to have compact support, and (v)~we require the social welfare function to be continuous. For an appropriate variant of Savage's theorem that addresses the first three differences, see Theorem~2.17 and Proposition~4.4 of \cite{Wakker1993}. The last two differences are easily translated into additional assumptions about preferences.} Here, the social welfare function, defined on the set of utility profiles, plays the role of the utility function in subjective expected utility; we use the term {\it subjective expected welfare} instead of {\it subjective expected utility} to avoid confusion.

\vspace{\baselineskip} \noindent \textsc{Definition:} Fix an environment. Let $\mathbb{A} \subseteq (\mathbb{R}^N)^V \times \mathbb{W}^V$ denote the set of endogenous shareholding auctions. We define the {\it auction-profile summary}, $\mathcal{U}^{\mathbb{A} \times V}: \mathbb{A} \times V \to \mathbb{U}$, to be the function such that for each $((\tau, \alpha), v) \in \mathbb{A} \times V$,
\begin{itemize}
\item $i \in N$ implies $[\mathcal{U}^{\mathbb{A} \times V}((\tau, \alpha), v)]_i = v_i \cdot \alpha_i(v) + \tau_i(v)$, and

\item $[\mathcal{U}^{\mathbb{A} \times V}((\tau, \alpha), v)]_{i_0} = [-\sum_{i \in N} \tau_i(v)] - C(|\alpha(v)|)$.
\end{itemize}
For each prior $\mu$ and each continuous social welfare function $\mathcal{W}$, we define the {\it $\mu$-expected $\mathcal{W}$-welfare functional}, $\mathbb{E}_\mu \mathcal{W}: \mathbb{A} \to \mathbb{R} \cup \{-\infty, \infty\}$, using the Lebesgue integral as follows: for each $(\tau, \alpha) \in \mathbb{A}$,
\begin{align*}
\mathbb{E}_\mu \mathcal{W}(\tau, \alpha) \equiv \int_{v \in V} \mathcal{W} \circ \mathcal{U}^{\mathbb{A} \times V}((\tau, \alpha), v) d\mu.
\end{align*}
We say that $(\tau, \alpha)$ is {\it $(\mu, \mathcal{W})$-optimal} if maximizes $\mathbb{E}_\mu \mathcal{W}$ across $\mathbb{A}$.

\vspace{\baselineskip} It turns out that the tie-breaking details do not matter: all of the relevant welfare information of an endogenous shareholding auction is contained in its funded subsidy curve.

\vspace{\baselineskip} \noindent \textsc{Definition:} Fix an environment. Let $\mathbb{S}$ denote the collection of funded subsidy curves. If $\mathsf{S}(1) \neq \infty$, then we define the {\it curve-profile summary}, $\mathcal{U}^{\mathbb{S} \times V}: \mathbb{S} \times V \to \mathbb{U}$, to be the function such that for each $(\varsigma, v) \in \mathbb{S} \times V$,
\begin{itemize}
\item $i \in N$ implies $[\mathcal{U}^{\mathbb{S} \times V}(\varsigma, v)]_i = \max\{v_i - \pgro_{v_{-i}}, 0\} + \varsigma(\pgro_{v_{-i}})$, and

\item $[\mathcal{U}^{\mathbb{S} \times V}(\varsigma, v)]_{i_0} = \max_{W \subseteq N} [\sum_{i \in W} v_i - C(|W|)] - \sum_{i \in N} [ \max\{v_i - \pgro_{v_{-i}}, 0\} + \varsigma(\pgro_{v_{-i}}) ]$.
\end{itemize}
By the \hyperlink{SummaryLemma}{Summary~Lemma} below, $\mathcal{U}^{\mathbb{S} \times V}(\varsigma, v)$ is the realized utility profile at $v$ for any endogenous shareholding auction supported by $\varsigma$.

\vspace{\baselineskip} This is convenient, because the collection of funded subsidy curves has an associated topology that makes it both metrizable and compact.

\hypertarget{CompactnessLemma}{}
\vspace{\baselineskip} \noindent \textsc{Compactness Lemma:} Fix an environment. When $\mathbb{S}$ is endowed with the topology of compact convergence, it is both metrizable and compact.

\vspace{\baselineskip} Moreover, with this same topology, the function that maps funded subsidy curves and valuation profiles to utility summaries is a continuous function.

\hypertarget{SummaryLemma}{}
\vspace{\baselineskip} \noindent \textsc{Summary Lemma:} Fix an environment such that $\mathsf{S}(1) \neq \infty$. For each $\varsigma \in \mathbb{S}$, each $(\tau, \alpha) \in \mathbb{A}$ that is supported by $\varsigma$, and each $v \in V$, $\mathcal{U}^{\mathbb{A} \times V}((\tau, \alpha), v) = \mathcal{U}^{\mathbb{S} \times V}(\varsigma, v)$. Moreover, the function $\mathcal{U}^{\mathbb{S} \times V}: \mathbb{S} \times V \to \mathbb{U}$ is continuous when (i)~$\mathbb{S}$ is endowed with its compact convergence topology, (ii)~$V$ is endowed with its Euclidean topology, (iii)~$\mathbb{S} \times V$ is endowed with the associated product topology, and (iv)~$\mathbb{U}$ is endowed with its Euclidean topology.

\vspace{\baselineskip} These lemmas can be used to establish that there is an optimal endogenous shareholding auction for the regulator according to subjective expected welfare.

\hypertarget{Theorem6}{}
\vspace{\baselineskip} \noindent \textsc{Theorem 6:} Fix an environment. For each compact support prior $\mu$ and each continuous social welfare function $\mathcal{W}$, the functional $\mathbb{E}_\mu \mathcal{W}$ maps each endogenous shareholding auction to a well-defined finite value. Moreover, there is an endogenous shareholding auction that is $(\mu, \mathcal{W})$-optimal.

\vspace{\baselineskip} For this particular proof, we provide a proof sketch in \hyperlink{Section6}{Section~6}. Two examples of continuous social welfare functions are {\it producer surplus}, $u \mapsto u_{i_0}$, and {\it consumer surplus}, $u \mapsto \sum_{i \in N} u_i$. It is not difficult to use \hyperlink{Theorem4}{Theorem~4} and \hyperlink{Theorem5}{Theorem~5} to show that for each compact support prior, (i)~each VCG mechanism maximizes expected producer surplus, and (ii)~there is a valvular auction that maximizes expected consumer surplus; we omit the details. We also remark that it should be possible to require only that $\mathcal{W}$ is upper semicontinuous and preserve the conclusion that there is an optimal auction, at the cost of losing the conclusion that each auction has finite subjective expected welfare, by using Fatou's lemma instead of the bounded convergence theorem and making other similar lemma substitutions; we omit the details.\footnote{We thank Rich McLean for this observation.}

\hypertarget{Section6}{}
\section{Proof sketches}

In this section, we sketch our two most difficult proofs: the proof of the  \hyperlink{InvarianceLemma}{Invariance~Lemma} and the proof of \hyperlink{Theorem6}{Theorem~6}.

\vspace{\baselineskip} \noindent \textsc{Proof sketch for the Invariance Lemma.} The \hyperlink{InvarianceLemma}{Invariance~Lemma} states that if a Groves mechanism satisfies {\it no-envy}, then a consumer's subsidy cannot vary arbitrarily with his peers' bids: it is invariant to any change of the peer valuation profile that preserves his Groves price. In the proof, we let $i \in N$ and $v^*_{-i}, v^{**}_{-i} \in V_{-i}$ be such that $\pgro_{v^*_{-i}} = \pgro_{v^{**}_{-i}}$, we let $p \in \mathbb{P}$ denote the common Groves price of $v^*_{-i}$ and $v^{**}_{-i}$, and we argue that $i$ is assigned the same subsidy at both $v^*_{-i}$ and $v^{**}_{-i}$ in two cases: when $p$ is one of the marginal costs in $\{\mathsf{S}(1), \mathsf{S}(2), ..., \mathsf{S}(n)\}$ and when it is not. Though we are ultimately concerned with only $v^*_{-i}$ and $v^{**}_{-i}$, our arguments involve the entire set of peer profiles at which $i$ is assigned Groves price $p$, which we denote by $V^p_{-i}$.

We first consider the case that $p$ is not one of the marginal costs. To begin, we take an arbitrary peer profile $v_{-i} \in V^p_{-i}$ and complete the profile by having $i$ report $p$. Since~$p$ is not one of the marginal costs, it follows that at least one consumer with valuation $p$ wins and at least one consumer with valuation $p$ loses; this allows us to adapt a similar argument from \cite{Ohseto2006}, who characterizes the envy-free Groves mechanisms in a model with mandatory consumption and no production. Indeed, at this initial profile, no-envy implies that there are a winning transfer and a losing transfer that differ by~$p$, and moreover we have that each consumer faces Groves price~$p$. If a peer whose valuation is not $p$ changes his valuation to $p$, then by the \hyperlink{GrovesLemma}{Groves~Lemma}, this peer receives the same bundle, or leaves his previous bundle by paying $p$ to win, or leaves his previous bundle by receiving $p$ to lose; because the resulting profile also has at least one winner with valuation $p$ and at least one loser with valuation $p$, it follows that the new profile has the same winning transfer and losing transfer as the initial profile. Moreover, at the new profile each consumer again has Groves price $p$, which allows us to repeat the argument; it follows that~$i$ is assigned the same subsidy at~$v_{-i}$ and at the peer profile where all of his peers report~$p$. Since $v_{-i} \in V^p_{-i}$ was arbitrary, this argument applies to both $v^*_{-i}$ and $v^{**}_{-i}$, so we are done.

Unfortunately, the above argument does not work when $p$ is one of the marginal costs, which is best illustrated in the extreme case of constant marginal costs $\mathsf{S}(1) = \mathsf{S}(2) = ... = \mathsf{S}(n)$. In this case, any consumer with valuation $p$ might either win or lose, so we cannot directly conclude that the initial profile has a winning transfer and a losing transfer that differ by~$p$. Instead, we use a completely separate argument with eight claims illustrated by \hyperlink{Figure3}{Figure~3}. The idea is that we partition $V_{-i}$ into classes $V_{-i}(n^>, n^=)$ based on (i)~the number $n^>$ of consumers with valuations greater than $p$, and (ii)~the number $n^=$ of consumers with valuations equal to $p$. First, we argue that a particular collection of these classes partitions~$V^p_{-i}$. This implies that there is at least one nonempty class $V_{-i}(n^>, 0) \subseteq V^p_{-i}$, and we further argue that the subsidy function $\sigma_i$ is constant across any such class. Next, we argue that if $V_{-i}(n^>, n^=) \subseteq V^p_{-i}$ and $V_{-i}(n^>, n^=+1) \subseteq V^p_{-i}$ are both nonempty, and if $\sigma_i$ is constant across the former class, then~$\sigma_i$ is constant across the union of these classes. Finally, we argue that the previous sentence is true if we replace $V_{-i}(n^>, n^=+1)$ with $V_{-i}(n^>-1, n^=+1)$. Altogether, this allows us to conclude that $\sigma_i$ is constant across~$V^p_{-i}$; since $v^*_{-i}$ and $v^{**}_{-i}$ both belong to $V^p_{-i}$, we are done. Notice that this argument does not work if $p$ is not a marginal cost, because in that case there is no nonempty class $V_{-i}(n^>, 0) \subseteq V^p_{-i}$; thus we use separate arguments for the two cases.

\hypertarget{Figure3}{}
\begin{figure}[]
\centering
\includegraphics[width=120mm]{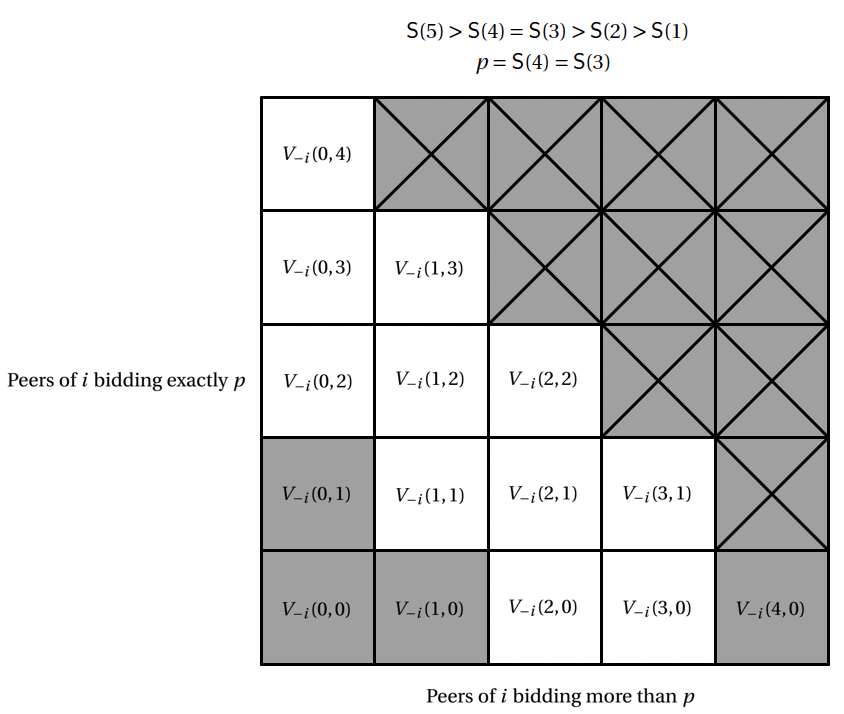}
\caption{{\it Proving the Invariance Lemma.} In this example, there are five consumers, the supply curve is such that $\mathsf{S}(5) > \mathsf{S}(4) = \mathsf{S}(3) > \mathsf{S}(2) > \mathsf{S}(1)$, and $p = \mathsf{S}(4) = \mathsf{S}(3)$. We want to show that the subsidy function $\sigma_i$ is constant across $V^p_{-i} \subseteq V_{-i}$: the collection of peer profiles that offer $i$ Groves price $p$. The horizontal axis measures the number of peers who report more than $p$, the vertical axis measures the number of peers who report exactly~$p$, and each box in the grid represents the corresponding set of peer profiles $V_{-i}(n^>, n^=)$; those crossed out are empty as $i$ only has four peers. By Claim~1 and Claim~2,~$V^p_{-i}$ is the union of the white boxes. Claim~3 is simply used to prove other claims. By Claim~4, Claim~5, and Claim~6,~$\sigma_i$ is constant over each white box in the bottom row. By Claim~7, if~$\sigma_i$ is constant over one white box with a second white box above it, then~$\sigma_i$ is constant over both. By Claim~8, if~$\sigma_i$ is constant over one white box with a second white box directly to its upper-left, then~$\sigma_i$ is constant over both. These observations together imply that $\sigma_i$ is constant over the union of the white boxes $V^p_{-i}$.}
\label{overflow}
\end{figure}

\vspace{\baselineskip} \noindent \textsc{Proof sketch for Theorem 6.} The difficult part of \hyperlink{Theorem6}{Theorem~6} is the claim that for each prior with compact support and each social welfare function that is continuous, there is an endogenous shareholding auction that is optimal in the sense that it maximizes expected welfare. This claim is trivial when $\mathsf{S}(1) = \infty$, as in this case there is only one endogenous shareholding auction; thus let us assume $\mathsf{S}(1) \neq \infty$, in which case $\mathbb{P} \subseteq \mathbb{R}$. Let $\mathbb{S}$ denote the collection of funded subsidy curves.

At a high level, since expected welfare is a functional that maps endogenous shareholding auctions to expected welfare assessments, it would be sufficient to prove that there is a topology on its domain that makes the domain compact and the functional continuous. That said, by the \hyperlink{SummaryLemma}{Summary~Lemma}, the expected welfare of an endogenous shareholding auction is determined by its funded subsidy curve, so there is a well-defined functional $\mathscr{W}: \mathbb{S} \to \mathbb{R} \cup \{-\infty, \infty\}$ that maps subsidy curves to expected welfare assessments, and it would {\it also} be sufficient to prove that there is a topology on $\mathbb{S}$ that makes $\mathbb{S}$ compact and makes $\mathscr{W}$ continuous. Because $\mathbb{S}$ is a collection of continuous functions from $\mathbb{P} \subseteq \mathbb{R}$ to $\mathbb{R}$, it has many well-known topologies, and because $\mathbb{P}$ need not be compact, it turns out that the most convenient topology to work with is the {\it topology of compact convergence}. Crucially, this topology is metrizable.

The \hyperlink{CompactnessLemma}{Compactness~Lemma} establishes that $\mathbb{S}$ is compact, and the proof involves exploiting the rich mathematical structure of funded subsidy curves. In particular, the proof (i)~uses a version of the Arzel\`{a}-Ascoli theorem to establish that $\mathbb{S}$ is contained in a compact subspace of the associated metrizable space of continuous functions from $\mathbb{P}$ to $\mathbb{R}$, and this theorem can be applied because each funded subsidy curve has Lipschitz constant one and because each price has a bounded interval into which each funded subsidy curve sends it, and (ii)~argues that the funded subsidy curves are moreover closed because containing all limits of convergent sequences is sufficient for closure in a metrizable space, and because no convergent sequence of funded subsidy curves has a limit that violates any of the requirements for a funded subsidy curve.

Establishing the continuity of $\mathscr{W}$ is made simpler by the fact that the chosen topology is metrizable. Indeed, for this reason it is sufficient to establish sequential continuity, and there are powerful theorems that help establish this for functionals that involve integration. At a high level, the idea is to start with the $\lim$ symbol inside the argument, then move it across a series of equalities all the way to the left, and the most difficult step is the final move under the integral symbol. For this step, we use the bounded convergence theorem, a corollary of Lebesgue's dominated convergence theorem, and most of the work involves establishing the hypotheses of the bounded convergence theorem. It is here that we use our assumption that the prior has compact support.

\hypertarget{AppendixA}{}
\setcounter{secnumdepth}{0}
\section{Appendix A: Proofs for Section 4}

In this appendix, we prove the \hyperlink{QuantityLemma}{Quantity~Lemma}, the \hyperlink{PriceLemma}{Price~Lemma}, the \hyperlink{IntervalLemma}{Interval~Lemma}, the \hyperlink{BudgetSetLemma}{Budget~Set~Lemma}, the \hyperlink{GrovesLemma}{Groves~Lemma}, and the \hyperlink{VCGLemma}{VCG~Lemma}. We begin with the \hyperlink{QuantityLemma}{Quantity~Lemma}.

\vspace{\baselineskip} \noindent \textsc{Proof of Quantity~Lemma:} Let $v \in V$. We prove the implications in sequence.

\vspace{\baselineskip} \noindent \textsc{[$\Rightarrow$]} Let $(t, W) \in X$ be $v$-efficient and define $q \equiv |W|$. Since $W \in \mathbb{W}$, thus $\mathsf{C}(q) = \sum_{q'\in \llbracket 1, q \rrbracket} \mathsf{S}(q')$.

First, we claim that for each $i \in W$ and each $j \in N \backslash W$, we have $v_i \geq v_j$. Indeed, we cannot have $i \in W$ and $j \in N \backslash W$ such that $v_j > v_i$, as otherwise $\mathsf{TS}_v((W \backslash \{i\}) \cup \{j\}) > \mathsf{TS}_v(W)$, contradicting that $(t, W)$ is $v$-efficient.

Second, we claim that $q \leq \qmax_v$. Indeed, assume by way of contradiction that $q > \qmax_v = \max \{q' \in Q | \mathsf{D}_v(q') \geq \mathsf{S}(q') \}$. Then $\mathsf{S}(q) > \mathsf{D}_v(q)$, and by the first claim there is $i \in W$ such that $v_i = \mathsf{D}_v(q)$. But then $\mathsf{TS}_v(W\backslash \{i\}) > \mathsf{TS}_v(W)$, contradicting that $(t, W)$ is $v$-efficient.

To conclude, we claim that $q \geq \qmin_v$. Indeed, assume by way of contradiction that $q < \qmin_v = \max \{q' \in Q | \mathsf{D}_v(q') > \mathsf{S}(q') \}$. Then $q+1 \leq \qmin_v$, so by monotonicity of both $\mathsf{D}_v$ and $\mathsf{S}$, we have $\mathsf{D}_v(q+1) \geq \mathsf{D}_v(\qmin_v) > \mathsf{S}(\qmin_v) \geq \mathsf{S}(q+1)$, and by the first claim there is $i \in N \backslash W$ such that $v_i = \mathsf{D}_v(q+1)$. Moreover, since $\mathsf{D}_v(q+1) > \mathsf{S}(q+1)$, thus $W \cup \{i\} \in \mathbb{W}$, so $\mathsf{TS}_v(W \cup \{i\})$ is well-defined. But then $\mathsf{TS}_v(W \cup \{i\}) > \mathsf{TS}_v(W)$, contradicting that $(t, W)$ is $v$-efficient.

\vspace{\baselineskip} \noindent \textsc{[$\Leftarrow$]} Let $(t, W) \in X$ satisfy the hypotheses. For each $q \in \llbracket 0, \qcap \rrbracket$, let $\mathsf{TS}^*_v(q) \equiv \sum_{q' \in \llbracket 1, q \rrbracket} ( \mathsf{D}_v(q') - \mathsf{S}(q') )$ denote the total surplus when the winners are $q$ highest-valuation consumers.

By monotonicity of both $\mathsf{D}_v$ and $\mathsf{S}$, the expression $(\mathsf{D}_v(q') - \mathsf{S}(q'))$ is non-increasing in $q'$; thus $\mathsf{TS}^*_v$ is maximized at $q$ if and only if $\mathsf{TS}^*_v(q)$ sums across all positive terms and no negative terms, or equivalently if and only if $q \in \llbracket \qmin_v, \qmax_v \rrbracket$. Since $|W| \in \llbracket \qmin_v, \qmax_v \rrbracket$, we have that $\mathsf{TS}^*_v$ is maximized at $|W|$, so since $(t, W)$ has no loser with a higher valuation than a winner, we have that for each $W' \in \mathbb{W}$, $\mathsf{TS}_v(W) = \mathsf{TS}^*_v(|W|) \geq \mathsf{TS}^*_v(|W'|) \geq \mathsf{TS}_v(W')$.~$\blacksquare$

\vspace{\baselineskip} Second, we prove the \hyperlink{PriceLemma}{Price~Lemma}.

\vspace{\baselineskip} \noindent \textsc{Proof of Price~Lemma:} Let $v \in V$. We prove the statements in sequence.

\vspace{\baselineskip} \noindent \textsc{Step 1:} Establish the first result stated by the lemma.

\vspace{\baselineskip} \noindent \textsc{[$\Rightarrow$]} Let $p \in \mathbb{R} \cup \{\infty\}$, let $s \in \mathbb{R}^N$, let $(t, W) \in X$ be a competitive equilibrium supported by $(s, p)$, and define $q \equiv |W|$. Then $\mathsf{D}_v(q) \geq p \geq \mathsf{S}(q)$ and $\mathsf{S}(q+1) \geq p \geq \mathsf{D}_v(q+1)$, so by monotonicity of $\mathsf{D}_v$ and $\mathsf{S}$ we have $q \in \llbracket \qmin_v, \qmax_v\rrbracket$.

First, we claim $p \geq \pmin_v$. Indeed, if $p < \pmin_v = \max \{ \mathsf{D}_v(\qmax_v+1), \mathsf{S}(\qmax_v)\}$, then either (i)~$\pmin_v = \mathsf{S}(\qmax_v)$, in which case $\mathsf{D}_v(\qmax_v) \geq \mathsf{S}(\qmax_v) = \pmin_v > p$ and thus at least $\qmax_v$ consumers demand an object while the monopolist supplies less than $\qmax_v$, contradicting that supply equals demand, or (ii)~$\pmin_v = \mathsf{D}_v(\qmax_v+1)$, in which case more than $\qmax_v$ consumers demand an object, contradicting $q \leq \qmax_v$.

Second, we claim $p \leq \pmax_v$. Indeed, if $p > \pmax_v = \min \{ \mathsf{D}_v(\qmax_v), \mathsf{S}(\qmax_v+1)\}$, then either (i)~$\pmax_v = \mathsf{D}_v(\qmax_v)$, in which case $p > \pmax_v = \mathsf{D}_v(\qmax_v) \geq \mathsf{S}(\qmax_v)$ and thus fewer than $\qmax_v$ consumers demand an object while the monopolist supplies at least $\qmax_v$, contradicting that supply equals demand, or (ii)~$\pmax_v = \mathsf{S}(\qmax_v+1)$, in which case the monopolist supplies more than $\qmax_v$, contradicting $q \leq \qmax_v$.

\vspace{\baselineskip} \noindent \textsc{[$\Leftarrow$]} Let $p \in [\pmin_v, \pmax_v]$. We construct $s \in \mathbb{R}^N$ and $(t, W) \in X$ such that $(t, W)$ is a competitive equilibrium supported by $(s, p)$ as follows. First, for each $i \in N$, define $s_i \equiv 0$. Second, let $W \subseteq N$ be any set of $\qmax_v$ consumers such that for each $i \in W$ and each $j \in N \backslash W$, we have $v_i \geq v_j$. Finally, let $t \in \mathbb{R}^N$ be such that (i)~for each $i \in W$, we have $t_i = -p$, and (ii)~for each $i \in N \backslash W$, we have $t_i = 0$. Since $p \in [\pmin_v, \pmax_v]$, thus $p \in [\mathsf{D}_v(\qmax_v+1), \mathsf{D}_v(\qmax_v)]$ and $p \in [\mathsf{S}(\qmax_v), \mathsf{S}(\qmax_v+1)]$, so $(t, W)$ is a competitive equilibrium supported by $(s, p)$, as desired.

\vspace{\baselineskip} \noindent \textsc{Step 2:} Establish the second result stated by the lemma.

\vspace{\baselineskip} Let $i \in N$. The proof consists of five sub-steps.

\vspace{\baselineskip} \noindent \textsc{Step 2.1:} $\mathsf{D}_v(\qmax_v) \geq \pmax_v \geq \pmaxdown_v \geq \pminup_v \geq \pmin_v \geq \mathsf{D}_v(\qmax_v+1)$.

\vspace{\baselineskip} First, we claim $\pmax_v = \min \{ \mathsf{D}_v(\qmax_v), \mathsf{S}(\qmax_v+1)\} \geq \max \{ \mathsf{D}_v(\qmax_v+1), \mathsf{S}(\qmax_v)\} = \pmin_v$. Indeed, the equalities hold by definition, and the inequality is due to four inequalities: (i)~by monotonicity of $\mathsf{D}_v$, $\mathsf{D}_v(\qmax_v) \geq \mathsf{D}_v(\qmax_v+1)$, (ii)~by definition of $\qmax_v$, $\mathsf{D}_v(\qmax_v) \geq \mathsf{S}(\qmax_v)$, (iii)~by definition of $\qmax_v$, $\mathsf{S}(\qmax_v+1) > \mathsf{D}_v(\qmax_v+1)$, and (iv)~by monotonicity of $\mathsf{S}$, $\mathsf{S}(\qmax_v+1) \geq \mathsf{S}(\qmax_v)$.

Second, we claim $\pmaxdown_v = \min\{\pmax_v, \mathsf{S}(n)\} \geq \max\{\pmin_v, \mathsf{S}(1)\} = \pminup_v$. Indeed, the equalities hold by definition, and the inequality is due to four inequalities: (i)~by the previous claim, $\pmax_v \geq \pmin_v$, (ii)~by the definitions of $\pmax_v$ and $\qmax_v$, and by monotonicity of $\mathsf{S}$, $\qmax_v \geq 1$ implies $\pmax_v = \min \{ \mathsf{D}_v(\qmax_v), \mathsf{S}(\qmax_v+1)\} \geq \min \{ \mathsf{D}_v(\qmax_v), \mathsf{S}(\qmax_v)\} = \mathsf{S}(\qmax_v) \geq \mathsf{S}(1)$ and $\qmax_v = 0$ implies $\pmax_v = \min \{ \mathsf{D}_v(\qmax_v), \mathsf{S}(\qmax_v+1)\} = \min\{\infty, \mathsf{S}(1)\} = \mathsf{S}(1)$, so in both cases $\pmax_v \geq \mathsf{S}(1)$, (iii)~by monotonicity of $\mathsf{S}$, and by the definitions of $\qmax_v$ and $\pmin_v$, $\qmax_v \leq n-1$ implies $\mathsf{S}(n) \geq \mathsf{S}(\qmax_v+1) = \max \{ \mathsf{D}_v(\qmax_v+1), \mathsf{S}(\qmax_v+1)\} \geq \max \{ \mathsf{D}_v(\qmax_v+1), \mathsf{S}(\qmax_v)\} = \pmin_v$ and $\qmax_v = n$ implies $\mathsf{S}(n) = \max\{-\infty, \mathsf{S}(n)\} = \max \{ \mathsf{D}_v(\qmax_v+1), \mathsf{S}(\qmax_v)\} = \pmin_v$, so in both cases $\mathsf{S}(n) \geq \pmin_v$, and (iv)~by monotonicity of $\mathsf{S}$, $\mathsf{S}(n) \geq \mathsf{S}(1)$.

To conclude, by the second claim and the definitions of the four prices, we have $\mathsf{D}_v(\qmax_v) \geq \min \{ \mathsf{D}_v(\qmax_v), \mathsf{S}(\qmax_v+1)\} = \pmax_v \geq \min\{\pmax_v, \mathsf{S}(n)\} = \pmaxdown_v \geq \pminup_v = \max\{\pmin_v, \mathsf{S}(1)\} \geq \pmin_v = \max \{ \mathsf{D}_v(\qmax_v+1), \mathsf{S}(\qmax_v)\} \geq \mathsf{D}_v(\qmax_v+1)$, as desired.

\vspace{\baselineskip} \noindent \textsc{Step 2.2:} $v_i \geq \mathsf{D}_v(\qmax_v)$ implies $\pminup_v = \pmin_v$ and $v_i \leq \mathsf{D}_v(\qmax_v+1)$ implies $\pmax_v = \pmaxdown_v$.

\vspace{\baselineskip} If $v_i \geq \mathsf{D}_v(\qmax_v)$, then $\qmax_v \geq 1$, so by monotonicity of $\mathsf{S}$ we have that $\pmin_v = \max \{ \mathsf{D}_v(\qmax_v+1), \mathsf{S}(\qmax_v)\} \geq \mathsf{S}(\qmax_v) \geq \mathsf{S}(1)$, so $\pmin_v = \max\{\pmin_v, \mathsf{S}(1)\} = \pminup_v$, as desired.

If $v_i \leq \mathsf{D}_v(\qmax_v+1)$, then $\qmax_v \leq n-1$, so by monotonicity of $\mathsf{S}$ we have that $\pmax_v = \min \{ \mathsf{D}_v(\qmax_v), \mathsf{S}(\qmax_v+1)\} \leq \mathsf{S}(\qmax_v+1) \leq \mathsf{S}(n)$, so $\pmax_v = \min\{\pmax_v, \mathsf{S}(n)\} = \pmaxdown_v$, as desired.

\vspace{\baselineskip} \noindent \textsc{Step 2.3:} If $v_i \geq \mathsf{D}_v(\qmax_v)$, then $\pgro_{v_{-i}} = \pmin_v$.

\vspace{\baselineskip} Assume the hypothesis. In this case, it is straightforward to show that (i)~$\qmax_v \geq 1$, (ii)~for each $q \in \llbracket 1, \qmax_v -1 \rrbracket$, $\mathsf{D}_{v_{-i}}(q) \geq \mathsf{D}_v(q+1) \geq \mathsf{D}_v(\qmax_v) \geq \mathsf{S}(\qmax_v) \geq \mathsf{S}(q)$, (iii)~$\mathsf{D}_{v_{-i}}(\qmax_v) = \mathsf{D}_v(\qmax_v+1)$, and (iv)~for each $q \in \llbracket \qmax_v + 1, n-1 \rrbracket$, $\mathsf{D}_{v_{-i}}(q) \leq \mathsf{D}_v(q) \leq \mathsf{D}_v(\qmax_v+1) < \mathsf{S}(\qmax_v+1) \leq \mathsf{S}(q)$. From here, we consider two cases whose arguments are similar but not identical; we verbally highlight where each uses an argument that the other does not.

If $\mathsf{D}_v(\qmax_v+1) \geq \mathsf{S}(\qmax_v)$, then $\mathsf{D}_{v_{-i}}(\qmax_v) = \mathsf{D}_v(\qmax_v+1) \geq \mathsf{S}(\qmax_v)$, so in this case $\qmax_{v_{-i}} = \qmax_v$. Altogether, then, using $\mathsf{D}_{v_{-i}}(\qmax_v) = \mathsf{D}_v(\qmax_v+1)$ from the previous sentence, we have $\pgro_{v_{-i}} = \min\{\mathsf{D}_{v_{-i}}(\qmax_{v_{-i}}), \mathsf{S}(\qmax_{v_{-i}}+1)\} = \min\{\mathsf{D}_{v_{-i}}(\qmax_v), \mathsf{S}(\qmax_v+1)\} = \min \{\mathsf{D}_v(\qmax_v+1), \mathsf{S}(\qmax_v+1)\} = \mathsf{D}_v(\qmax_v+1) = \max \{ \mathsf{D}_v(\qmax_v+1), \mathsf{S}(\qmax_v)\} = \pmin_v$.

If $\mathsf{D}_v(\qmax_v+1) < \mathsf{S}(\qmax_v)$, then $\mathsf{D}_{v_{-i}}(\qmax_v) = \mathsf{D}_v(\qmax_v+1) < \mathsf{S}(\qmax_v)$, so in this case $\qmax_{v_{-i}} = \qmax_v-1$. Altogether, then, using $\mathsf{D}_{v_{-i}}(\qmax_v-1) \geq \mathsf{S}(\qmax_v)$ from the second inequality of this step (for $q=\qmax_v-1$), we have $\pgro_{v_{-i}} = \min\{\mathsf{D}_{v_{-i}}(\qmax_{v_{-i}}), \mathsf{S}(\qmax_{v_{-i}}+1)\} = \min\{\mathsf{D}_{v_{-i}}(\qmax_v-1), \mathsf{S}(\qmax_v)\} = \mathsf{S}(\qmax_v) = \max \{ \mathsf{D}_v(\qmax_v+1), \mathsf{S}(\qmax_v)\} = \pmin_v$.

\vspace{\baselineskip} \noindent \textsc{Step 2.4:} If $v_i \leq \mathsf{D}_v(\qmax_v+1)$, then $\pgro_{v_{-i}} = \pmax_v$.

\vspace{\baselineskip} Assume the hypothesis. In this case, it is straightforward to show that (i)~$\qmax_v \leq n-1$, (ii)~for each $q \in \llbracket 1, \qmax_v \rrbracket$, $\mathsf{D}_{v_{-i}}(q) = \mathsf{D}_v(q) \geq \mathsf{D}_v(\qmax_v) \geq \mathsf{S}(\qmax_v) \geq \mathsf{S}(q)$, and (iii)~for each $q \in \llbracket \qmax_v+1, n-1 \rrbracket$, $\mathsf{D}_{v_{-i}}(q) \leq \mathsf{D}_v(q) \leq \mathsf{D}_v(\qmax_v+1) < \mathsf{S}(\qmax_v+1) \leq \mathsf{S}(q)$. Altogether, then, $\qmax_{v_{-i}} = \qmax_v$.

To conclude, $\pgro_{v_{-i}} = \min\{\mathsf{D}_{v_{-i}}(\qmax_{v_{-i}}), \mathsf{S}(\qmax_{v_{-i}}+1)\} = \min\{\mathsf{D}_{v_{-i}}(\qmax_v), \mathsf{S}(\qmax_v+1)\} = \min\{\mathsf{D}_v(\qmax_v), \mathsf{S}(\qmax_v+1)\} = \pmax_v$, as desired.

\vspace{\baselineskip} \noindent \textsc{Step 2.5:} Conclude.

\vspace{\baselineskip} If $v_i \geq \mathsf{D}_v(\qmax_v)$, then by Step~2.1, Step~2.2, and Step~2.3, we have $v_i \geq \mathsf{D}_v(\qmax_v) \geq \pmax_v \geq \pmaxdown_v \geq \pminup_v = \pmin_v = \pgro_{v_{-i}} = \pmin_v \geq \mathsf{D}_v(\qmax_v+1)$. If $v_i \leq \mathsf{D}_v(\qmax_v+1)$, then by Step~2.1, Step~2.2, and Step~2.4, we have $\mathsf{D}_v(\qmax_v) \geq \pmax_v = \pgro_{v_{-i}} = \pmax_v = \pmaxdown_v \geq \pminup_v \geq \pmin_v \geq \mathsf{D}_v(\qmax_v+1) \geq v_i$. Since either $v_i \geq \mathsf{D}_v(\qmax_v)$ or $v_i \leq \mathsf{D}_v(\qmax_v+1)$, we are done.

\vspace{\baselineskip} \noindent \textsc{Step 3:} Establish the last two results stated by the lemma.

\vspace{\baselineskip} Let $(t, W) \in X$ be $v$-efficient and let $i \in N$. If $\mathsf{S}(1) = \infty$, then $\qmax_v = 0$ and $\qmax_{v_{-i}} = 0$, so (i)~by the \hyperlink{QuantityLemma}{Quantity~Lemma}, we have $|W| = 0$ and thus $i \not \in W$, and (ii)~$\pgro_{v_{-i}} = \infty > v_i$; thus we have the desired conclusion. Let us assume then that $\mathsf{S}(1) \neq \infty$.

Define $\mathbb{W}_{-i} \equiv \{W' \in \mathbb{W} | i \not \in W'\}$ and $\mathbb{W}_i \equiv \{W' \in \mathbb{W} | i \in W'\}$. Moreover, define the supply curve $\mathsf{S}_\leftarrow$ by (i)~$\mathsf{S}_\leftarrow(0) = -\infty$, and (ii)~for each $q \in \{1, 2, ...\}$, we have $\mathsf{S}_\leftarrow(q) = \mathsf{S}(q+1)$. Intuitively, the plan is to (i)~apply the \hyperlink{QuantityLemma}{Quantity~Lemma} to the environment $(N \backslash \{i\}, \mathsf{S})$ to obtain $W_{-i} \subseteq N \backslash \{i\}$ that maximizes total surplus across $\mathbb{W}_{-i}$, (ii)~apply the \hyperlink{QuantityLemma}{Quantity~Lemma} to the environment $(N \backslash \{i\}, \mathsf{S}_\leftarrow)$ to obtain $W_i \subseteq N \backslash \{i\}$ such that $W_i \cup \{i\}$ maximizes total surplus across $\mathbb{W}_i$, and (iii)~use $W_{-i}$ and $W_i \cup \{i\}$ to show that the difference in total surplus between an optimizer in $\mathbb{W}_i$ and an optimizer in $\mathbb{W}_{-i}$ is precisely $v_i - \pgro_{v_{-i}}$.

First, let $W_{-i} \subseteq N \backslash \{i\}$ be a set of $\qmax_{v_{-i}}$ peers of $i$ whose valuations are highest: for each $j \in W_{-i}$ and each $k \in (N \backslash \{i\}) \backslash W_{-i}$, $v_j \geq v_k$. By applying the \hyperlink{QuantityLemma}{Quantity~Lemma} to $(\mathsf{S}, N \backslash \{i\})$, we have that $W_{-i} \in \argmax_{W' \in \mathbb{W}_{-i}} \mathsf{TS}_v(W')$.

Second, define $\mathbb{W}^*_{-i|i} \subseteq \mathbb{W}_{-i}$ by $\mathbb{W}^*_{-i|i} \equiv \argmax_{W' \in \mathbb{W}_{-i}} [ \sum_{j \in W'} v_j - \sum_{q \in \llbracket 1, |W'| \rrbracket} \mathsf{S}_\leftarrow(q) ]$. Intuitively, this gathers groups of peers of $i$ that maximize surplus subject to the constraint that $i$ join the group. Indeed, $\mathbb{W}^*_{-i|i} = \argmax_{W' \subseteq N \backslash \{i\}} \mathsf{TS}_v(W' \cup \{i\})$, because
\begin{align*}
\mathbb{W}^*_{-i|i} &= \argmax_{W' \in \mathbb{W}_{-i}} \Big[ \sum_{j \in W'} v_j - \sum_{q \in \llbracket 1, |W'| \rrbracket} \mathsf{S}_\leftarrow(q) \Big]
\\ &=  \argmax_{W' \in \mathbb{W}_{-i}} \Big[ \sum_{j \in W'} v_j - \sum_{q \in \llbracket 1, |W'| \rrbracket} \mathsf{S}(q+1) \Big]
\\ &= \argmax_{W' \in \mathbb{W}_{-i}} \Big[ \sum_{j \in W'} v_j - \sum_{q \in \llbracket 1, |W'| \rrbracket} \mathsf{S}(q+1) + (v_i - \mathsf{S}(1)) \Big]
\\ &= \argmax_{W' \subseteq N \backslash \{i\}} \mathsf{TS}_v(W' \cup \{i\}).
\end{align*}
Thus total surplus is maximized across $\mathbb{W}_i$ by the union of $\{i\}$ with a member of $\mathbb{W}^*_{-i|i}$. From here, we consider two cases. In the first case, replacing a member of $W_{-i}$ is either impossible or at least as costly as joining $W_{-i}$, and $W_{-i} \in \mathbb{W}^*_{-i|i}$. In the second case, joining $W_{-i}$ is more costly than replacing some $j^* \in W_{-i}$, and $W_{-i} \backslash \{j^*\} \in \mathbb{W}^*_{-i|i}$.

\vspace{\baselineskip} \textsc{Case 1:} $\mathsf{D}_{v_{-i}}(\qmax_{v_{-i}}) \geq \mathsf{S}(\qmax_{v_{-i}}+1) = \pgro_{v_{-i}}$. In this case, we have $\mathsf{D}_{v_{-i}}(\qmax_{v_{-i}}) \geq \mathsf{S}(\qmax_{v_{-i}}+1) = \mathsf{S}_\leftarrow(\qmax_{v_{-i}})$ and $\mathsf{D}_{v_{-i}}(\qmax_{v_{-i}}+1) < \mathsf{S}(\qmax_{v_{-i}}+1) \leq \mathsf{S}(\qmax_{v_{-i}}+2) = \mathsf{S}_\leftarrow(\qmax_{v_{-i}}+1)$, so by applying the \hyperlink{QuantityLemma}{Quantity~Lemma} to $(\mathsf{S}_\leftarrow, N \backslash \{i\})$ we have $W_{-i} \in \mathbb{W}^*_{-i|i}$. Thus
\begin{align*}
\max_{W' \in \mathbb{W}_i} \mathsf{TS}_v(W') - \max_{W'' \in \mathbb{W}_{-i}} \mathsf{TS}_v(W'') &= \mathsf{TS}_v(W_{-i} \cup \{i\}) - \mathsf{TS}_v(W_{-i})
\\ &= v_i - \mathsf{S}(\qmax_{v_{-i}}+1)
\\ &= v_i - \pgro_{v_{-i}}.
\end{align*}
From here, the desired conclusions follow directly from $v$-efficiency.

\vspace{\baselineskip} \textsc{Case 2:} $\mathsf{S}(\qmax_{v_{-i}}+1) > \mathsf{D}_{v_{-i}}(\qmax_{v_{-i}}) = \pgro_{v_{-i}}$. In this case, we have $\qmax_{v_{-i}} \geq 1$, so by construction of $W_{-i}$ there is $j \in W_{-i}$ such that $v_j = \mathsf{D}_{v_{-i}}(\qmax_{v_{-i}})$; let $j^*$ be any such consumer. Moreover, (i)~$\mathsf{D}_{v_{-i}}(\qmax_{v_{-i}}) < \mathsf{S}(\qmax_{v_{-i}}+1) = \mathsf{S}_\leftarrow(\qmax_{v_{-i}})$, (ii)~$\qmax_{v_{-i}} = 1$ implies $\mathsf{D}_{v_{-i}}(\qmax_{v_{-i}}-1) \geq \mathsf{S}_\leftarrow(\qmax_{v_{-i}}-1)$, and (iii)~$\qmax_{v_{-i}} >1$ implies $\mathsf{D}_{v_{-i}}(\qmax_{v_{-i}}-1) \geq \mathsf{D}_{v_{-i}}(\qmax_{v_{-i}}) \geq \mathsf{S}(\qmax_{v_{-i}}) = \mathsf{S}_\leftarrow(\qmax_{v_{-i}}-1)$, so by applying the \hyperlink{QuantityLemma}{Quantity~Lemma} to $(\mathsf{S}_\leftarrow, N \backslash \{i\})$ we have $W_{-i} \backslash \{j^*\} \in \mathbb{W}^*_{-i|i}$. Thus
\begin{align*}
\max_{W' \in \mathbb{W}_i} \mathsf{TS}_v(W') - \max_{W'' \in \mathbb{W}_{-i}} \mathsf{TS}_v(W'') &= \mathsf{TS}_v((W_{-i} \backslash \{j^*\}) \cup \{i\}) - \mathsf{TS}_v(W_{-i})
\\ &= v_i - \mathsf{D}_{v_{-i}}(\qmax_{v_{-i}})
\\ &= v_i - \pgro_{v_{-i}}.
\end{align*}
From here, the desired conclusions follow directly from $v$-efficiency.~$\blacksquare$

\vspace{\baselineskip} Third, we prove the \hyperlink{IntervalLemma}{Interval~Lemma}.

\vspace{\baselineskip} \noindent \textsc{Proof of Interval~Lemma:} First, let $p \in \mathbb{P}$ and let $v \in V$ be such that each consumer has valuation $p$. Then $\mathbb{P} \neq \{\infty\}$, so $\mathsf{S}(1) \neq \infty$ and $\mathbb{P} = [\mathsf{S}(1), \mathsf{S}(n)] \backslash \{\infty\}$. We claim that $\pmin_v = \pmax_v = p$. Indeed, $\mathsf{S}(n) \geq p \geq \mathsf{S}(1)$, $\qmax_v \geq 1$, and $\mathsf{D}_v(1) = \mathsf{D}_v(\qmax_v) = \mathsf{D}_v(n) = p$, so
\begin{itemize}
\item if $\mathsf{D}_v(\qmax_v+1) \geq \mathsf{S}(\qmax_v)$, then $\qmax_v < n$, so $\pmin_v = \mathsf{D}_v(\qmax_v+1) = p$;

\item if $\mathsf{S}(\qmax_v) > \mathsf{D}_v(\qmax_v+1)$, then $p = \mathsf{D}_v(\qmax_v) \geq \mathsf{S}(\qmax_v) > \mathsf{D}_v(\qmax_v+1)$, so $\qmax_v = n$, so $\mathsf{S}(n) \geq p \geq \mathsf{S}(\qmax_v) = \mathsf{S}(n)$, so $\pmin_v = \mathsf{S}(\qmax_v) = p$;

\item if $\mathsf{S}(\qmax_v+1) \geq \mathsf{D}_v(\qmax_v)$, then $\pmax_v = \mathsf{D}_v(\qmax_v) = p$; and

\item we cannot have $\mathsf{D}_v(\qmax_v) > \mathsf{S}(\qmax_v+1)$, as otherwise $\mathsf{D}_v(n) = p = \mathsf{D}_v(\qmax_v) > \mathsf{S}(\qmax_v+1) > \mathsf{D}_v(\qmax_v+1)$, so $\qmax_v = n$, so $\mathsf{S}(\qmax_v+1) \geq \mathsf{S}(n) \geq p > \mathsf{S}(\qmax_v+1)$, contradicting $\mathsf{S}(\qmax_v+1) = \mathsf{S}(\qmax_v+1)$.
\end{itemize}
Altogether, then, $\pmin_v = \pmax_v = p$. Thus by the \hyperlink{PriceLemma}{Price~Lemma}, for each $i \in N$ we have $\pgro_{v_{-i}} = p$, as desired.

Second, let $i \in N$. If $\mathsf{S}(1) = \infty$, then (i)~$\mathbb{P} = \{\infty\}$, and (ii)~for each $v_{-i} \in V_{-i}$ we have $\qmax_{v_{-i}} = 0$ and thus $\pgro_{v_{-i}} = \infty$, so altogether $\mathbb{P} = \{\infty\} = \{\pgro_{v_{-i}} | v_{-i} \in V_{-i}\}$, as desired; thus let us assume $\mathsf{S}(1) \neq \infty$. Then $\mathbb{P} = [\mathsf{S}(1), \mathsf{S}(n)] \backslash \{\infty\}$. By the first result stated by the lemma, established in the previous paragraph, we have $\mathbb{P} \subseteq \{\pgro_{v_{-i}} | v_{-i} \in V_{-i}\}$. Moreover, for each $v_{-i} \in V_{-i}$, (i)~$\pgro_{v_{-i}} \leq \mathsf{S}(\qmax_{v_{-i}}+1) \leq \mathsf{S}(n)$, (ii)~$\qmax_{v_{-i}} = 0$ implies $\pgro_{v_{-i}} = \mathsf{S}(1)$, (iii)~$\qmax_{v_{-i}} > 0$ implies $\mathsf{D}_{v_{-i}}(\qmax_{v_{-i}}) \geq \mathsf{S}(\qmax_{v_{-i}}) \geq \mathsf{S}(1)$ and $\mathsf{S}(\qmax_{v_{-i}}+1) \geq \mathsf{S}(1)$, and thus $\pgro_{v_{-i}} = \min\{\mathsf{D}_{v_{-i}}(\qmax_{v_{-i}}), \mathsf{S}(\qmax_{v_{-i}}+1)\} \geq \mathsf{S}(1)$, and (iv)~$\qmax_{v_{-i}} > 0$ implies $\pgro_{v_{-i}} \leq \mathsf{D}_{v_{-i}}(\qmax_{v_{-i}}) < \infty$; thus altogether we have $\pgro_{v_{-i}} \in [\mathsf{S}(1), \mathsf{S}(n)] \backslash \{\infty\}$. Since $v_{-i} \in V_{-i}$ was arbitrary, thus $\{\pgro_{v_{-i}} | v_{-i} \in V_{-i}\} \subseteq [\mathsf{S}(1), \mathsf{S}(n)] \backslash \{\infty\} = \mathbb{P}$. Altogether, then, $\mathbb{P} = \{\pgro_{v_{-i}} | v_{-i} \in V_{-i}\}$, as desired.

Finally, assume $\mathsf{S}(1) \neq \infty$. Then $\mathbb{P} = [\mathsf{S}(1), \mathsf{S}(n)] \backslash \{\infty\}$. If $\qcap = n$, then $\cup_{q \in \llbracket 1, \qcap \rrbracket} \mathbb{P}_q = [\mathsf{S}(1), \mathsf{S}(\qcap)] = [\mathsf{S}(1), \mathsf{S}(n)] \backslash \{\infty\} = \mathbb{P}$, as desired. If $\qcap < n$, then $\cup_{q \in \llbracket 1, \qcap \rrbracket} \mathbb{P}_q = [\mathsf{S}(1), \infty) = [\mathsf{S}(1), \mathsf{S}(n)] \backslash \{\infty\} = \mathbb{P}$, as desired.~$\blacksquare$

\vspace{\baselineskip} Fourth, we prove the \hyperlink{BudgetSetLemma}{Budget~Set~Lemma}.

\vspace{\baselineskip} \noindent \textsc{Proof of Budget Set Lemma:} We prove the three items in sequence, and for each item we prove the implications in sequence.

\vspace{\baselineskip} \noindent \textsc{[(}i\textsc{)$\Rightarrow$]} Let $v \in V$ and define $q \equiv |\alpha(v)|$. By the \hyperlink{QuantityLemma}{Quantity~Lemma}, $q \in \llbracket \qmin_v, \qmax_v \rrbracket$, and moreover for each $i \in \alpha(v)$ and each $j \in N \backslash \alpha(v)$ we have $v_i \geq v_j$.

Define $p_{N_0}(v) \equiv \min\{\mathsf{D}_v(q), \mathsf{S}(q+1)\}$. First, $\mathsf{D}_v(q) \geq \mathsf{D}_v(q+1)$ and $\mathsf{S}(q+1) \geq \mathsf{S}(q)$. Second, since $q \in \llbracket \qmin_v, \qmax_v \rrbracket$, thus by monotonicity of $\mathsf{D}_v$ and $\mathsf{S}$ we have $\mathsf{D}_v(q) \geq \mathsf{S}(q)$ and $\mathsf{S}(q+1) \geq \mathsf{D}_v(q+1)$. It follows that whether $p_{N_0}(v) = \mathsf{D}_v(q)$ or $p_{N_0}(v) = \mathsf{S}(q+1)$, we have $p_{N_0}(v) \in [\mathsf{D}_v(q+1), \mathsf{D}_v(q)]$ and $p_{N_0}(v) \in [\mathsf{S}(q), \mathsf{S}(q+1)]$, so $p_{N_0}(v) \in [\pmin_v, \pmax_v]$. For each $i \in \alpha(v)$, define $s_i(v) \equiv \tau_i(v) + p_{N_0}(v)$, and for each $i \in N \backslash \alpha(v)$, define $s_i(v) \equiv \tau_i(v)$; it is easy to verify that $(\tau(v), \alpha(v))$ is a competitive equilibrium supported by $((s_i(v))_{i \in N}, p_{N_0}(v))$.

\vspace{\baselineskip} \noindent \textsc{[(}i\textsc{)$\Leftarrow$]} Let $v \in V$. By hypothesis, there are $(s_i(v))_{i \in N} \in \mathbb{R}^N$ and $p_{N_0}(v) \in [\pmin_v, \pmax_v]$ such that $(\tau(v), \alpha(v))$ is a competitive equilibrium supported by $((s_i(v))_{i \in N}, p_{N_0}(v))$. Define $q \equiv |\alpha(v)|$. We have $\mathsf{D}_v(q) \geq p_{N_0}(v) \geq \mathsf{S}(q)$ and $\mathsf{S}(q+1) \geq p_{N_0}(v) \geq \mathsf{D}_v(q+1)$, so by monotonicity of $\mathsf{D}_v$ and $\mathsf{S}$ we have $q \in \llbracket \qmin_v, \qmax_v\rrbracket$. Moreover, for each $i \in \alpha(v)$ and each $j \in N \backslash \alpha(v)$, we have $v_i \geq p_{N_0}(v) \geq v_j$. Altogether, then, by the \hyperlink{QuantityLemma}{Quantity~Lemma} we have that $(\tau(v), \alpha(v))$ is $v$-efficient. Since $v \in V$ was arbitrary, we are done.

\vspace{\baselineskip} \noindent \textsc{[(}ii\textsc{)$\Rightarrow$]} Let $i \in N$ and let $v_{-i} \in V_{-i}$. There cannot be $v^+_i, v^-_i \in V_i$ such that $\alpha_i(v^+_i, v_{-i}) = \alpha_i(v^-_i, v_{-i})$ and $\tau_i(v^+_i, v_{-i}) > \tau_i(v^-_i, v_{-i})$, as otherwise $i$ benefits by misreporting $v^+_i$ when his true valuation is $v^-_i$, contradicting {\it strategy-proofness}. Thus there are at most two bundles that $i$ can achieve when his peers report $v_{-i}$. If there is only one such bundle, then we are in the personalized bundle case and we are done; thus let us assume that there is one such winning bundle and one such losing bundle. Let $s_i(v_{-i})$ be the transfer associated with the losing bundle and let $p_i(v_{-i})$ be such that $s_i(v_{-i}) - p_i(v_{-i})$ is the transfer associated with the winning bundle. It follows directly from the definition of {\it strategy-proofness} that we are in the personalized subsidy and personalized price case.

\vspace{\baselineskip} \noindent \textsc{[(}ii\textsc{)$\Leftarrow$]} This follows directly from the definition of {\it strategy-proofness}.

\vspace{\baselineskip} \noindent \textsc{[(}iii\textsc{)$\Rightarrow$]} Let $v \in V$. We consider three cases.

First, if $\alpha(v) = \emptyset$, then by {\it no-envy} there is $t_0 \in \mathbb{R}$ such that for each $i \in N$ we have $\tau_i(v) = t_0$. Define $p_N(v) \equiv \max_{i \in N} v_i$ and define $s_N(v) \equiv t_0$. It is easy to verify that we have the desired conclusion.

Second, if $\alpha(v) = N$, then by {\it no-envy} there is $t_0 \in \mathbb{R}$ such that for each $i \in N$ we have $\tau_i(v) = t_0$. Define $p_N(v) \equiv \min_{i \in N} v_i$ and define $s_N(v) \equiv t_0 + p_N(v)$. It is easy to verify that we have the desired conclusion.

Finally, if $\alpha(v) \not \in \{ \emptyset, N\}$, then by {\it no-envy} there are $t_\text{W}, t_\text{L} \in \mathbb{R}$ such that for each $i \in N$, $i \in \alpha(v)$ implies $\tau_i(v) = t_\text{W}$ and $i \not \in \alpha(v)$ implies $\tau_i(v) = t_\text{L}$. Define $p_N(v) \equiv t_\text{L} - t_\text{W}$ and define $s_N(v) \equiv t_\text{L}$. By {\it no-envy}, we have the desired conclusion.

\vspace{\baselineskip} \noindent \textsc{[(}iii\textsc{)$\Leftarrow$]} This follows directly from the definition of {\it no-envy}.~$\blacksquare$

\vspace{\baselineskip} Fifth, we prove the \hyperlink{GrovesLemma}{Groves~Lemma}.

\vspace{\baselineskip} \noindent \textsc{Proof of Groves Lemma:} By a superficial modification of \cite{Holmstrom1979} to allow for cost functions, we have (i) if and only if (ii). Moreover, it follows directly from the definition of {\it strategy-proofness} that (iii) implies (i). To complete the proof, we show that (i) implies (iii). Let $(\tau, \alpha)$ be a mechanism satisfying {\it efficiency} and {\it strategy-proofness}.

\vspace{\baselineskip} \textsc{Case 1:} $\mathsf{S}(1) = \infty$. Let $i \in N$ and let $v_{-i} \in V_{-i}$. In this case, $\pgro_{v_{-i}} = \infty$. Moreover, for each $v_i \in V_i$, we have $\qmax_v = 0$, so by the \hyperlink{QuantityLemma}{Quantity~Lemma} we have $\alpha(v) = \emptyset$; thus by the \hyperlink{BudgetSetLemma}{Budget~Set~Lemma}, there is $x_i(v_{-i}) \in X_i$ such that for each $v_i \in V_i$, we have $(\tau_i(v), \alpha_i(v)) = x_i(v_{-i})$. Define $\sigma_i(v_{-i})$ to be the transfer at $x_i(v_{-i})$. Since $v_{-i} \in V_{-i}$ was arbitrary, we have defined $\sigma_i$. Since $i \in N$ was arbitrary, we have defined $(\sigma_i)_{i \in N}$. It is straightforward to verify that we have the desired conclusion.

\vspace{\baselineskip} \textsc{Case 2:} $\mathsf{S}(1) \in \mathbb{R}$. Let $i \in N$ and let $v_{-i} \in V_{-i}$. In this case, $\pgro_{v_{-i}} \in \mathbb{R}$, so by the \hyperlink{PriceLemma}{Price~Lemma}, we have $\alpha_i(\pgro_{v_{-i}}+1, v_{-i}) = 1$ and $\alpha_i(\pgro_{v_{-i}}-1, v_{-i}) = 0$; thus by the \hyperlink{BudgetSetLemma}{Budget~Set~Lemma}, there are $s_i(v_{-i}) \in \mathbb{R}$ and $p_i(v_{-i}) \in \mathbb{R}$ such that for each $v_i \in V_i$, we have $(\tau_i(v), \alpha_i(v)) \in B^\delta_i(s_i(v_{-i}), p_i(v_{-i})|v_i)$.

We claim $p_i(v_{-i}) = \pgro_{v_{-i}}$. Indeed, let $\varepsilon > 0$. By the \hyperlink{PriceLemma}{Price~Lemma}, $\alpha_i(\pgro_{v_{-i}} + \varepsilon, v_{-i})=1$ and $\alpha_i(\pgro_{v_{-i}} - \varepsilon, v_{-i})=0$, so by the requirement on $s_i(v_{-i})$ and $p_i(v_{-i})$,
\begin{align*}
[\pgro_{v_{-i}} + \varepsilon] + [s_i(v_{-i}) - p_i(v_{-i})] &\geq s_i(v_{-i}), \text{ and}
\\ s_i(v_{-i}) &\geq [\pgro_{v_{-i}} - \varepsilon] + [s_i(v_{-i}) - p_i(v_{-i})];
\end{align*}
thus $p_i(v_{-i}) \in [\pgro_{v_{-i}} - \varepsilon, \pgro_{v_{-i}} + \varepsilon]$. Since $\varepsilon > 0$ was arbitrary, thus $p_i(v_{-i}) = \pgro_{v_{-i}}$, as desired.

Define $\sigma_i(v_{-i}) \equiv s_i(v_{-i})$. Since $v_{-i} \in V_{-i}$ was arbitrary, we have defined $\sigma_i$. Since $i \in N$ was arbitrary, we have defined $(\sigma_i)_{i \in N}$. It is straightforward to verify that we have the desired conclusion.~$\blacksquare$

\vspace{\baselineskip} To conclude this appendix, we prove the \hyperlink{VCGLemma}{VCG~Lemma}.

\vspace{\baselineskip} \noindent \textsc{Proof of VCG Lemma:} Since $(\tau, \alpha)$ is a Groves mechanism, thus by the \hyperlink{GrovesLemma}{Groves~Lemma} we have that for each $i \in N$, there is a subsidy function $\sigma_i:V_{-i} \to \mathbb{R}$ such that for each $v \in V$, we have $(\tau_i(v), \alpha_i(v)) \in B^\delta_i(\sigma_i(v_{-i}), \pgro_{v_{-i}} | v_i)$. We prove (i) implies (ii), (ii) implies (i), and (ii) is equivalent to (iii).

\vspace{\baselineskip} \noindent \textsc{[(}i\textsc{)$\Rightarrow$(}ii\textsc{)]} Let $i \in N$, let $v \in V$, and let $v^-_i \in V_i$ be such that $v^-_i < \mathsf{S}(1)$. Then by {\it efficiency} we have $i \not \in \alpha(v^-_i, v_{-i})$, so $\tau_i(v^-_i, v_{-i}) = \sigma_i(v_{-i})$, and moreover we have $\max_{W \subseteq N} \mathsf{TS}_{(v^-_i, v_{-i})}(W) = \max_{W \subseteq N \backslash \{i\}} \mathsf{TS}_{(0, v_{-i})}(W)$, so by hypothesis $\tau_i(v^-_i, v_{-i}) = 0$. Altogether, then, $\sigma_i(v_{-i}) = 0$, so $(\tau_i(v), \alpha_i(v)) \in B^\delta_i(0, \pgro_{v_{-i}} | v_i)$, as desired.

\vspace{\baselineskip} \noindent \textsc{[(}ii\textsc{)$\Rightarrow$(}i\textsc{)]} Let $i \in N$. Since $(\tau, \alpha)$ is a Groves mechanism, thus there is $\beta_i: V_{-i} \to \mathbb{R}$ such that for each $v \in V$, we have
\begin{align*}
\tau_i(v) = \big( [\max_{W \subseteq N} \mathsf{TS}_v(W)] - \alpha_i(v) v_i \big) + \beta_i(v_{-i}).
\end{align*}
Let $v_{-i} \in V_{-i}$ and let $v^-_i \in V_i$ be such that $v^-_i < \mathsf{S}(1)$. Then $v^-_i < \pgro_{v_{-i}}$, so by hypothesis $\tau_i(v^-_i, v_{-i}) = 0$ and $\alpha_i(v^-_i, v_{-i}) = 0$, and moreover we have $\max_{W \subseteq N} \mathsf{TS}_{(v^-_i, v_{-i})}(W) = \max_{W \subseteq N \backslash \{i\}} \mathsf{TS}_{(0, v_{-i})}(W)$. Altogether, then, $\beta_i(v_{-i}) = -[\max_{W \subseteq N \backslash \{i\}} \mathsf{TS}_{(0, v_{-i})}(W)]$, as desired.

\vspace{\baselineskip} \noindent \textsc{[(}ii\textsc{)$\Leftrightarrow$(}iii\textsc{)]} Let $i \in N$ and let $v \in V$. If $v_i \geq \pgro_{v_{-i}}$, then by the \hyperlink{PriceLemma}{Price~Lemma} we have $\pgro_{v_{-i}} = \pmin_v$, from which the desired conclusion follows. If $\pgro_{v_{-i}} > v_i$, then by the \hyperlink{PriceLemma}{Price~Lemma} we have $\pgro_{v_{-i}} \geq \pmin_v \geq v_i$ and $i \not \in \alpha(v)$, from which the desired conclusion follows.~$\blacksquare$

\hypertarget{AppendixB}{}
\setcounter{secnumdepth}{0}
\section{Appendix B: Proofs for Section 5.1}

In this appendix, we prove the \hyperlink{InvarianceLemma}{Invariance~Lemma}, \hyperlink{Theorem1}{Theorem~1}, and \hyperlink{Theorem2}{Theorem~2}. To follow the order in which these results are presented, we begin with the \hyperlink{InvarianceLemma}{Invariance~Lemma}, though we note that we do not use this lemma to prove \hyperlink{Theorem1}{Theorem~1}. Note that \hyperlink{Section6}{Section~6} contains a proof sketch for the \hyperlink{InvarianceLemma}{Invariance~Lemma}.

\vspace{\baselineskip} \noindent \textsc{Proof of Invariance~Lemma:} Let $(\tau, \alpha)$ satisfy {\it efficiency}, {\it strategy-proofness}, and {\it no-envy}, and let $i \in N$. By the \hyperlink{GrovesLemma}{Groves~Lemma}, there is a subsidy function $\sigma_i: V_{-i} \to \mathbb{R}$ that satisfies the first desired condition. Let $v^*_{-i}, v^{**}_{-i} \in V_{-i}$ such that $\pgro_{v^*_{-i}} = \pgro_{v^{**}_{-i}}$; it remains to show that $\sigma_i(v^*_{-i}) = \sigma_i(v^{**}_{-i})$. Define $p \equiv \pgro_{v^*_{-i}} = \pgro_{v^{**}_{-i}}$ and define $V^p_{-i} \subseteq V_{-i}$ by
\begin{align*}
V^p_{-i} \equiv \{v_{-i} \in V_{-i} | \pgro_{v_{-i}} = p\}.
\end{align*}
By the \hyperlink{IntervalLemma}{Interval~Lemma}, $p \in \mathbb{P}$. We consider two cases, and remark that our arguments involve members of $V^p_{-i}$ that need not be either $v^*_{-i}$ and $v^{**}_{-i}$; in both cases we refer to $v^*_{-i}$ and $v^{**}_{-i}$ only to conclude.

\vspace{\baselineskip} \noindent \textsc{Case~1:} $p \in \mathbb{P} \backslash \{\mathsf{S}(1), \mathsf{S}(2), ..., \mathsf{S}(n)\}$.

\vspace{\baselineskip} Let $v_{-i} \in V^p_{-i}$, define $q_i \equiv \qmax_{v_{-i}}$, and let $v^p \in V$ be the profile where each consumer reports $p$. By definition of $q_i$, we have $\mathsf{D}_{v_{-i}}(q_i) \geq \mathsf{S}(q_i)$. Since $p = \min \{ \mathsf{D}_{v_{-i}}(q_i), \mathsf{S}(q_i+1)\}$ and $p \neq \mathsf{S}(q_i+1)$, thus $\mathsf{S}(q_i+1) > p = \mathsf{D}_{v_{-i}}(q_i) \geq \mathsf{S}(q_i)$, so (i)~$p \neq \infty$, (ii)~$q_i > 0$, and (iii)~since $p \neq \mathsf{S}(q_i)$ we have $p \in (\mathsf{S}(q_i), \mathsf{S}(q_i+1))$.

Let $(i_t)_{t \in \llbracket 1, n-1 \rrbracket}$ index the peers of $i$; thus we have $\{i_t\}_{t \in \llbracket 1, n-1 \rrbracket} = N \backslash \{i\}$. Define $v^0 \equiv (p, v_{-i})$, and for each $t \in \llbracket 1, n-1\rrbracket$ define $v^t \in V$ by $v^t \equiv (p, v^{t-1}_{-i_t})$; observe that $v^{n-1} = v^p$. Finally, by {\it no-envy}, (i)~for each $v' \in V$ such that $\alpha(v') \neq \emptyset$, there is a winning transfer $\tau_{\text{W}}(v') \in \mathbb{R}$ such that for each $j \in \alpha(v')$, we have $\tau_j(v') = \tau_{\text{W}}(v')$; and (ii)~for each $v' \in V$ such that $\alpha(v') \neq N$, there is a losing transfer~$\tau_{\text{L}}(v') \in \mathbb{R}$ such that for each $j \in N \backslash \alpha(v')$, we have $\tau_j(v') = \tau_{\text{L}}(v')$. From here, we make five observations, then make two claims, then conclude.

We begin by making five observations that hold for each profile in $\{v^t\}_{t \in \llbracket 0, n-1 \rrbracket}$. Indeed, let $t \in \llbracket 0, n-1 \rrbracket$. For brevity, we use earlier observations to establish later ones without reference. First, since (i)~$\mathsf{D}_{v_{-i}}(q_i) = p$, (ii)~at $v^t$ each peer of $i$ either reports as at $v_{-i}$ or reports $p$, and (iii)~$v^t_i = p$, thus at $v^t$ we have that more than $q_i$ consumers report at least $p$ and less than $q_i$ consumers report more than $p$. Second, since $p \in (\mathsf{S}(q_i), \mathsf{S}(q_i+1))$, thus $\qmin_{v^t} = \qmax_{v^t} = q_i$, so by the \hyperlink{QuantityLemma}{Quantity~Lemma} we have $|\alpha(v^t)|=q_i$. Third, since $q_i \in \llbracket 1, n-1\rrbracket$, thus $\alpha(v^t) \neq \emptyset$ and $\alpha(v^t) \neq N$, so $\tau_{\text{W}}(v^t)$ and $\tau_{\text{L}}(v^t)$ are well-defined. Fourth, by the \hyperlink{QuantityLemma}{Quantity~Lemma}, there are $j \in \alpha(v^t)$ and $j' \in N \backslash \alpha(v^t)$ such that $v^t_j = v^t_{j'} = p$, so by two applications of {\it no-envy} for $j$ and~$j'$ we have $p + \tau_{\text{W}}(v^t) = \tau_{\text{L}}(v^t)$. Finally, for each $j \in N$, at $v^t_{-j}$ we have that at least $q_i$ consumers report at least $p$ and less than $q_i$ consumers report more than $p$, so since $p \in (\mathsf{S}(q_i), \mathsf{S}(q_i+1))$ we have $\qmax_{v^t_{-j}} = q_i$ and thus $\pgro_{v^t_{-j}} = \min \{ \mathsf{D}_{v^t_{-j}}(q_i), \mathsf{S}(q_i+1)\} = p$.

Our first claim is that for each $t \in \llbracket 0, n-2 \rrbracket$, we have $\tau_{\text{W}}(v^t) = \tau_{\text{W}}(v^{t+1})$. Indeed, let $t \in \llbracket 0, n-2 \rrbracket$ and define $j \equiv i_{t+1}$. By construction we have $v^t_{-j} = v^{t+1}_{-j}$, and by the fifth observation we have $\pgro_{v^t_{-j}} = p$. We consider four cases. First, if $j \in \alpha(v^{t+1})$ and $j \in \alpha(v^t)$, then by {\it strategy-proofness} we have $\tau_{\text{W}}(v^{t+1}) = \tau_{\text{W}}(v^t)$, as desired. Second, if $j \in \alpha(v^{t+1})$ and $j \not \in \alpha(v^t)$, then by the \hyperlink{GrovesLemma}{Groves~Lemma} both $(\tau_{\text{W}}(v^{t+1}), 1)$ and $(\tau_{\text{L}}(v^t), 0)$ belong to a budget set with price $\pgro_{v^{t+1}_{-j}} = \pgro_{v^t_{-j}} = p$, so $\tau_{\text{W}}(v^{t+1}) = \tau_{\text{L}}(v^t) - p$, so by the fourth observation we have $\tau_{\text{W}}(v^{t+1}) = \tau_{\text{W}}(v^t)$, as desired. Third, if $j \not \in \alpha(v^{t+1})$ and $j \in \alpha(v^t)$, then by the \hyperlink{GrovesLemma}{Groves~Lemma} both $(\tau_{\text{L}}(v^{t+1}), 0)$ and $(\tau_{\text{W}}(v^t), 1)$ belong to a budget set with price $\pgro_{v^{t+1}_{-j}} = \pgro_{v^t_{-j}} = p$, so we have $\tau_{\text{L}}(v^{t+1}) = \tau_{\text{W}}(v^t) + p$, so by the fourth observation we have $\tau_{\text{W}}(v^{t+1}) = \tau_{\text{W}}(v^t)$, as desired. Finally, if $j \not \in \alpha(v^{t+1})$ and $j \not \in \alpha(v^t)$, then by {\it strategy-proofness} we have $\tau_{\text{L}}(v^{t+1}) = \tau_{\text{L}}(v^t)$, so by the fourth observation we have $\tau_{\text{W}}(v^{t+1}) = \tau_{\text{W}}(v^t)$, as desired.

Define $v^+_i \equiv \mathsf{D}_{v_{-i}}(1)+1$. Our second claim is that for each $t \in \llbracket 0, n-1\rrbracket$, we have $i \in \alpha(v^+_i, v^t_{-i})$ and $\tau_i(v^+_i, v^t_{-i}) = \tau_{\text{W}}(v^t)$. Indeed, since $q_i > 0$ and $\mathsf{D}_{v_{-i}}(q_i) = p$, thus by the fifth observation we have $v^+_i > \mathsf{D}_{v_{-i}}(1) \geq \mathsf{D}_{v_{-i}}(q_i) = p = \pgro_{v^t_{-i}}$, so by the \hyperlink{PriceLemma}{Price~Lemma} we have $i \in \alpha(v^+_i, v^t_{-i})$. Moreover, (i)~by the fourth observation, either $i \in \alpha(v^t)$ and $\tau_i(v^t) = \tau_{\text{W}}(v^t)$ or $i \not \in \alpha(v^t)$ and $\tau_i(v^t) = \tau_{\text{W}}(v^t)+p$, and (ii)~by the \hyperlink{GrovesLemma}{Groves~Lemma}, both $(\tau_i(v^+_i, v^t_{-i}), \alpha_i(v^+_i, v^t_{-i}))$ and $(\tau_i(v^t), \alpha_i(v^t))$ belong to a budget set with price $\pgro_{v^t_{-i}} = p$; thus in either case we have $\tau_i(v^+_i, v^t_{-i}) = \tau_{\text{W}}(v^t)$, as desired.

To conclude, by the two claims we have (i)~$\tau_i(v^+_i, v_{-i}) = \tau_i(v^+_i, v^0_{-i}) = \tau_{\text{W}}(v^0) = \tau_{\text{W}}(v^{n-1}) = \tau_i(v^+_i, v^{n-1}_{-i}) = \tau_i(v^+_i, v^p_{-i})$, (ii)~$i \in \alpha(v^+_i, v_{-i})$, and (iii)~$i \in \alpha(v^+_i, v^p_{-i})$. Moreover, by the fifth observation we have $\pgro_{v_{-i}} = \pgro_{v^0_{-i}} = p = \pgro_{v^{n-1}_{-i}} = \pgro_{v^p_{-i}}$. Altogether, then, by construction of $\sigma_i$ we have $\sigma_i(v_{-i}) - p = \sigma_i(v_{-i}) - \pgro_{v_{-i}} = \tau_i(v^+_i, v_{-i}) = \tau_i(v^+_i, v^p_{-i}) = \sigma_i(v^p_{-i}) - \pgro_{v^p_{-i}} = \sigma_i(v^p_{-i}) - p$, so $\sigma_i(v_{-i}) = \sigma_i(v^p_{-i})$. Since $v_{-i} \in V^p_{-i}$ was arbitrary, thus $\sigma_i(v^*_{-i}) = \sigma_i(v^p_{-i}) = \sigma_i(v^{**}_{-i})$, as desired.

\vspace{\baselineskip} \noindent \textsc{Case~2:} $p \in \{\mathsf{S}(1), \mathsf{S}(2), ..., \mathsf{S}(n)\}$.

\vspace{\baselineskip} In this case, we begin by partitioning $V_{-i}$ into subsets that we refer to as {\it classes} based on (i) the number $n^>$ of peers with valuations higher than $p$, and (ii) the number $n^=$ of peers with valuations equal to $p$. First, for each $v_{-i} \in V_{-i}$, define
\begin{align*}
\mathsf{n}^>(v_{-i}) &\equiv |\{ j \in N \backslash \{i\} | v_j > p \}|, \text{ and}
\\ \mathsf{n}^=(v_{-i}) &\equiv |\{ j \in N \backslash \{i\} | v_j = p \}|.
\end{align*}
Second, for each $(n^>, n^=) \in \llbracket 0, n-1 \rrbracket^2$, define the class $V_{-i}(n^>, n^=) \subseteq V_{-i}$ by
\begin{align*}
V_{-i}(n^>, n^=) \equiv \{ v_{-i} \in V_{-i} | \mathsf{n}^>(v_{-i}) = n^> \text{ and } \mathsf{n}^=(v_{-i}) = n^= \}.
\end{align*}
Clearly, the nonempty classes partition $V_{-i}$.

Next, we gather the nonempty classes over which the Groves price is constantly~$p$. Formally, let $\mathcal{V} \subseteq \{V_{-i}(n^>, n^=) | (n^>, n^=) \in \llbracket 0, n-1 \rrbracket^2 \}$ be defined as follows: for each $(n^>, n^=) \in \llbracket 0, n-1 \rrbracket^2$, $V_{-i}(n^>, n^=) \in \mathcal{V}$ if and only if $V_{-i}(n^>, n^=)$ is a nonempty subset of~$V^p_{-i}$. To complete the proof for this case, we state and prove eight claims, after which we conclude that $\sigma_i$ is constant over $\bigcup \mathcal{V}$ and moreover that $\bigcup \mathcal{V} = V^p_{-i}$, with our first two claims providing conditions on $n^>$ and $n^=$ that guarantee $V_{-i}(n^>, n^=) \subseteq V^p_{-i}$.

Before proceeding, we introduce some useful notation: since $p$ may be the marginal cost for several different quantities, let us define $\qmin_{\mathsf{S}=p} \equiv \min \{ q \in \llbracket 1, n \rrbracket | \mathsf{S}(q) = p \}$ and $\qmax_{\mathsf{S}=p} \equiv \max \{ q \in \llbracket 1, n \rrbracket | \mathsf{S}(q) = p \}$.

\vspace{\baselineskip} \noindent \textsc{Claim~1:} For each $v_{-i} \in V_{-i}$, $\pgro_{v_{-i}} > p$ if and only if $\mathsf{n}^>(v_{-i}) \geq \qmax_{\mathsf{S}=p}$.

\vspace{\baselineskip} Let $v_{-i} \in V_{-i}$ and define $q_i \equiv \qmax_{v_{-i}}$. We prove the two implications in sequence.

\vspace{\baselineskip} \noindent \textsc{[$\Rightarrow$]} We prove the contrapositive. Indeed, assume $\mathsf{n}^>(v_{-i}) < \qmax_{\mathsf{S}=p}$. If $\qmax_{\mathsf{S}=p} = n$, then $p = \mathsf{S}(n) \geq \pgro_{v_{-i}}$ as desired, so assume $\qmax_{\mathsf{S}=p} \in \llbracket 1, n-1 \rrbracket$. Since $\mathsf{n}^>(v_{-i}) < \qmax_{\mathsf{S}=p}$, thus we have $p \geq \mathsf{D}_{v_{-i}}(\qmax_{\mathsf{S}=p})$, so $\mathsf{S}(\qmax_{\mathsf{S}=p} + 1) > p \geq \mathsf{D}_{v_{-i}}(\qmax_{\mathsf{S}=p}) \geq \mathsf{D}_{v_{-i}}(\qmax_{\mathsf{S}=p} + 1)$, so $\qmax_{\mathsf{S}=p} +1 > q_i$ and thus $\qmax_{\mathsf{S}=p} \geq q_i$. If $\qmax_{\mathsf{S}=p} > q_i$, then $\qmax_{\mathsf{S}=p} \geq q_i + 1$, so $p = \mathsf{S}(\qmax_{\mathsf{S}=p}) \geq \mathsf{S}(q_i+1) \geq \pgro_{v_{-i}}$; if $\qmax_{\mathsf{S}=p} = q_i$, then $p \geq \mathsf{D}_{v_{-i}}(\qmax_{\mathsf{S}=p}) = \mathsf{D}_{v_{-i}}(q_i) \geq \pgro_{v_{-i}}$; thus in both cases we have $p \geq \pgro_{v_{-i}}$, as desired.

\vspace{\baselineskip} \noindent \textsc{[$\Leftarrow$]} Assume $\mathsf{n}^>(v_{-i}) \geq \qmax_{\mathsf{S}=p}$. Then (i)~since $n > \mathsf{n}^>(v_{-i})$ we have $n > \qmax_{\mathsf{S}=p}$, and (ii)~$\mathsf{D}_{v_{-i}}(\qmax_{\mathsf{S}=p}) > p = \mathsf{S}(\qmax_{\mathsf{S}=p})$ and thus $q_i \geq \qmax_{\mathsf{S}=p}$. If $\pgro_{v_{-i}} = \mathsf{D}_{v_{-i}}(q_i)$ and $q_i > \qmax_{\mathsf{S}=p}$, then $q_i \geq \qmax_{\mathsf{S}=p} + 1$, so $\pgro_{v_{-i}} = \mathsf{D}_{v_{-i}}(q_i) \geq \mathsf{S}(q_i) \geq \mathsf{S}(\qmax_{\mathsf{S}=p} +1) > p$; if $\pgro_{v_{-i}} = \mathsf{D}_{v_{-i}}(q_i)$ and $q_i = \qmax_{\mathsf{S}=p}$, then $\pgro_{v_{-i}} = \mathsf{D}_{v_{-i}}(q_i) = \mathsf{D}_{v_{-i}}(\qmax_{\mathsf{S}=p}) > p$; if $\pgro_{v_{-i}} = \mathsf{S}(q_i+1)$, then $\pgro_{v_{-i}} = \mathsf{S}(q_i+1) \geq \mathsf{S}(\qmax_{\mathsf{S}=p} + 1) > p$; thus in all cases, $\pgro_{v_{-i}} > p$, as desired.

\vspace{\baselineskip} \noindent \textsc{Claim~2:} For each $v_{-i} \in V_{-i}$, $\pgro_{v_{-i}} < p$ if and only if $\mathsf{n}^>(v_{-i}) + \mathsf{n}^=(v_{-i}) < \qmin_{\mathsf{S}=p} - 1$.

\vspace{\baselineskip} Let $v_{-i} \in V_{-i}$, define $q_i \equiv \qmax_{v_{-i}}$, and define $\mathsf{n}^\geq(v_{-i}) \equiv \mathsf{n}^>(v_{-i}) + \mathsf{n}^=(v_{-i})$. We prove the two implications in sequence.

\vspace{\baselineskip} \noindent \textsc{[$\Rightarrow$]} We prove the contrapositive. Assume $\mathsf{n}^\geq(v_{-i}) \geq \qmin_{\mathsf{S}=p} - 1$. If $\qmin_{\mathsf{S}=p} = 1$, then $\pgro_{v_{-i}} \geq \mathsf{S}(1) = p$ as desired, so assume $\qmin_{\mathsf{S}=p} \in \llbracket 2, n \rrbracket$. Since $\mathsf{n}^\geq(v_{-i}) \geq \qmin_{\mathsf{S}=p} - 1$, thus we have $\mathsf{D}_{v_{-i}}(\qmin_{\mathsf{S}=p} - 1) \geq p$, so $\mathsf{D}_{v_{-i}}(\qmin_{\mathsf{S}=p} - 1) \geq p = \mathsf{S}(\qmin_{\mathsf{S}=p}) > \mathsf{S}(\qmin_{\mathsf{S}=p} - 1)$, so $q_i \geq \qmin_{\mathsf{S}=p} - 1$. If $\pgro_{v_{-i}} = \mathsf{D}_{v_{-i}}(q_i)$ and $q_i > \qmin_{\mathsf{S}=p} - 1$, then $q_i \geq \qmin_{\mathsf{S}=p}$, so $\pgro_{v_{-i}} = \mathsf{D}_{v_{-i}}(q_i) \geq \mathsf{S}(q_i) \geq \mathsf{S}(\qmin_{\mathsf{S}=p}) = p$; if $\pgro_{v_{-i}} = \mathsf{D}_{v_{-i}}(q_i)$ and $q_i = \qmin_{\mathsf{S}=p} - 1$, then $\pgro_{v_{-i}} = \mathsf{D}_{v_{-i}}(q_i) = \mathsf{D}_{v_{-i}}(\qmin_{\mathsf{S}=p} - 1) \geq p$; if $\pgro_{v_{-i}} = \mathsf{S}(q_i+1)$, then $\pgro_{v_{-i}} = \mathsf{S}(q_i+1) \geq \mathsf{S}(\qmin_{\mathsf{S}=p}) = p$; thus in all cases, $\pgro_{v_{-i}} \geq p$, as desired.

\vspace{\baselineskip} \noindent \textsc{[$\Leftarrow$]} Assume $\mathsf{n}^\geq(v_{-i}) < \qmin_{\mathsf{S}=p} - 1$. Then (i)~since $\mathsf{n}^\geq(v_{-i}) \geq 0$ we have $\qmin_{\mathsf{S}=p} - 1 \geq 1$, and (ii)~$\mathsf{S}(\qmin_{\mathsf{S}=p}) = p > \mathsf{D}_{v_{-i}}(\qmin_{\mathsf{S}=p} - 1) \geq \mathsf{D}_{v_{-i}}(\qmin_{\mathsf{S}=p})$ and thus $\qmin_{\mathsf{S}=p} -1 \geq q_i$. If $\qmin_{\mathsf{S}=p} - 1 > q_i$, then $\qmin_{\mathsf{S}=p} - 1 \geq q_i + 1$, so $p = \mathsf{S}(\qmin_{\mathsf{S}=p}) > \mathsf{S}(\qmin_{\mathsf{S}=p} - 1) \geq \mathsf{S}(q_i + 1) \geq \pgro_{v_{-i}}$; if $\qmin_{\mathsf{S}=p} - 1 = q_i$, then $p = \mathsf{S}(\qmin_{\mathsf{S}=p}) > \mathsf{D}_{v_{-i}}(\qmin_{\mathsf{S}=p} - 1) = \mathsf{D}_{v_{-i}}(q_i) \geq \pgro_{v_{-i}}$; thus in both cases we have $p > \pgro_{v_{-i}}$, as desired.

\vspace{\baselineskip} \noindent \textsc{Claim~3:} For each $n^> \in \llbracket 0, n-1\rrbracket$ such that $V_{-i}(n^>, 0) \in \mathcal{V}$ and each $v_{-i} \in V_{-i}(n^>, 0)$, we have $\qmax_{v_{-i}} = n^>$ and $p = \mathsf{S}(n^>+1)$.

\vspace{\baselineskip} Let $n^> \in \llbracket 0, n-1\rrbracket$ such that $V_{-i}(n^>, 0) \in \mathcal{V}$, let $v_{-i} \in V_{-i}(n^>, 0)$, and define $q_i \equiv \qmax_{v_{-i}}$. Since $V_{-i}(n^>, 0) \in \mathcal{V}$, thus $\pgro_{v_{-i}} = p$. If $\mathsf{S}(1) = \infty$, then $p = \pgro_{v_{-i}} = \infty$, so we have $q_i = 0 = n^>$ and $p = \infty = \mathsf{S}(1) = \mathsf{S}(n^>+1)$ as desired; thus let us assume $\mathsf{S}(1) \neq \infty$. If $q_i = 0$, then $\mathsf{D}_{v_{-i}}(q_i) = \infty \neq \pgro_{v_{-i}} = p$; if $q_i \in \llbracket 1, n-1\rrbracket$, then since $\mathsf{n}^=(v_{-i}) = 0$ we have $\mathsf{D}_{v_{-i}}(q_i) \neq p$; thus in both cases we have $\mathsf{D}_{v_{-i}}(q_i) \neq p$. Then $\mathsf{D}_{v_{-i}}(q_i) > \pgro_{v_{-i}} = p = \mathsf{S}(q_i+1) > \mathsf{D}_{v_{-i}}(q_i+1)$, so $q_i = n^>$ and $p = \mathsf{S}(n^>+1)$, as desired.

\vspace{\baselineskip} \noindent \textsc{Claim~4:} If $V_{-i}(0, 0) \in \mathcal{V}$, then $\sigma_i$ is constant on $V_{-i}(0, 0)$.

\vspace{\baselineskip} Assume that $V_{-i}(0, 0) \in \mathcal{V}$ and let $v_{-i}, v_{-i}' \in V_{-i}(0, 0)$. Let $(i_t)_{t \in \llbracket 1, n-1 \rrbracket}$ index the peers of $i$; thus we have $\{i_t\}_{t \in \llbracket 1, n-1 \rrbracket} = N \backslash \{i\}$. Let $v^-_i \in V_i$ such that $v^-_i < \mathsf{S}(1)$, define $v^0 \equiv (v^-_i, v_{-i})$, and for each $t \in \llbracket 1, n-1\rrbracket$ define $v^t \in V$ by $v^t \equiv (v'_{i_t}, v^{t-1}_{-i_t})$; observe that $v^{n-1} = (v^-_i, v'_{-i})$.

Since $V_{-i}(0, 0) \in \mathcal{V}$ and $v_{-i}, v'_{-i} \in V_{-i}(0, 0)$, thus by Claim~3 we have $\qmax_{v_{-i}} = \qmax_{v'_{-i}} = 0$ and $p = \mathsf{S}(1)$, so for each $j \in N \backslash \{i\}$ we have $v_j < \mathsf{S}(1)$ and $v'_j < \mathsf{S}(1)$. Moreover, we have $v^-_i < \mathsf{S}(1)$. Then for each $t \in \llbracket 0, n \rrbracket$, at $v^t$ we have that all consumers report less than $\mathsf{S}(1)$, and thus by {\it efficiency} and {\it no-envy} that all consumers lose and receive a common losing transfer. By {\it strategy-proofness}, for each $t \in \llbracket 1, n-1 \rrbracket$ we have $\tau_{i_t}(v^t) = \tau_{i_t}(v^{t-1})$. Altogether, then, $\tau_i(v^-_i, v_{-i}) = \tau_i(v^0) = \tau_i(v^{n-1}) = \tau_i(v^-_i, v'_{-i})$ and $\alpha_i(v^-_i, v_{-i}) = 0 = \alpha_i(v^-_i, v'_{-i})$, so by construction of $\sigma_i$ we have $\sigma_i(v_{-i}) = \sigma_i(v_{-i}')$, as desired.

\vspace{\baselineskip} \noindent \textsc{Claim~5:} If $V_{-i}(n-1, 0) \in \mathcal{V}$, then $\sigma_i$ is constant on $V_{-i}(n-1, 0)$.

\vspace{\baselineskip} Assume that $V_{-i}(n-1, 0) \in \mathcal{V}$. If $n = 1$, then we are done by Claim~4; thus let us assume $n > 1$. By Claim~3, $p = \mathsf{S}(n)$, so there is a profile in $V_{-i}(n-1, 0)$, so it is possible for a peer of $i$ to have a valuation greater than $p = \mathsf{S}(n)$; thus $\mathsf{S}(n) \neq \infty$. Let $v_{-i}, v_{-i}' \in V_{-i}(n-1, 0)$.

Let $(i_t)_{t \in \llbracket 1, n-1 \rrbracket}$ index the peers of $i$; thus we have $\{i_t\}_{t \in \llbracket 1, n-1 \rrbracket} = N \backslash \{i\}$. Let $v^+_i \in V_i$ such that $v^+_i > \mathsf{S}(n)$, define $v^0 \equiv (v^+_i, v_{-i})$, and for each $t \in \llbracket 1, n-1\rrbracket$ define $v^t \in V$ by $v^t \equiv (v'_{i_t}, v^{t-1}_{-i_t})$; observe that $v^{n-1} = (v^+_i, v'_{-i})$.

Since $V_{-i}(n-1, 0) \in \mathcal{V}$ and $v_{-i}, v'_{-i} \in V_{-i}(n-1, 0)$, thus by Claim~3 we have $\qmax_{v_{-i}} = \qmax_{v'_{-i}} = n-1$ and $p = \mathsf{S}(n)$, so for each $j \in N \backslash \{i\}$ we have $v_j > \mathsf{S}(n)$ and $v'_j > \mathsf{S}(n)$. Moreover, we have $v^+_i > \mathsf{S}(n)$. Then for each $t \in \llbracket 0, n \rrbracket$, at $v^t$ we have that all consumers report more than $\mathsf{S}(n)$, and thus by {\it efficiency} and {\it no-envy} that all consumers win and receive a common winning transfer. By {\it strategy-proofness}, for each $t \in \llbracket 1, n-1 \rrbracket$ we have $\tau_{i_t}(v^t) = \tau_{i_t}(v^{t-1})$. Altogether, then, $\tau_i(v^+_i, v_{-i}) = \tau_i(v^0) = \tau_i(v^{n-1}) = \tau_i(v^+_i, v'_{-i})$, $\alpha_i(v^+_i, v_{-i}) = 1 = \alpha_i(v^+_i, v'_{-i})$, and $\pgro_{v_{-i}} = p = \pgro_{v'_{-i}}$, so by construction of $\sigma_i$ we have $\sigma_i(v_{-i}) = \sigma_i(v_{-i}')$, as desired.

\vspace{\baselineskip} \noindent \textsc{Claim~6:} For each $n^> \in \llbracket 1, n-2 \rrbracket$ such that $V_{-i}(n^>, 0) \in \mathcal{V}$, $\sigma_i$ is constant on $V_{-i}(n^>, 0)$.

\vspace{\baselineskip} Let $n^> \in \llbracket 1, n-2 \rrbracket$ be such that $V_{-i}(n^>, 0) \in \mathcal{V}$. The proof of this claim involves two subclaims.

The first subclaim is that for each $v_{-i} \in V_{-i}(n^>, 0)$, each $v_i \in V_i$ such that $v_i < p$, and each $j \in N \backslash \{i\}$ such that $v_j < p$, we have $\pgro_{v_{-j}} = p$. Indeed, let $v_{-i} \in V_{-i}(n^>, 0)$, $v_i \in V_i$, and $j \in N \backslash \{i\}$ satisfy the hypotheses. By Claim~3, we have $p = \mathsf{S}(n^>+1)$; thus at $v$, there are $n^>$ consumers with bids over $p = \mathsf{S}(n^>+1)$ and the others (including $i$ and $j$) have bids under $p = \mathsf{S}(n^>+1)$. It follows that $\qmax_{v_{-j}} = n^>$, and moreover that $\mathsf{D}_{v_{-j}}(\qmax_{v_{-j}}) = \mathsf{D}_{v_{-j}}(n^>) > p = \mathsf{S}(n^>+1) = \mathsf{S}(\qmax_{v_{-j}}+1)$, so $\pgro_{v_{-j}} = \min\{\mathsf{D}_{v_{-j}}(\qmax_{v_{-j}}), \mathsf{S}(\qmax_{v_{-j}}+1)\} = p$, as desired.

To state our second subclaim, we first introduce a definition. In particular, for each pair $v_{-i}, v_{-i}'' \in V_{-i}(n^>, 0)$, we say that $v_{-i}''$ is a {\it swap of $v_{-i}$} if and only if there are $j, k \in N \backslash \{i\}$ such that (i)~$v_k > p > v_j$, (ii)~$v_j'' > p > v_k''$, and (iii)~for each $i' \in N \backslash \{i, j, k\}$, we have $v_{i'} = v''_{i'}$. Observe that if $v_{-i}''$ is a swap of $v_{-i}$, then $v_{-i}$ is a swap of $v_{-i}''$.

The second subclaim is that for each pair $v_{-i}, v_{-i}'' \in V_{-i}(n^>, 0)$ such that $v_{-i}''$ is a swap of~$v_{-i}$, we have $\sigma_i(v_{-i}) = \sigma_i(v_{-i}'')$. Indeed, let $v_{-i}, v_{-i}'' \in V_{-i}(n^>, 0)$ satisfy the hypothesis; then there are $j, k \in N \backslash \{i\}$ such that (i)~$v_k > p > v_j$, (ii)~$v_j'' > p > v_k''$, and (iii)~for each $i' \in N \backslash \{i, j, k\}$, we have $v_{i'} = v''_{i'}$. Let $v^-_i \in V_i$ such that $v^-_i < \mathsf{S}(1)$, and define $v, v', v'' \in V$ by (i)~$v \equiv (v^-_i, v_{-i})$, (ii)~$v'' \equiv (v^-_i, v_{-i}'')$, and (iii)~$v' \equiv (v''_j, v_{-j}) = (v_k, v''_{-k})$. Since $V_{-i}(n^>, 0) \in \mathcal{V}$ we have $\pgro_{v_{-i}} = p = \pgro_{v''_{-i}}$, and by the first subclaim we have $\pgro_{v'_{-j}} = \pgro_{v_{-j}} = p = \pgro_{v''_{-k}} = \pgro_{v'_{-k}}$. Then (i)~$\pgro_{v_{-i}} = p > v_i$ and $\pgro_{v_{-j}} = p > v_j$, (ii)~$v'_j > p = \pgro_{v'_{-j}}$ and $v'_k > p = \pgro_{v'_{-k}}$, and (iii)~$\pgro_{v''_{-i}} = p > v''_i$ and $\pgro_{v''_{-k}} = p > v''_k$, so by the \hyperlink{GrovesLemma}{Groves~Lemma}, (i)~at $v$ both $i$ and $j$ lose, (ii)~at $v'$ both $j$ and $k$ win, and (iii)~at $v''$ both $i$ and $k$ lose. Then by (i)~{\it no-envy} at $v$, (ii)~the \hyperlink{GrovesLemma}{Groves~Lemma} and $\pgro_{v_{-j}} = \pgro_{v'_{-j}} = p$, (iii)~{\it no-envy} at $v'$, (iv)~the \hyperlink{GrovesLemma}{Groves~Lemma} and $\pgro_{v'_{-k}} = \pgro_{v''_{-k}} = p$, and (v)~{\it no-envy} at $v''$, we have $\tau_i(v) = \tau_j(v) = \tau_j(v') + p = \tau_k(v') + p = \tau_k(v'') = \tau_i(v'')$. Since $\alpha_i(v) = 0 = \alpha_i(v'')$, thus by construction of $\sigma_i$ we have $\sigma_i(v_{-i}) = \sigma_i(v_{-i}'')$, as desired.

To conclude, let $v_{-i}, v_{-i}' \in V_{-i}(n^>, 0)$. Since $n^> \in \llbracket 1, n-2 \rrbracket$, each peer profile in $V_{-i}(n^>,0)$ has at least one peer who reports more than $p$ and at least one peer who reports less than $p$. Let $(i_t)_{t \in \llbracket 1, n-1 \rrbracket}$ index the peers of $i$ in order of their bids at $v'_{-i}$; thus we have $\{i_t\}_{t \in \llbracket 1, n-1 \rrbracket} = N \backslash \{i\}$ and $v'_{i_1} \geq v'_{i_2} \geq ... \geq v'_{i_{n-1}}$. Clearly, there is $v^1_{-i} \in V_{-i}(n^>, 0)$ such that (i)~at $v^1_{-i}$, $i_1$ reports as at $v'_{-i}$, and (ii)~$v^1_{-i}$ is either a swap of $v_{-i}$ (if $v_{i_1} < p$) or a swap of {\it a swap of} $v_{-i}$ (if $v_{i_1} > p$), so by the second subclaim we have $\sigma_i(v_{-i}) = \sigma_i(v^1_{-i})$. Similarly, if $n^> \geq 2$, then there is $v^2_{-i} \in V_{-i}(n^>, 0)$ such that (i)~at $v^2_{-i}$, both $i_1$ and $i_2$ report as at~$v'_{-i}$, and (ii)~$v^2_{-i}$ is either a swap of $v^1_{-i}$ or a swap of {\it a swap of} $v^1_{-i}$, so by the second subclaim we have $\sigma_i(v_{-i}) = \sigma_i(v^1_{-i}) = \sigma_i(v^2_{-i})$. Proceeding in this fashion, there is $v^{n^>}_{-i} \in V_{-i}(n^>, 0)$ such that (i)~at $v^{n^>}_{-i}$, all consumers in $\{i_1, i_2, ..., i_{n^>}\}$ report as at $v'_{-i}$, and (ii)~$\sigma_i(v_{-i}) = \sigma_i(v^{n^>}_{-i})$. From here, let $v^-_i \in V_i$ such that $v^-_i < \mathsf{S}(1)$, define $v^{n^>} \equiv (v^-_i, v^{n^>}_{-i})$, and for each $t \in \llbracket n^>+1, n-1\rrbracket$ define $v^t \equiv (v'_{i_t}, v^{t-1}_{-i_t})$; observe that $v^{n-1} = (v^-_i, v'_{-i})$. For each $t \in \llbracket n^>, n-1\rrbracket$, (i)~since $v'_{-i}$ and $v^{n^>}_{-i}$ both belong to $V_{-i}(n^>, 0)$, thus by construction we have that $v^t_{-i} \in V_{-i}(n^>, 0)$ and each consumer in $\{i\} \cup \{i_{n^>+1}, i_{n^>+2}, ..., i_{n-1}\}$ reports less than $p$ at $v^t$, so (ii)~using $V_{-i}(n^>, 0) \in \mathcal{V}$ for $i$ and using the first subclaim for the consumers in $\{i_{n^>+1}, i_{n^>+2}, ..., i_{n-1}\}$ we have that each consumer in $\{i\} \cup \{i_{n^>+1}, i_{n^>+2}, ..., i_{n-1}\}$ faces Groves price $p$ at $v^t$; thus by the \hyperlink{GrovesLemma}{Groves~Lemma}, we have that each consumer in $\{i\} \cup \{i_{n^>+1}, i_{n^>+2}, ..., i_{n-1}\}$ loses at $v^t$. Since $t \in \llbracket n^>, n-1\rrbracket$ was arbitrary, it follows from iterative applications of {\it no-envy} and {\it strategy-proofness} that $\tau_i(v^{n^>}) = \tau_i(v^{n-1})$. Since moreover we have $\alpha_i(v^{n^>}) = 0 = \alpha_i(v^{n-1})$ and $v^{n-1}_{-i} = v'_{-i}$, altogether by construction of $\sigma_i$ we have $\sigma_i(v_{-i}) = \sigma_i(v^{n^>}_{-i}) = \sigma_i(v^{n-1}_{-i}) = \sigma_i(v'_{-i})$, as desired.

\vspace{\baselineskip} \noindent \textsc{Claim~7:} For each pair $n^>, n^= \in \llbracket 0, n-1 \rrbracket$, if both $V_{-i}(n^>, n^=)$ and $V_{-i}(n^>, n^=+1)$ belong to $\mathcal{V}$, and if $\sigma_i$ is constant on $V_{-i}(n^>, n^=)$, then $\sigma_i$ is constant on the union $V_{-i}(n^>, n^=+1) \cup V_{-i}(n^>, n^=)$.

\vspace{\baselineskip} Let $n^>, n^= \in \llbracket 0, n-1\rrbracket$ satisfy the hypotheses. Then there is $s \in \mathbb{R}$ such that for each $v_{-i} \in V_{-i}(n^>, n^=)$, $\sigma_i(v_{-i}) = s$.

Let $v_{-i} \in V_{-i}(n^>, n^=+1)$. Since $V_{-i}(n^>, n^=+1) \in \mathcal{V}$, thus $\pgro_{v_{-i}} = p$. Define $v \in V$ by $v \equiv (p, v_{-i})$. Since $v_{-i} \in V_{-i}(n^>, n^=+1)$, thus there is $j \in N \backslash \{i\}$ such that $v_j = p$. By the \hyperlink{GrovesLemma}{Groves~Lemma}, there is $\sigma_j(v_{-j}) \in \mathbb{R}$ such that for each $v'_j \in V_j$, we have $(\tau_j(v'_j, v_{-j}), \alpha_j(v'_j, v_{-j})) \in B^\delta_j(\sigma_j(v_{-j}), \pgro_{v_{-j}} | v'_j)$, and since $v_i = v_j$ we have $\pgro_{v_{-j}} = \pgro_{v_{-i}} = p$. Thus by {\it no-envy} at $v$, regardless of whether both $i$ and $j$ win, or both lose, or one wins and the other loses, we have $\sigma_i(v_{-i}) = \sigma_j(v_{-j})$.

Let $\varepsilon > 0$ and define $v^\varepsilon \in V$ by $v^\varepsilon \equiv (p - \varepsilon, v_{-j})$; thus we modify $v$ by having $j$ report less than $p$, so $v^\varepsilon_{-i} \in V_{-i}(n^>, n^=)$. Since $V_{-i}(n^>, n^=) \in \mathcal{V}$, thus $\pgro_{v^\varepsilon_{-i}} = p$, so by the \hyperlink{GrovesLemma}{Groves~Lemma}, at $v^\varepsilon$ either $i$ wins and receives $s - p$ or $i$ loses and receives~$s$. Moreover, $v^\varepsilon_{-j} = v_{-j}$, so $(\tau_j(v^\varepsilon), \alpha_j(v^\varepsilon)) \in B^\delta_j(\sigma_j(v_{-j}), p | p - \varepsilon)$, so at $v^\varepsilon$ we have that $j$ loses and receives $\sigma_j(v_{-j})$. If $i$ wins at $v^\varepsilon$, then by {\it no-envy}, $\sigma_j(v_{-j}) \geq (p - \varepsilon) + (s - p)$ and $p + (s - p) \geq \sigma_j(v_{-j})$. If~$i$ loses at $v^\varepsilon$, then by {\it no-envy}, $\sigma_j(v_{-j}) = s$. Thus in both cases, $\sigma_j(v_{-j}) \in [s - \varepsilon, s]$.

Since $\varepsilon > 0$ was arbitrary, thus $\sigma_j(v_{-j}) = s$, so $\sigma_i(v_{-i}) = s$. Since $v_{-i}$ was an arbitrary member of $V_{-i}(n^>, n^=+1)$, thus $\sigma_i$ is constant on $V_{-i}(n^>, n^=+1) \cup V_{-i}(n^>, n^=)$, as desired.

\vspace{\baselineskip} \noindent \textsc{Claim~8:} For each pair $n^>, n^= \in \llbracket 0, n-1 \rrbracket$, if both $V_{-i}(n^>, n^=)$ and $V_{-i}(n^>-1, n^=+1)$ belong to $\mathcal{V}$, and if $\sigma_i$ is constant on $V_{-i}(n^>, n^=)$, then $\sigma_i$ is constant on the union $V_{-i}(n^>-1, n^=+1) \cup V_{-i}(n^>, n^=)$.

\vspace{\baselineskip} Let $n^>, n^= \in \llbracket 0, n-1\rrbracket$ satisfy the hypotheses. Then there is $s \in \mathbb{R}$ such that for each $v_{-i} \in V_{-i}(n^>, n^=)$, $\sigma_i(v_{-i}) = s$.

Let $v_{-i} \in V_{-i}(n^>-1, n^=+1)$. Since $V_{-i}(n^>-1, n^=+1) \in \mathcal{V}$, thus $\pgro_{v_{-i}} = p$. Define $v \in V$ by $v \equiv (p, v_{-i})$. Since $v_{-i} \in V_{-i}(n^>-1, n^=+1)$, thus there is $j \in N \backslash \{i\}$ such that $v_j = p$. By the \hyperlink{GrovesLemma}{Groves~Lemma}, there is $\sigma_j(v_{-j}) \in \mathbb{R}$ such that for each $v'_j \in V_j$, we have $(\tau_j(v'_j, v_{-j}), \alpha_j(v'_j, v_{-j})) \in B^\delta_j(\sigma_j(v_{-j}), \pgro_{v_{-j}} | v'_j)$, and since $v_i = v_j$ we have $\pgro_{v_{-j}} = \pgro_{v_{-i}} = p$. Thus by {\it no-envy} at $v$, regardless of whether both $i$ and $j$ win, or both lose, or one wins and the other loses, we have $\sigma_i(v_{-i}) = \sigma_j(v_{-j})$.

Let $\varepsilon > 0$ and define $v^\varepsilon \in V$ by $v^\varepsilon \equiv (p + \varepsilon, v_{-j})$; thus we modify $v$ by having $j$ report more than $p$, so $v^\varepsilon_{-i} \in V_{-i}(n^>, n^=)$. Since $V_{-i}(n^>, n^=) \in \mathcal{V}$, thus $\pgro_{v^\varepsilon_{-i}} = p$, so by the \hyperlink{GrovesLemma}{Groves~Lemma}, at $v^\varepsilon$ either $i$ wins and receives $s - p$ or $i$ loses and receives~$s$. Moreover, $v^\varepsilon_{-j} = v_{-j}$, so $(\tau_j(v^\varepsilon), \alpha_j(v^\varepsilon)) \in B^\delta_j(\sigma_j(v_{-j}), p | p + \varepsilon)$, so at $v^\varepsilon$ we have that $j$ wins and receives $\sigma_j(v_{-j})-p$. If $i$ wins at $v^\varepsilon$, then by {\it no-envy}, $\sigma_j(v_{-j}) = s$. If $i$ loses at $v^\varepsilon$, then by {\it no-envy}, $(p + \varepsilon) + (\sigma_j(v_{-j}) - p) \geq s$ and $s \geq p + (\sigma_j(v_{-j}) - p)$. Thus in both cases, $\sigma_j(v_{-j}) \in [s - \varepsilon, s]$.

Since $\varepsilon > 0$ was arbitrary, thus $\sigma_j(v_{-j}) = s$, so $\sigma_i(v_{-i}) = s$. Since $v_{-i}$ was an arbitrary member of $V_{-i}(n^>-1, n^=+1)$, thus $\sigma_i$ is constant on $V_{-i}(n^>-1, n^=+1) \cup V_{-i}(n^>, n^=)$, as desired.

\vspace{\baselineskip} This completes the proof of the eight claims. To conclude, by Claim~1 and Claim~2, we have that $V^p_{-i} = \bigcup \mathcal{V}$, and moreover that $\mathcal{V}$ is the collection of nonempty classes $V_{-i}(n^>, n^=)$ such that (i)~$n^> \leq \qmax_{\mathsf{S}=p} - 1$, (ii)~$n^> + n^= \geq \qmin_{\mathsf{S}=p} - 1$, and (iii)~$n^> + n^= \leq n-1$. From here, we argue that $\sigma_i$ is constant on $\bigcup \mathcal{V}$. Indeed, if $\mathsf{S}(1) = \infty$, then $p = \infty$, so the only member of $\mathcal{V}$ is $V_{-i}(0, 0)$, so by Claim~4 we have that~$\sigma_i$ is constant on $\bigcup \mathcal{V}$, as desired. Thus let us assume $\mathsf{S}(1) \neq \infty$. In this case we have $p \neq \infty$, so each $V_{-i}(n^>, n^=)$ satisfying the above constraints is nonempty. Then $V_{-i}(\qmin_{\mathsf{S}=p} - 1, 0) \in \mathcal{V}$; thus by Claim~4, Claim~5, or Claim~6, we have that $\sigma_i$ is constant on $V_{-i}(\qmin_{\mathsf{S}=p} - 1, 0)$, so by Claim~7, $\sigma_i$ is constant on
\begin{align*}
\bigcup\{V_{-i}(n^>, n^=) \in \mathcal{V} | n^> = \qmin_{\mathsf{S}=p} -1\},
\end{align*}
so by Claim~8, $\sigma_i$ is constant on
\begin{align*}
\bigcup \{V_{-i}(n^>, n^=) \in \mathcal{V} | n^> \leq \qmin_{\mathsf{S}=p} - 1 \}.
\end{align*}
Suppose there are $n^>, n^= \in \llbracket 0, n-1 \rrbracket$ such that $V_{-i}(n^>, n^=) \in \mathcal{V}$ and $n^> > \qmin_{\mathsf{S}=p} - 1$. Then $V_{-i}(n^>, 0) \in \mathcal{V}$; thus by Claim~4, Claim~5, or Claim~6, $\sigma_i$ is constant on $V_{-i}(n^>, 0)$, so (i)~by Claim~7, $\sigma_i$ is constant on $V_{-i}(n^>, 0) \cup V_{-i}(n^>, n^=)$, and (ii)~by Claim~8, $\sigma_i$ is constant on $V_{-i}(n^>, 0) \cup V_{-i}(\qmin_{\mathsf{S}=p} - 1, n^> - (\qmin_{\mathsf{S}=p} - 1))$. Altogether, then, $\sigma_i$ is constant on $V_{-i}(\qmin_{\mathsf{S}=p} - 1, 0) \cup V_{-i}(n^>, n^=)$. Since $n^>, n^= \in \llbracket 0, n-1 \rrbracket$ were an arbitrary pair such that $V_{-i}(n^>, n^=) \in \mathcal{V}$ and $n^> > \qmin_{\mathsf{S}=p} - 1$, thus $\sigma_i$ is constant on $\bigcup \mathcal{V}$, as desired.

Since $v^*_{-i}, v^{**}_{-i} \in V^p_{-i}$, since $V^p_{-i} = \bigcup \mathcal{V}$, and since $\sigma_i$ is constant on $\bigcup \mathcal{V}$, thus we have $\sigma_i(v^*_{-i}) = \sigma_i(v^{**}_{-i})$, as desired.~$\blacksquare$

\vspace{\baselineskip} Second, we prove \hyperlink{Theorem1}{Theorem~1}.

\vspace{\baselineskip} \noindent \textsc{Proof of Theorem 1:} Since $(\tau, \alpha)$ is a Groves mechanism, thus by the \hyperlink{GrovesLemma}{Groves~Lemma} we have that for each $i \in N$, there is a subsidy function $\sigma_i:V_{-i} \to \mathbb{R}$ such that for each $v \in V$, we have $(\tau_i(v), \alpha_i(v)) \in B^\delta_i(\sigma_i(v_{-i}), \pgro_{v_{-i}} | v_i)$. We prove (i) implies (ii), (ii) implies (i), and (ii) is equivalent to (iii).

\begin{sloppypar}
\vspace{\baselineskip} \noindent \textsc{[(}i\textsc{)$\Rightarrow$(}ii\textsc{)]} Let $i \in N$, let $v \in V$, and let $v^-_i \in V_i$ be such that $v^-_i < \mathsf{S}(1)$. Then by {\it efficiency} we have $i \not \in \alpha(v^-_i, v_{-i})$, so $\tau_i(v^-_i, v_{-i}) = \sigma_i(v_{-i})$, and moreover we have $\max_{W \subseteq N} \mathsf{TS}_{(v^-_i, v_{-i})}(W) = \max_{W \subseteq N \backslash \{i\}} \mathsf{TS}_{(0, v_{-i})}(W)$, so by hypothesis $\tau_i(v^-_i, v_{-i}) = \varsigma(\pgro_{v_{-i}})$. Altogether, then, $\sigma_i(v_{-i}) = \varsigma(\pgro_{v_{-i}})$, so $(\tau_i(v), \alpha_i(v)) \in B^\delta_i(\varsigma(\pgro_{v_{-i}}), \pgro_{v_{-i}} | v_i)$, as desired.
\end{sloppypar}

\vspace{\baselineskip} \noindent \textsc{[(}ii\textsc{)$\Rightarrow$(}i\textsc{)]} Let $i \in N$. Since $(\tau, \alpha)$ is a Groves mechanism, thus there is $\beta_i: V_{-i} \to \mathbb{R}$ such that for each $v \in V$, we have
\begin{align*}
\tau_i(v) = \big( [\max_{W \subseteq N} \mathsf{TS}_v(W)] - \alpha_i(v) v_i \big) + \beta_i(v_{-i}).
\end{align*}
Let $v_{-i} \in V_{-i}$ and let $v^-_i \in V_i$ be such that $v^-_i < \mathsf{S}(1)$. Then $v^-_i < \pgro_{v_{-i}}$, so by hypothesis $\tau_i(v^-_i, v_{-i}) = \varsigma(\pgro_{v_{-i}})$ and $\alpha_i(v^-_i, v_{-i}) = 0$, and moreover we have $\max_{W \subseteq N} \mathsf{TS}_{(v^-_i, v_{-i})}(W) = \max_{W \subseteq N \backslash \{i\}} \mathsf{TS}_{(0, v_{-i})}(W)$. Altogether, then,
\begin{align*}
\beta_i(v_{-i}) = -[\max_{W \subseteq N \backslash \{i\}} \mathsf{TS}_{(0, v_{-i})}(W)] + [\varsigma(\pgro_{v_{-i}}) ].
\end{align*}
Since $i \in N$ and $v_{-i} \in V_{-i}$ were arbitrary, thus $(\tau, \alpha)$ is an equal subsidy auction supported by $\varsigma$, as desired.

\vspace{\baselineskip} \noindent \textsc{[(}ii\textsc{)$\Leftrightarrow$(}iii\textsc{)]} Let $i \in N$ and let $v \in V$. Observe that $\varsigma^\leftrightarrow(\pmin_v) = \varsigma(\pminup_v)$ and $\varsigma^\leftrightarrow(\pmax_v) = \varsigma(\pmaxdown_v)$; thus it suffices to prove the two statements are equivalent while ignoring the equality in statement (iii). We do so in each of three cases.

First, suppose $v_i \geq \pmax_v \geq \pmaxdown_v \geq \pminup_v = \pmin_v = \pgro_{v_{-i}}$ and $i \in \alpha(v)$. Since $\varsigma$ is a subsidy curve, thus either $\pmaxdown_v = \pminup_v$ or $1 \geq \frac{\varsigma(\pmaxdown_v) - \varsigma(\pminup_v)}{\pmaxdown_v-\pminup_v}$, so in either case $v_i \geq \pmaxdown_v = \pmin_v + [\pmaxdown_v - \pmin_v] \geq \pmin_v + [\pmaxdown_v - \pminup_v] \geq \pmin_v + [\varsigma(\pmaxdown_v) - \varsigma(\pminup_v)]$. Moreover,
\begin{align*}
\varsigma(\pgro_{v_{-i}}) - \pgro_{v_{-i}} &= \varsigma(\pminup_v) - \pmin_v
\\ &= \varsigma(\pmaxdown_v) - \Big(\pmin_v + [\varsigma(\pmaxdown_v)-\varsigma(\pminup_v)]\Big).
\end{align*}
The desired equivalence follows immediately.

Second, suppose $\pgro_{v_{-i}} = \pmax_v = \pmaxdown_v \geq \pminup_v \geq \pmin_v \geq v_i$ and $i \not \in \alpha(v)$. Since $\varsigma$ is a subsidy curve, thus either $\pmaxdown_v = \pminup_v$ or $\frac{\varsigma(\pmaxdown_v) - \varsigma(\pminup_v)}{\pmaxdown_v-\pminup_v} \geq 0$, so in either case $\pmin_v + [\varsigma(\pmaxdown_v) - \varsigma(\pminup_v)] \geq \pmin_v \geq v_i$. Moreover, $\varsigma(\pgro_{v_{-i}}) = \varsigma(\pmaxdown_v)$. The desired equivalence follows immediately.

Third, suppose we are in neither of the previous two cases. By the \hyperlink{PriceLemma}{Price~Lemma}, we cannot have $v_i > \pgro_{v_{-i}}$ (else we would be in the first case) or $v_i < \pgro_{v_{-i}}$ (else we would be in the second case), so again using the \hyperlink{PriceLemma}{Price~Lemma}, we have $\pmax_v = \pmaxdown_v = \pminup_v = \pmin_v = \pgro_{v_{-i}}$. Then (i)~$\varsigma(\pgro_{v_{-i}}) = \varsigma(\pmaxdown_v)$, (ii)~$\pgro_{v_{-i}} = \pmin_v$, and (iii)~$[\varsigma(\pmaxdown_v) - \varsigma(\pminup_v)] = 0$; thus the two budget sets are the same, and the desired conclusion follows immediately.~$\blacksquare$

\vspace{\baselineskip} To conclude this appendix, we prove \hyperlink{Theorem2}{Theorem~2}.

\vspace{\baselineskip} \noindent \textsc{Proof of Theorem 2:} We first establish the two implications for the first statement, then conclude by establishing the second statement.

\vspace{\baselineskip} \noindent \textsc{[$\Rightarrow$]} Let $(\tau, \alpha)$ satisfy the axioms. By the \hyperlink{InvarianceLemma}{Invariance~Lemma}, for each $i \in N$, there is $\varsigma_i: \mathbb{P} \to \mathbb{R}$ such that for each $v \in V$, $(\tau_i(v), \alpha_i(v)) \in B^\delta_i(\varsigma_i(\pgro_{v_{-i}}), \pgro_{v_{-i}} | v_i)$. We first claim that for each pair $i, j \in N$ we have $\varsigma_i = \varsigma_j$, which we prove in two cases.

First, if $\mathsf{S}(1) = \infty$, then let $v \in V$. For each $i \in N$, $\pgro_{v_{-i}} = \infty$, so $\tau_i(v) = \varsigma_i(\infty)$ and $\alpha_i(v) = 0$; thus by {\it no-envy}, for each pair $i, j \in N$ we have $\varsigma_i(\infty) = \varsigma_j(\infty)$, and thus as $\mathbb{P} = \{\infty\}$ we have $\varsigma_i = \varsigma_j$, as desired.

Second, if $\mathsf{S}(1) < \infty$, then $\mathbb{P} = [\mathsf{S}(1), \mathsf{S}(n)] \backslash \{\infty\}$. Let $p \in \mathbb{P}$ and let $v \in V$ be the profile where each consumer has valuation $p$. By the \hyperlink{IntervalLemma}{Interval~Lemma}, for each $i \in N$ we have $\pgro_{v_{-i}} = p$. Then at $v$, each winner $i$ receives transfer $\varsigma_i(p) - p$ and each loser $j$ receives transfer $\varsigma_j(p)$; thus by {\it no-envy}, for each pair $i, j \in N$ we have $\varsigma_i(p) = \varsigma_j(p)$. Since $p$ was arbitrary, thus for each pair $i, j \in N$ we have $\varsigma_i = \varsigma_j$, as desired.

By the claim, there is $\varsigma: \mathbb{P} \to \mathbb{R}$ such that for each $i \in N$ we have $\varsigma_i = \varsigma$. To conclude, we claim that $\varsigma$ is a subsidy curve. If $\mathbb{P}$ is a singleton then we are done; thus let us assume $\mathbb{P}$ is not a singleton. Let $q \in \llbracket 1, n-1 \rrbracket$ be such that $\mathsf{S}(q+1) > \mathsf{S}(q)$, let $p, p' \in [\mathsf{S}(q), \mathsf{S}(q+1)]$ be such that $p' > p$, let $v$ be a profile where $q$ consumers have valuation $p'$ and the others have valuation $p$, let $i \in N$ be such that $v_i = p'$, and let $j \in N$ be such that $v_j = p$. Then $\pgro_{v_{-i}} = p$ and $\pgro_{v_{-j}} = p'$, so by the \hyperlink{PriceLemma}{Price~Lemma}, $i$ wins and receives $\varsigma(p)-p$ and $j$ loses and receives $\varsigma(p')$. By {\it no-envy}, $p' + (\varsigma(p) - p) \geq \varsigma(p')$ and $\varsigma(p') \geq p + (\varsigma(p)-p)$, so $\frac{\varsigma(p') - \varsigma(p)}{p' - p} \in [0, 1]$. Since $p, p' \in [\mathsf{S}(q), \mathsf{S}(q+1)]$ were arbitrary, thus $\varsigma$ has the desired restriction for each price pair in $[\mathsf{S}(q), \mathsf{S}(q+1)]$. Since $q \in \llbracket 1, n-1\rrbracket$ such that $\mathsf{S}(q+1) > \mathsf{S}(q)$ was arbitrary, thus $\varsigma$ has the desired restriction for each price pair in a non-degenerate sub-interval of $\mathbb{P}$ bounded by consecutive marginal costs, and it is straightforward to show that this implies the desired restriction holds for each price pair in $\mathbb{P}$; we omit the details. Altogether, then, $\varsigma$ is a subsidy curve, so by \hyperlink{Theorem1}{Theorem~1} we have that $(\tau, \alpha)$ is an endogenous subsidy auction, as desired.

\vspace{\baselineskip} \noindent \textsc{[$\Leftarrow$]} Suppose $(\tau, \alpha)$ is an endogenous subsidy auction. Then $(\tau, \alpha)$ is a Groves mechanism, so it satisfies {\it efficiency} and {\it strategy-proofness}. Moreover, for each $v \in V$, by \hyperlink{Theorem1}{Theorem~1} we have that each consumer receives his most preferred bundle from a common budget set, so $(\tau, \alpha)$ is {\it envy-free}.

\vspace{\baselineskip} \noindent \textsc{Conclusion.} Let $\varsigma$ denote the subsidy curve and let $v \in V$. To conclude, we argue that the allocation selected at $v$ is an equal subsidy equilibrium in two cases.

First, suppose $\mathsf{S}(1) = \infty$. Then by the \hyperlink{QuantityLemma}{Quantity~Lemma}, we have $\alpha(v) = \emptyset$. Define $s_N \equiv \varsigma(\infty)$ and $p = \infty$; it is easy to verify that $(\tau(v), \alpha(v))$ is an equal subsidy equilibrium supported by $(s_N, p)$.

Second, suppose $\mathsf{S}(1) < \infty$. Then by the \hyperlink{QuantityLemma}{Quantity~Lemma}, we have $|\alpha(v)| \in \llbracket \qmin_v, \qmax_v \rrbracket$, and by \hyperlink{Theorem1}{Theorem~1}, we have that each consumer receives his most preferred bundle from the common budget set with subsidy $s_N = \varsigma(\pmaxdown_v)$ and price $p = \pmin_v + [\varsigma(\pmaxdown_v) - \varsigma(\pminup_v)]$. We claim that $p \in [\pmin_v, \pmax_v]$. Indeed, (i)~if $\pmaxdown_v = \pminup_v$, then $p = \pmin_v$, and (ii)~if $\pmaxdown_v \neq \pminup_v$, then by the \hyperlink{PriceLemma}{Price~Lemma} we have $\pmax_v \geq \pmaxdown_v > \pminup_v \geq \pmin_v$, so since $\varsigma$ is a subsidy curve we have $\pmax_v - \pmin_v \geq \pmaxdown_v - \pminup_v \geq \varsigma(\pmaxdown_v) - \varsigma(\pminup_v) \geq 0$, so $p \in [\pmin_v, \pmax_v]$; thus in both cases we have $p \in [\pmin_v, \pmax_v]$, as desired. Then $\mathsf{S}(\qmax_v+1) \geq \pmax_v \geq p \geq \pmin_v \geq \mathsf{S}(\qmax_v)$, so $\qmax_v$ is a profit-maximizing response to price $p$. Moreover, if $\qmax_v > \qmin_v$, then for each $q \in \llbracket \qmin_v+1, \qmax_v \rrbracket$ we have $\mathsf{D}_v(q) = \mathsf{S}(q) = \mathsf{D}_v(\qmax_v) = \mathsf{S}(\qmax_v)$, so $\pmin_v \geq \mathsf{S}(\qmax_v) = \mathsf{D}_v(\qmax_v) \geq \pmax_v$, so by the \hyperlink{PriceLemma}{Price~Lemma} $\pmax_v = \pmin_v = \mathsf{S}(\qmax_v)$, so since $p \in [\pmin_v, \pmax_v]$ we have $p = \mathsf{S}(\qmax_v) = \mathsf{S}(q)$. Altogether, then, any quantity in $\llbracket \qmin_v, \qmax_v \rrbracket$ is profit-maximizing in response to price $p$, so $|\alpha(v)|$ is profit-maximizing in response to price $p$, so $(\tau(v), \alpha(v))$ is an equal subsidy equilibrium supported by $(s_N, p)$, as desired.~$\blacksquare$

\hypertarget{AppendixC}{}
\setcounter{secnumdepth}{0}
\section{Appendix C: Proofs for Section 5.2}

In this appendix, we prove the \hyperlink{CeilingLemma}{Ceiling~Lemma} and \hyperlink{Theorem3}{Theorem~3}. We begin with the \hyperlink{CeilingLemma}{Ceiling~Lemma}.

\vspace{\baselineskip} \noindent \textsc{Proof of Ceiling~Lemma:} Let $\varsigma$ be a funded subsidy curve and let $p \in \mathbb{P}$. If $\mathsf{S}(1) = \infty$, then $p = \infty$ and $\piavg(p) = 0$, so by $[\mathcal{F}_1]$ we are done; thus let us assume $\mathsf{S}(1) \neq \infty$. Then $\qcap \geq 1$, and by the \hyperlink{IntervalLemma}{Interval~Lemma} there is $q \in \llbracket 1, \qcap \rrbracket$ such that $p \in \mathbb{P}_q$.

We first claim that $\piavg(\mathsf{S}(q)) + \tfrac{q}{n} \cdot [p - \mathsf{S}(q)] \leq \piavg(p)$. Indeed,
\begin{align*}
\piavg(\mathsf{S}(q)) + \tfrac{q}{n} \cdot [p - \mathsf{S}(q)] & = [\tfrac{1}{n}] \cdot [q \mathsf{S}(q) - \mathsf{C}(q)] + \tfrac{q}{n} \cdot [p - \mathsf{S}(q)]
\\ &= [\tfrac{1}{n}] \cdot [qp - \mathsf{C}(q)]
\\ &\leq \piavg(p),
\end{align*}
as desired. From here, we consider three cases.

First, if $q \in \llbracket 1, \qcap - 1 \rrbracket$, then $p \in \mathbb{P}_q = [\mathsf{S}(q), \mathsf{S}(q+1)]$. Then by the fact that $\varsigma$ is a subsidy curve, $[\mathcal{F}_2]$, and the claim,
\begin{align*}
\varsigma(p) &= [\tfrac{q}{n}] \cdot \varsigma(p) + [1-\tfrac{q}{n}]\varsigma(p)
\\ &\leq [\tfrac{q}{n}] \cdot [\varsigma(\mathsf{S}(q)) + p - \mathsf{S}(q)] + [1- \tfrac{q}{n}] \cdot \varsigma(\mathsf{S}(q+1))
\\ &= [\tfrac{q}{n}] \cdot \varsigma(\mathsf{S}(q)) + [1- \tfrac{q}{n}] \cdot \varsigma(\mathsf{S}(q+1)) + [\tfrac{q}{n}] \cdot [p - \mathsf{S}(q)]
\\ &\leq \piavg(\mathsf{S}(q)) + \tfrac{q}{n} \cdot [p - \mathsf{S}(q)]
\\ &\leq \piavg(p),
\end{align*}
as desired.

Second, if $q = \qcap = n$, then $p \in \mathbb{P}_q = \{\mathsf{S}(\qcap)\}$, so $p = \mathsf{S}(n)$. If $n = 1$, then $\piavg(p) = 0$, so we are done by $[\mathcal{F}_1]$. If $n>1$, then $n-1 \in \llbracket 1, \qcap - 1\rrbracket$ and $p \in \mathbb{P}_{n-1}$, so we are done by the previous case.

Third, if $q = \qcap \in \llbracket 1, n-1 \rrbracket$, then $p \in \mathbb{P}_q = [\mathsf{S}(\qcap), \infty)$. If $p = \mathsf{S}(\qcap)$, then (i)~$\qcap = 1$ implies that by $[\mathcal{F}_1]$ we have $\varsigma(p) = 0 = \piavg(p)$ as desired, and (ii)~$\qcap > 1$ implies that $p \in \mathbb{P}_{\qcap-1}$ and therefore we are done by the first case; thus let us assume $p \in (\mathsf{S}(\qcap), \infty)$. Then by the fact that $\varsigma$ is a subsidy curve, $[\mathcal{F}_3]$, and the claim,
\begin{align*}
\varsigma(p) &= [\tfrac{q}{n}] \cdot \varsigma(p) + [1-\tfrac{q}{n}]\varsigma(p)
\\ &\leq [\tfrac{q}{n}] \cdot [\varsigma(\mathsf{S}(q)) + p - \mathsf{S}(q)] + [1- \tfrac{q}{n}] \cdot \varsigma(p)
\\ &= [\tfrac{q}{n}] \cdot \varsigma(\mathsf{S}(q)) + [1- \tfrac{q}{n}] \cdot \varsigma(p) + [\tfrac{q}{n}] \cdot [p - \mathsf{S}(q)]
\\ &\leq \piavg(\mathsf{S}(q)) + \tfrac{q}{n} \cdot [p - \mathsf{S}(q)]
\\ &\leq \piavg(p),
\end{align*}
as desired.~$\blacksquare$

\vspace{\baselineskip} To conclude this appendix, we prove \hyperlink{Theorem3}{Theorem~3}.

\vspace{\baselineskip} \noindent \textsc{Proof of Theorem 3:} We first establish the two implications for the first statement, then conclude by establishing the second statement.

\vspace{\baselineskip} \noindent \textsc{[$\Rightarrow$]} Let $(\tau, \alpha)$ satisfy the axioms. By \hyperlink{Theorem2}{Theorem~2}, $(\tau, \alpha)$ is an endogenous subsidy auction; let $\varsigma$ be its subsidy curve. It remains to show that $\varsigma$ is funded, or that it satisfies properties $[\mathcal{F}_1]$, $[\mathcal{F}_2]$, and $[\mathcal{F}_3]$. We prove that each of these properties holds in sequence.

\vspace{\baselineskip} \noindent $[\mathcal{F}_1]$ Let $v \in V$ be defined in two cases: (i)~if $\mathsf{S}(1) \neq \infty$, then each agent has valuation $\mathsf{S}(1)-1$, and (ii)~if $\mathsf{S}(1) = \infty$, then each agent has valuation $0$. By the \hyperlink{QuantityLemma}{Quantity~Lemma}, $\alpha(v) = \emptyset$. Moreover, for each $i \in N$, $\qmax_{v_{-i}} = 0$, so $\pgro_{v_{-i}} = \mathsf{S}(1)$. Thus (i)~by {\it consumer voluntary participation}, for each $i \in N$ we have $\varsigma(\mathsf{S}(1)) = \tau_i(v) \geq 0$, and (ii)~by {\it producer voluntary participation} we have $0 = -\mathsf{C}(|\alpha(v)|) \geq \sum \tau_i(v) = n \cdot \varsigma(\mathsf{S}(1))$, so altogether we have $\varsigma(\mathsf{S}(1)) = 0$, as desired.

\vspace{\baselineskip} \noindent $[\mathcal{F}_2]$ Let $q \in \llbracket 1, \qcap-1 \rrbracket$. Let $v \in V$ be such that $q$ consumers have valuation $\mathsf{S}(q+1) + 1$ and the others have valuation $\mathsf{S}(1) -1$. By the \hyperlink{QuantityLemma}{Quantity~Lemma}, $|\alpha(v)| = q$. Moreover, $\qmax_v = q$, $\pmin_v = \mathsf{S}(q)$, and $\pmax_v = \mathsf{S}(q+1)$. By the \hyperlink{PriceLemma}{Price~Lemma}, (i)~for each $i \in \alpha(v)$ we have $\pgro_{v_{-i}} = \mathsf{S}(q)$, and (ii)~for each $i \in N \backslash \alpha(v)$ we have $\pgro_{v_{-i}} = \mathsf{S}(q+1)$. Altogether, then, by {\it producer voluntary participation} we have $-\mathsf{C}(q) = -\mathsf{C}(|\alpha(v)|) \geq \sum \tau_i(v) = q \cdot ( \varsigma(\mathsf{S}(q)) - \mathsf{S}(q)) + (n-q) \cdot \varsigma(\mathsf{S}(q+1))$, from which the desired inequality follows: simply add $q \cdot \mathsf{S}(q)$ to both sides and then divide both sides by $n$.

\vspace{\baselineskip} \noindent $[\mathcal{F}_3]$ If $\qcap \in \llbracket 1, n-1 \rrbracket$ and thus $\mathsf{S}(1) \neq \infty$, then let $p \in (\mathsf{S}(\qcap), \infty)$ and let $v \in V$ be such that $\qcap$ consumers have valuation $p$ and the others have valuation $\mathsf{S}(1)-1$. By the \hyperlink{QuantityLemma}{Quantity~Lemma}, $|\alpha(v)| = \qcap$. Moreoever, $\qmax_v = \qcap$, $\pmin_v = \mathsf{S}(\qcap)$, and $\pmax_v = p$. By the \hyperlink{PriceLemma}{Price~Lemma}, (i)~for each $i \in \alpha(v)$ we have $\pgro_{v_{-i}} = \mathsf{S}(\qcap)$, and (ii)~for each $i \in N \backslash \alpha(v)$ we have $\pgro_{v_{-i}} = p$. Altogether, then, by {\it producer voluntary participation} we have $-\mathsf{C}(\qcap) = -\mathsf{C}(|\alpha(v)|) \geq \sum \tau_i(v) = \qcap \cdot ( \varsigma(\mathsf{S}(\qcap)) - \mathsf{S}(\qcap)) + (n-\qcap) \cdot \varsigma(p)$, from which the desired inequality follows: simply add $\qcap \cdot \mathsf{S}(\qcap)$ to both sides and then divide both sides by $n$.

\vspace{\baselineskip} \noindent \textsc{[$\Leftarrow$]} Let $(\tau, \alpha)$ be an endogenous shareholding auction and let $\varsigma$ be its funded subsidy curve. By \hyperlink{Theorem2}{Theorem~2}, $(\tau, \alpha)$ satisfies {\it production efficiency}, {\it strategy-proofness}, and {\it no-envy}. For each $i \in N$ and each $v \in V$, $\varsigma(\pgro_{v_{-i}}) \geq \varsigma(\mathsf{S}(1)) = 0$; it follows immediately from \hyperlink{Theorem1}{Theorem~1} that $(\tau, \alpha)$ satisfies {\it consumer voluntary participation}.

It remains to show that $(\tau, \alpha)$ satisfies {\it producer voluntary participation}. Let $v \in V$ and define $q \equiv |\alpha(v)|$. By \hyperlink{Theorem1}{Theorem~1} and the \hyperlink{PriceLemma}{Price~Lemma}, (i)~for each $i \in \alpha(v)$ we have $\tau_i(v) = \varsigma(\pmin_v)-\pmin_v$, and (ii)~for each $i \in N \backslash \alpha(v)$ we have $\tau_i(v) = \varsigma(\pmax_v)$. By {\it efficiency}, $q \in \llbracket 0, \qcap \rrbracket$. We consider four cases.

\vspace{\baselineskip} \noindent \textsc{Case 1:} $q = 0$. Then by the \hyperlink{QuantityLemma}{Quantity~Lemma}, $\mathsf{S}(1) \geq \mathsf{D}_v(1)$. If $\qmax_v = 0$, then $\pmax_v = \mathsf{S}(1)$. If $\qmax_v > 0$, then by the \hyperlink{PriceLemma}{Price~Lemma}, $\mathsf{S}(1) \geq \mathsf{D}_v(1) \geq \mathsf{D}_v(\qmax_v) \geq \pmax_v \geq \pmin_v \geq \mathsf{S}(\qmax_v) \geq \mathsf{S}(1)$, so $\pmax_v = \mathsf{S}(1)$. Thus in both cases, $\pmax_v = \mathsf{S}(1)$, so by $[\mathcal{F}_1]$ we have
\begin{align*}
\sum \tau_i(v) &= n \cdot \varsigma(\mathsf{S}(1))
\\ &= 0
\\ &= -\mathsf{C}(|\alpha(v)|),
\end{align*}
as desired.

\vspace{\baselineskip} \noindent \textsc{Case 2:} $q \in \llbracket 1, \qcap-1 \rrbracket$. By the \hyperlink{QuantityLemma}{Quantity~Lemma}, $q \leq \qmax_v$. We make two claims, then conclude.

First, we claim $\varsigma(\pmin_v) - \pmin_v \leq \varsigma(\mathsf{S}(q)) - \mathsf{S}(q)$. Indeed, $\mathsf{S}(q) \leq \mathsf{S}(\qmax_v) \leq \pmin_v$, and if $\mathsf{S}(q) = \pmin_v$ then we are done, so assume $\mathsf{S}(q) < \pmin_v$. Then since $\varsigma$ is a subsidy curve, thus $\frac{\varsigma(\pmin_v) - \varsigma(\mathsf{S}(q))}{\pmin_v - \mathsf{S}(q)} \leq 1$, so $\varsigma(\pmin_v) - \pmin_v \leq \varsigma(\mathsf{S}(q)) - \mathsf{S}(q)$, as desired.

Second, we claim $\varsigma(\pmax_v) \leq \varsigma(\mathsf{S}(q+1))$. Indeed, if $q = \qmax_v$, then $\pmax_v \leq \mathsf{S}(\qmax_v+1) = \mathsf{S}(q+1)$, so since $\varsigma$ is a subsidy curve we are done. If instead $q < \qmax_v$, then by the \hyperlink{QuantityLemma}{Quantity~Lemma} and the \hyperlink{PriceLemma}{Price~Lemma}, $\mathsf{S}(q+1) \geq \mathsf{D}_v(q+1) \geq \mathsf{D}_v(\qmax_v) \geq \pmax_v \geq \pmin_v \geq \mathsf{S}(\qmax_v) \geq \mathsf{S}(q+1)$, so $\pmax_v = \mathsf{S}(q+1)$ and we are done. Thus in both cases we have the desired inequality.

To conclude, by the two claims and $[\mathcal{F}_2]$, we have
\begin{align*}
\sum \tau_i(v) &= q \cdot [\varsigma(\pmin_v) - \pmin_v] + (n-q) \cdot \varsigma(\pmax_v)
\\ &\leq q \cdot [\varsigma(\mathsf{S}(q)) - \mathsf{S}(q)] + (n-q) \cdot \varsigma(\mathsf{S}(q+1))
\\ &= n \cdot \Big( [\tfrac{q}{n}] \cdot \varsigma(\mathsf{S}(q)) + [1-\tfrac{q}{n}] \cdot \varsigma(\mathsf{S}(q+1)) \Big) - q \cdot \mathsf{S}(q)
\\ & \leq n \cdot \piavg(\mathsf{S}(q)) - q \cdot \mathsf{S}(q)
\\ &= n \cdot \tfrac{1}{n} \cdot [q \cdot \mathsf{S}(q) - \mathsf{C}(q)] - q \cdot \mathsf{S}(q)
\\ &= -\mathsf{C}(q)
\\ &= -\mathsf{C}(|\alpha(v)|),
\end{align*}
as desired.

\vspace{\baselineskip} \noindent \textsc{Case 3:} $q = \qcap$ and $q \not \in \{0, n\}$. In this case, $\mathsf{S}(1) \neq \infty$ and $\mathsf{S}(n) = \infty$, so (i)~$\mathbb{P} = [\mathsf{S}(1), \infty)$ and thus $\pmax_v+1 \in \mathbb{P}$, and (ii)~$\qcap = q \leq \qmax_v \leq \qcap$ and thus $\qcap = \qmax_v$. Then by the \hyperlink{PriceLemma}{Price~Lemma} we have $\mathsf{S}(\qcap) = \mathsf{S}(\qmax_v) \leq \pmin_v \leq \pmax_v < \pmax_v + 1$. From here, we make two claims similar to those in the previous case, then conclude.

First, we claim $\varsigma(\pmin_v) - \pmin_v \leq \varsigma(\mathsf{S}(\qcap)) - \mathsf{S}(\qcap)$. Indeed, $\mathsf{S}(\qcap) \leq \pmin_v$, and if $\mathsf{S}(\qcap) = \pmin_v$ then we are done, so assume $\mathsf{S}(\qcap) < \pmin_v$. Then since $\varsigma$ is a subsidy curve, thus $\frac{\varsigma(\pmin_v) - \varsigma(\mathsf{S}(\qcap))}{\pmin_v - \mathsf{S}(\qcap)} \leq 1$, so $\varsigma(\pmin_v) - \pmin_v \leq \varsigma(\mathsf{S}(\qcap)) - \mathsf{S}(\qcap)$, as desired.

Second, we claim $\varsigma(\pmax_v) \leq \varsigma(\pmax_v+1)$. Indeed, since $\varsigma$ is a subsidy curve we are done.

To conclude, by the two claims and $[\mathcal{F}_3]$, we have
\begin{align*}
\sum \tau_i(v) &= \qcap \cdot [\varsigma(\pmin_v) - \pmin_v] + (n-\qcap) \cdot \varsigma(\pmax_v)
\\ & \leq \qcap \cdot [\varsigma(\mathsf{S}(\qcap)] - \mathsf{S}(\qcap)] + (n-\qcap) \cdot \varsigma(\pmax_v + 1)
\\ &= n \cdot \Big( [\tfrac{\qcap}{n}] \cdot \varsigma(\mathsf{S}(\qcap)) + [1-\tfrac{\qcap}{n}] \cdot \varsigma(\pmax_v + 1) \Big) - \qcap \cdot \mathsf{S}(\qcap)
\\ & \leq n \cdot \piavg(\mathsf{S}(\qcap)) - \qcap \cdot \mathsf{S}(\qcap)
\\ &= n \cdot \tfrac{1}{n} \cdot [\qcap \cdot \mathsf{S}(\qcap) - \mathsf{C}(\qcap)] - \qcap \cdot \mathsf{S}(\qcap)
\\ &= -\mathsf{C}(\qcap)
\\ & = -\mathsf{C}(|\alpha(v)|),
\end{align*}
as desired.

\vspace{\baselineskip} \noindent \textsc{Case 4:} $q = \qcap = n$. Then by the \hyperlink{QuantityLemma}{Quantity~Lemma}, $\mathsf{D}_v(n) \geq \mathsf{S}(n)$, so $\qmax_v = n$, so $\pmin_v = \mathsf{S}(n)$. By the \hyperlink{CeilingLemma}{Ceiling~Lemma}, $\varsigma(\mathsf{S}(n)) \leq \piavg(\mathsf{S}(n))$, so $n \cdot \varsigma(\mathsf{S}(n)) \leq n \cdot \mathsf{S}(n) - \mathsf{C}(n)$, so
\begin{align*}
\sum \tau_i(v) &= n \cdot [\varsigma(\pmin_v) - \pmin_v]
\\ &= n \cdot [\varsigma(\mathsf{S}(n)) - \mathsf{S}(n)]
\\ &\leq - \mathsf{C}(n)
\\ &= -\mathsf{C}(|\alpha(v)|),
\end{align*}
as desired.

\vspace{\baselineskip} \noindent \textsc{Conclusion.} Let $v \in V$. To conclude, we argue that the allocation selected at $v$ is an equal shareholding equilibrium. We consider two cases.

First, suppose $\mathsf{S}(1) = \infty$. As argued in the conclusion of the proof of \hyperlink{Theorem2}{Theorem~2}, we have that $(\tau(v), \alpha(v))$ is an equal subsidy equilibrium supported by $(\varsigma(\infty), \infty)$. Then $|\alpha(v)| = 0$, and moreover by $[\mathcal{F}_1]$ we have $\varsigma(\infty) = \varsigma(\mathsf{S}(1)) = 0$. Thus for each $\textfrak{s:}_N \in [0, \frac{1}{n}]$, $(\tau(v), \alpha(v))$ is an equal shareholding equilibrium supported by $(\textfrak{s:}_N, \infty)$.

Second, suppose $\mathsf{S}(1) < \infty$. Define $s_0 \equiv \varsigma(\pmaxdown_v)$ and $p \equiv \pmin_v + [\varsigma(\pmaxdown_v) - \varsigma(\pminup_v)]$. As argued in the conclusion of the proof of \hyperlink{Theorem2}{Theorem~2}, we have that $(\tau(v), \alpha(v))$ is an equal subsidy equilibrium supported by $(s_0, p)$. Since $\varsigma$ is a funded subsidy curve, thus $s_0 = \varsigma(\pmaxdown_v) \geq \varsigma(\mathsf{S}(1)) = 0$. Moreover, by {\it producer voluntary participation}, we have that $n \cdot s_0 - |\alpha(v)| \cdot p = \sum \tau_i(v) \leq - \mathsf{C}(|\alpha(v)|)$, so $s_0 \leq \frac{1}{n} \cdot (p \cdot |\alpha(v)| - \mathsf{C}(|\alpha(v)|))$. If the profit $p \cdot |\alpha(v)| - \mathsf{C}(|\alpha(v)|)$ is zero, then $s_0 = 0$, so for each $\textfrak{s:}_N \in [0, \frac{1}{n}]$ we have that $(\tau(v), \alpha(v))$ is an equal shareholding equilibrium supported by $(\textfrak{s:}_N, p)$, as desired. If the profit is positive, then define $\textfrak{s:}_N \equiv \frac{s_0}{p \cdot |\alpha(v)| - \mathsf{C}(|\alpha(v)|)}$; then $\textfrak{s:}_N \in [0, \frac{1}{n}]$ and $(\tau(v), \alpha(v))$ is an equal shareholding equilibrium supported by $(\textfrak{s:}_N, p)$, as desired.~$\blacksquare$

\hypertarget{AppendixD}{}
\setcounter{secnumdepth}{0}
\section{Appendix D: Proofs for Section 5.3}

In this appendix, we prove \hyperlink{Theorem4}{Theorem~4}, the \hyperlink{DominationLemma}{Domination~Lemma}, and \hyperlink{Theorem5}{Theorem~5}. We begin with \hyperlink{Theorem4}{Theorem~4}.

\vspace{\baselineskip} \noindent \textsc{Proof of Theorem 4:} We prove both implications in sequence.

\vspace{\baselineskip} \noindent \textsc{[$\Rightarrow$]} If $\mathsf{S}(1) = \infty$, then each endogenous shareholding auction is a VCG mechanism, so we are done; thus let us assume $\mathsf{S}(1) < \infty$. Let $(\tau^*, \alpha^*)$ be a weakly producer-dominant endogenous shareholding auction, let $(\tau, \alpha)$ be a VCG mechanism, let $\varsigma^*$ denote the funded subsidy curve of $(\tau^*, \alpha^*)$, let $p \in \mathbb{P}$, and let $v \in V$ be such that each consumer has valuation $p$.

Since $(\tau^*, \alpha^*)$ weakly producer-dominates $(\tau, \alpha)$, thus
\begin{align*}
\mathsf{TS}_v(\alpha^*(v)) - \mathsf{CS}_v(\tau^*(v), \alpha^*(v)) &= \mathsf{PS}(\tau^*(v), \alpha^*(v))
\\ &\geq \mathsf{PS}(\tau(v), \alpha(v))
\\ &= \mathsf{TS}_v(\alpha(v)) - \mathsf{CS}_v(\tau(v), \alpha(v)).
\end{align*}
Moreover, since both $(\tau^*, \alpha^*)$ and $(\tau, \alpha)$ are endogenous shareholding auctions, thus we have $\mathsf{TS}_v(\alpha^*(v)) = \mathsf{TS}_v(\alpha(v))$, so by \hyperlink{Theorem1}{Theorem~1} and the \hyperlink{IntervalLemma}{Interval~Lemma},
\begin{align*}
0 &= \sum_{i \in \alpha(v)} [v_i - p]
\\&= \sum_{i \in N} [v_i \cdot \alpha_i(v) + \tau_i(v)]
\\ &= \mathsf{CS}_v(\tau(v), \alpha(v))
\\ &\geq \mathsf{CS}_v(\tau^*(v), \alpha^*(v))
\\ &= \sum_{i \in N} [v_i \cdot \alpha^*_i(v) + \tau^*_i(v)]
\\ &= \sum_{i \in \alpha^*(v)} [v_i - p] + n \cdot \varsigma^*(p)
\\ &= n \cdot \varsigma^*(p).
\end{align*}
Thus $\varsigma^*(p) \leq 0$, and since $\varsigma^*$ is a funded subsidy curve we have $\varsigma^*(p) \geq 0$, so altogether we have $\varsigma^*(p) = 0$. Since $p \in \mathbb{P}$ was arbitrary, thus $\varsigma^*$ assigns zero to each price in $\mathbb{P}$, so by the \hyperlink{VCGLemma}{VCG~Lemma} we have that $(\tau^*, \alpha^*)$ is a VCG mechanism, as desired.

\vspace{\baselineskip} \noindent \textsc{[$\Leftarrow$]} If $\mathsf{S}(1) = \infty$, then each endogenous shareholding auction is a VCG mechanism. Moreover, it is straightforward to show that in this case, each endogenous shareholding auction always selects no winners and assigns no transfers, and thus always selects an allocation with zero producer surplus, so we are done. Thus let us assume $\mathsf{S}(1) < \infty$.

Let $(\tau^*, \alpha^*)$ be a VCG mechanism and assume, by way of contradiction, that $(\tau^*, \alpha^*)$ is not producer-optimal. Then there are endogenous shareholding auction $(\tau, \alpha)$ and $v \in V$ such that
\begin{align*}
\mathsf{TS}_v(\alpha(v)) - \mathsf{CS}_v(\tau(v), \alpha(v)) &= \mathsf{PS}(\tau(v), \alpha(v))
\\ &> \mathsf{PS}(\tau^*(v), \alpha^*(v))
\\ &= \mathsf{TS}_v(\alpha^*(v)) - \mathsf{CS}_v(\tau^*(v), \alpha^*(v)).
\end{align*}
Define $q \equiv |\alpha(v)|$. Since both $(\tau^*, \alpha^*)$ and $(\tau, \alpha)$ are endogenous shareholding auctions, thus we have $\mathsf{TS}_v(\alpha^*(v)) = \mathsf{TS}_v(\alpha(v))$, so by the \hyperlink{VCGLemma}{VCG~Lemma}, \hyperlink{Theorem1}{Theorem~1}, and the \hyperlink{PriceLemma}{Price~Lemma},
\begin{align*}
\sum_{i \in \alpha^*(v)} [v_i - \pmin_v] &= \sum_{i \in N} [v_i \cdot \alpha^*_i(v) + \tau^*_i(v)]
\\ &= \mathsf{CS}_v(\tau^*(v), \alpha^*(v))
\\ &> \mathsf{CS}_v(\tau(v), \alpha(v))
\\ &= \sum_{i \in N} [v_i \cdot \alpha_i(v) + \tau_i(v)]
\\ &= \sum_{i \in \alpha(v)} [v_i + \varsigma(\pminup_v) - \pmin_v] + (n-q) \cdot \varsigma(\pmaxdown_v)
\\ &= \sum_{i \in \alpha(v)} [v_i - \pmin_v] + q \cdot \varsigma(\pminup_v) + (n-q) \cdot \varsigma(\pmaxdown_v).
\end{align*}
Since $(\tau^*, \alpha^*)$ and $(\tau, \alpha)$ are {\it efficient}, thus by the \hyperlink{PriceLemma}{Price~Lemma}, for each $i \in N$ we have that (i)~$v_i > \pmin_v$ implies $i \in \alpha(v)$ and $i \in \alpha^*(v)$, and (ii)~$\pmin_v > v_i$ implies $i \not \in \alpha(v)$ and $i \not \in \alpha^*(v)$; thus $\sum_{i \in \alpha^*(v)} [v_i - \pmin_v] = \sum_{i \in \alpha(v)} [v_i - \pmin_v]$. But then we have $q \cdot \varsigma(\pminup_v) + (n-q) \cdot \varsigma(\pmaxdown_v) < 0$, so either $\varsigma(\pminup_v) < 0$ or $\varsigma(\pmaxdown_v) < 0$, contradicting that $\varsigma$ is a funded subsidy curve.~$\blacksquare$

\vspace{\baselineskip} Second, we prove the \hyperlink{DominationLemma}{Domination~Lemma}.

\vspace{\baselineskip} \noindent \textsc{Proof of Domination~Lemma:} We prove both implications in sequence.

\vspace{\baselineskip} \noindent \textsc{[$\Rightarrow$]} If $\mathsf{S}(1) = \infty$, then both $\alpha^*$ and $\alpha$ always select the empty set, so there are $v \in V$ and $i \in N$ such that $\varsigma^*(\infty) = \tau^*_i(v) > \tau_i(v) = \varsigma(\infty)$, so as $\mathbb{P} = \{\infty\}$ we are done; thus let us assume $\mathsf{S}(1) < \infty$. First, let $p \in \mathbb{P}$ and let $v \in V$ be such that each consumer has valuation $p$. By the \hyperlink{IntervalLemma}{Interval~Lemma} and \hyperlink{Theorem1}{Theorem~1}, for each $i \in N$, $\varsigma^*(p) = v_i \cdot \alpha^*_i(v) + \tau^*_i(v) \geq v_i \cdot \alpha_i(v) + \tau_i(v) = \varsigma(p)$, as desired. Second, by \hyperlink{Theorem1}{Theorem~1}, there are $v \in V$ and $i \in N$ such that $\varsigma^*(\pgro_{v_{-i}}) + \max\{v_i - \pgro_{v_{-i}}, 0\} = v_i \cdot \alpha^*_i(v) + \tau^*_i(v) > v_i \cdot \alpha_i(v) + \tau_i(v) = \varsigma(\pgro_{v_{-i}}) + \max\{v_i - \pgro_{v_{-i}}, 0\}$, so $\varsigma^*(\pgro_{v_{-i}}) > \varsigma(\pgro_{v_{-i}})$, as desired.

\vspace{\baselineskip} \noindent \textsc{[$\Leftarrow$]} If $\mathsf{S}(1) = \infty$, then both $\alpha^*$ and $\alpha$ always select the empty set, so for each $v \in V$ and each $i \in N$ we have $\tau^*_i(v) = \varsigma^*(\infty) > \varsigma(\infty) = \tau_i(v)$ and we are done; thus let us assume $\mathsf{S}(1) < \infty$. First, by \hyperlink{Theorem1}{Theorem~1}, for each $v \in V$ and each $i \in N$ we have $v_i \cdot \alpha^*_i(v) + \tau^*_i(v) = \varsigma^*(\pgro_{v_{-i}}) + \max\{v_i - \pgro_{v_{-i}}, 0\} \geq \varsigma(\pgro_{v_{-i}}) + \max\{v_i - \pgro_{v_{-i}}, 0\} = v_i \cdot \alpha_i(v) + \tau_i(v)$, as desired. Second, there is $p \in \mathbb{P}$ such that $\varsigma^*(p) > \varsigma(p)$, so by the \hyperlink{IntervalLemma}{Interval~Lemma} and \hyperlink{Theorem1}{Theorem~1}, for each $i \in N$ and for the $v \in V$ such that each consumer has valuation $p$, we have $v_i \cdot \alpha^*_i(v) + \tau^*_i(v) = \varsigma^*(p) > \varsigma(p) = v_i \cdot \alpha_i(v) + \tau_i(v)$, as desired.~$\blacksquare$

\vspace{\baselineskip} To conclude this appendix, we prove \hyperlink{Theorem5}{Theorem~5}.

\vspace{\baselineskip} \noindent \textsc{Proof of Theorem 5:} Let us say that a subsidy curve $\varsigma$ is {\it on-off} if there is a vector of cutoff prices $\kappa \in \mathbb{P}^{\qcap}$ such that $(\varsigma, \kappa)$ satisfies $[\mathcal{V}_1]$ and $[\mathcal{V}_2]$, and {\it pre-valvular} if there is a vector of cutoff prices $\kappa \in \mathbb{P}^{\qcap}$ such that $(\varsigma, \kappa)$ satisfies $[\mathcal{V}_1]$, $[\mathcal{V}_2]$ and $[\mathcal{V}_3]$. The proof consists of three steps.

\vspace{\baselineskip} \noindent \textsc{Step 1:} If a funded subsidy curve is not on-off, then it is dominated by a pre-valvular subsidy curve.

\vspace{\baselineskip} Let $\varsigma$ be a funded subsidy curve that is not on-off. Then $\mathsf{S}(1) < \infty$, so $\qcap \geq 1$. In the next three paragraphs, we define a vector of cutoff prices $\kappa \in \mathbb{P}^{\qcap}$ such that for each $q \in \llbracket 1, \qcap \rrbracket$, $\kappa_q \in \mathbb{P}_q$.

First, for each $q \in \llbracket 1, \qcap - 1\rrbracket$, define $\kappa_q \equiv \mathsf{S}(q) + [\varsigma(\mathsf{S}(q+1)) - \varsigma(\mathsf{S}(q))]$. Since $\varsigma$ is a subsidy curve, thus for each $q \in \llbracket 1, \qcap - 1\rrbracket$ we have $\kappa_q \in [\mathsf{S}(q), \mathsf{S}(q+1)] = \mathbb{P}_q$, as desired.

Second, if $\qcap = n$, then define $\kappa_{\qcap} = \mathsf{S}(\qcap)$. In this case, we have $\kappa_{\qcap} \in \{\mathsf{S}(\qcap)\} = \mathbb{P}_{\qcap}$, as desired.

Third, if $\qcap < n$, then by the \hyperlink{CeilingLemma}{Ceiling~Lemma} we have $\varsigma(\mathsf{S}(\qcap)) \leq \piavg(\mathsf{S}(\qcap))$. Moreover, on $\mathbb{P}_{\qcap} = [\mathsf{S}(\qcap), \infty)$, the function $p \mapsto \varsigma(\mathsf{S}(\qcap)) + (p - \mathsf{S}(\qcap)) - \piavg(p)$ has a derivative with respect to $p$ of $1 - \frac{\qcap}{n} > 0$; thus this is a line with positive slope that is non-positive at $\mathsf{S}(\qcap)$, so
\begin{align*}
\kappa_{\qcap} \equiv \max \{p \in [\mathsf{S}(\qcap), \infty) | \varsigma(\mathsf{S}(\qcap)) + (p - \mathsf{S}(\qcap)) \leq \piavg(p)\}
\end{align*}
is well-defined. By construction, $\kappa_{\qcap} \in [\mathsf{S}(\qcap), \infty) = \mathbb{P}_{\qcap}$, as desired. Moreover, observe that by construction we have $\varsigma(\mathsf{S}(\qcap)) + (\kappa_{\qcap} - \mathsf{S}(\qcap)) = \piavg(\kappa_{\qcap})$.

Define $\varsigma^*: \mathbb{P} \to \mathbb{R}$ to be the function such that $(\varsigma^*, \kappa)$ satisfies $[\mathcal{V}_1]$ and $[\mathcal{V}_2]$. It is straightforward to show that $\varsigma^*$ is a subsidy curve. Moreover, (i)~$\varsigma^*(\mathsf{S}(1)) = 0 = \varsigma(\mathsf{S}(1))$, and (ii)~for each $q \in \llbracket 1, \qcap-1 \rrbracket$ such that $\varsigma^*(\mathsf{S}(q)) = \varsigma(\mathsf{S}(q))$, we have
\begin{align*}
\varsigma^*(\mathsf{S}(q+1)) &= \varsigma^*(\mathsf{S}(q)) + (\kappa_q - \mathsf{S}(q))
\\ &= \varsigma^*(\mathsf{S}(q)) + (\mathsf{S}(q) + [\varsigma(\mathsf{S}(q+1)) - \varsigma(\mathsf{S}(q))] - \mathsf{S}(q))
\\ &= \varsigma^*(\mathsf{S}(q)) + (\mathsf{S}(q) + [\varsigma(\mathsf{S}(q+1)) - \varsigma^*(\mathsf{S}(q))] - \mathsf{S}(q))
\\ &= \varsigma(\mathsf{S}(q+1));
\end{align*}
thus by induction, $\varsigma^*$ and $\varsigma$ agree at all finite marginal costs.

We claim that $\varsigma^*$ dominates $\varsigma$. Indeed, since $\varsigma^*$ and $\varsigma$ are both subsidy curves, since $\varsigma$ is funded, and since $\varsigma^*$ is on-off while $\varsigma$ is not, it is straightforward to show that for each $p \in [\mathsf{S}(1), \kappa_{\qcap}]$, we have $\varsigma^*(p) \geq \varsigma(p)$; we omit the details. To complete the proof that $\varsigma^*$ dominates $\varsigma$, suppose there is $p \in \mathbb{P}$ such that $p > \kappa_{\qcap}$. Then $\qcap \in \llbracket 1, n-1 \rrbracket$, so it follows from $[\mathcal{F}_3]$ that
\begin{align*}
\piavg(\mathsf{S}(\qcap)) - \varsigma(\mathsf{S}(\qcap)) &\geq \big[ \tfrac{\qcap}{n} - 1 \big] \cdot \varsigma(\mathsf{S}(\qcap)) + \big[ 1 - \tfrac{\qcap}{n} \big] \cdot \varsigma(p)
\\ &= \big[ 1 - \tfrac{\qcap}{n} \big] \cdot [\varsigma(p) - \varsigma(\mathsf{S}(\qcap))].
\end{align*}
By construction, $\varsigma(\mathsf{S}(\qcap)) + (\kappa_{\qcap} - \mathsf{S}(\qcap)) = \piavg(\kappa_{\qcap})$, and by definition of $\piavg$ we have $\piavg(\kappa_{\qcap}) = \piavg(\mathsf{S}(\qcap)) + \tfrac{\qcap}{n} [\kappa_{\qcap} - \mathsf{S}(\qcap)]$, so
\begin{align*}
\big[ 1 - \tfrac{\qcap}{n} \big] \cdot [\kappa_{\qcap} - \mathsf{S}(\qcap)] = \piavg(\mathsf{S}(\qcap)) - \varsigma(\mathsf{S}(\qcap)).
\end{align*}
Altogether, then, we have
\begin{align*}
\big[ 1 - \tfrac{\qcap}{n} \big] \cdot [\varsigma^*(p) - \varsigma^*(\mathsf{S}(\qcap))]  &= \big[ 1 - \tfrac{\qcap}{n} \big] \cdot [\varsigma^*(\kappa_{\qcap}) - \varsigma^*(\mathsf{S}(\qcap))]  \\ &= \big[ 1 - \tfrac{\qcap}{n} \big] \cdot [\kappa_{\qcap} - \mathsf{S}(\qcap)] 
\\ &= \piavg(\mathsf{S}(\qcap)) - \varsigma(\mathsf{S}(\qcap))
\\ &\geq \big[ 1 - \tfrac{\qcap}{n} \big] \cdot [\varsigma(p) - \varsigma(\mathsf{S}(\qcap))]
\\ &= \big[ 1 - \tfrac{\qcap}{n} \big] \cdot [\varsigma(p) - \varsigma^*(\mathsf{S}(\qcap))],
\end{align*}
so $\varsigma^*(p) \geq \varsigma(p)$, as desired.

To conclude, we claim that $(\varsigma^*, \kappa)$ satisfies $[\mathcal{V}_3]$. First, for each $q \in \llbracket 1, \qcap-1 \rrbracket$,
\begin{align*}
\varsigma^*(\kappa_q) &= [\tfrac{q}{n}] \cdot \varsigma^*(\kappa_q) + [1 - \tfrac{q}{n}] \cdot \varsigma^*(\kappa_q)
\\ &= [\tfrac{q}{n}] \cdot [\varsigma^*(\mathsf{S}(q)) + \kappa_q - \mathsf{S}(q)] + [1 - \tfrac{q}{n}] \cdot \varsigma^*(\mathsf{S}(q+1))
\\ &= [\tfrac{q}{n}] \cdot [\varsigma(\mathsf{S}(q)) + \kappa_q - \mathsf{S}(q)] + [1 - \tfrac{q}{n}] \cdot \varsigma(\mathsf{S}(q+1))
\\ &= [\tfrac{q}{n}] \cdot \varsigma(\mathsf{S}(q)) + [1 - \tfrac{q}{n}] \cdot \varsigma(\mathsf{S}(q+1)) + [\tfrac{q}{n}] \cdot [\kappa_q - \mathsf{S}(q)]
\\ &\leq \piavg(\mathsf{S}(q)) + [\tfrac{q}{n}] \cdot [\kappa_q - \mathsf{S}(q)]
\\ &= [\tfrac{q}{n} \cdot \mathsf{S}(q) - \tfrac{1}{n} \mathsf{C}(q)] + [\tfrac{q}{n}] \cdot [\kappa_q - \mathsf{S}(q)]
\\ &= [\tfrac{q}{n} \cdot \kappa_q - \tfrac{1}{n} \mathsf{C}(q)]
\\ &= \piavg(\kappa_q).
\end{align*}
Second, if $\qcap = n$, then $\kappa_{\qcap} = \mathsf{S}(\qcap)$, so by the \hyperlink{CeilingLemma}{Ceiling~Lemma},
\begin{align*}
\varsigma^*(\kappa_{\qcap}) &= \varsigma^*(\mathsf{S}(\qcap))
\\ &= \varsigma(\mathsf{S}(\qcap))
\\ & \leq \piavg(\mathsf{S}(\qcap))
\\ &= \piavg(\kappa_{\qcap}).
\end{align*}
Finally, if $\qcap < n$, then by construction $\kappa_{\qcap} \leq \piavg(\kappa_{\qcap}) + \mathsf{S}(\qcap) - \varsigma(\mathsf{S}(\qcap))$, so
\begin{align*}
\varsigma^*(\kappa_{\qcap}) &= \varsigma^*(\mathsf{S}(\qcap)) + \kappa_{\qcap} - \mathsf{S}(\qcap)
\\ &\leq \varsigma(\mathsf{S}(\qcap)) + [\piavg(\kappa_{\qcap}) + \mathsf{S}(\qcap) - \varsigma(\mathsf{S}(\qcap))]  - \mathsf{S}(\qcap)
\\ &= \piavg(\kappa_{\qcap}),
\end{align*}
as desired.

\vspace{\baselineskip} \noindent \textsc{Step 2:} If a subsidy curve is on-off, then it is pre-valvular if and only if it is funded.

\vspace{\baselineskip} \noindent \textsc{[$\Rightarrow$]} Let $\varsigma$ be a pre-valvular subsidy curve and let $\kappa \in \mathbb{P}^{\qcap}$ be the associated vector of cutoff prices. Then $(\varsigma, \kappa)$ satisfies $[\mathcal{V}_1]$, or equivalently $\varsigma$ satisfies $[\mathcal{F}_1]$; thus it remains to show that $\varsigma$ satisfies $[\mathcal{F}_2]$ and $[\mathcal{F}_3]$. In what follows, we freely use $[\mathcal{V}_2]$ and $[\mathcal{V}_3]$.

First, let $q \in \llbracket 1, \qcap - 1 \rrbracket$. Then
\begin{align*}
[\tfrac{q}{n}] \cdot \varsigma(\mathsf{S}(q)) + [1 - \tfrac{q}{n}] \cdot \varsigma(\mathsf{S}(q+1)) &= [\tfrac{q}{n}] \cdot [\varsigma(\kappa_q) - (\kappa_q - \mathsf{S}(q))] + [1 - \tfrac{q}{n}] \cdot \varsigma(\kappa_q)
\\ &= \varsigma(\kappa_q) - [\tfrac{q}{n}] \cdot (\kappa_q - \mathsf{S}(q))
\\ &\leq \piavg(\kappa_q) - [\tfrac{q}{n}] \cdot (\kappa_q - \mathsf{S}(q))
\\ &= [\tfrac{1}{n}] \cdot [ (q \cdot \kappa_q  - \mathsf{C}(q)) - q \cdot (\kappa_q - \mathsf{S}(q)) ]
\\ &= \piavg(\mathsf{S}(q)).
\end{align*}
Since $q \in \llbracket 1, \qcap - 1 \rrbracket$ was arbitrary, thus $\varsigma$ satisfies $[\mathcal{F}_2]$, as desired.

Second, assume that $\qcap \in \llbracket 1, n-1 \rrbracket$ and let $p \in (\mathsf{S}(\qcap), \infty)$. If $p \geq \kappa_{\qcap}$, then
\begin{align*}
\begin{split}
[\tfrac{\qcap}{n}] \cdot \varsigma(\mathsf{S}(\qcap)) + [1 - \tfrac{\qcap}{n}] \cdot \varsigma(p) ={}&[\tfrac{\qcap}{n}] \cdot [\varsigma(\kappa_{\qcap}) - (\kappa_{\qcap} - \mathsf{S}(\qcap))] \\ {}& + [1 - \tfrac{\qcap}{n}] \cdot \varsigma(\kappa_{\qcap})
\end{split}
\\ ={}& \varsigma(\kappa_{\qcap}) - [\tfrac{\qcap}{n}] \cdot (\kappa_{\qcap} - \mathsf{S}(\qcap))
\\ \leq{}& \piavg(\kappa_{\qcap}) - [\tfrac{\qcap}{n}] \cdot (\kappa_{\qcap} - \mathsf{S}(\qcap))
\\ ={}& [\tfrac{1}{n}] \cdot [ (\qcap \cdot \kappa_{\qcap} - \mathsf{C}(\qcap)) - \qcap \cdot (\kappa_{\qcap} - \mathsf{S}(\qcap)) ]
\\ ={}& \piavg(\mathsf{S}(\qcap)).
\end{align*}
If $p < \kappa_{\qcap}$, then by the previous sentence, we have
\begin{align*}
[\tfrac{\qcap}{n}] \cdot \varsigma(\mathsf{S}(\qcap)) + [1 - \tfrac{\qcap}{n}] \cdot \varsigma(p) &\leq [\tfrac{\qcap}{n}] \cdot \varsigma(\mathsf{S}(\qcap)) + [1 - \tfrac{\qcap}{n}] \cdot \varsigma(\kappa_{\qcap})
\\ &\leq \piavg(\mathsf{S}(\qcap)).
\end{align*}
Thus in both cases we have the desired inequality. Since $p \in (\mathsf{S}(\qcap), \infty)$ was arbitrary, thus $\varsigma$ satisfies $[\mathcal{F}_3]$, as desired.

\vspace{\baselineskip} \noindent \textsc{[$\Leftarrow$]} Let $\varsigma$ be an on-off curve that is funded. Then there is $\kappa \in \mathbb{P}^{\qcap}$ such that $(\varsigma, \kappa)$ satisfies $[\mathcal{V}_1]$ and $[\mathcal{V}_2]$, and by the \hyperlink{CeilingLemma}{Ceiling~Lemma} we have that $(\varsigma, \kappa)$ satisfies $[\mathcal{V}_3]$; thus $\varsigma$ is pre-valvular, as desired.

\vspace{\baselineskip} \noindent \textsc{Step 3:} Prove the two implications in sequence.

\vspace{\baselineskip} \noindent \textsc{[$\Rightarrow$]} Let $(\tau, \alpha)$ be a consumer-optimal endogenous shareholding auction and let $\varsigma$ be its funded subsidy curve. Then $\varsigma$ is on-off; else (i)~by Step~1, $\varsigma$ is dominated by a pre-valvular subsidy curve $\varsigma^*$, and (ii)~by Step~2, $\varsigma^*$ is funded, so by the \hyperlink{DominationLemma}{Domination~Lemma}, if $(\tau^*, \alpha)$ is the endogenous subsidy auction that is supported by $\varsigma^*$, then $(\tau^*, \alpha)$ is an endogenous shareholding auction that consumer-dominates $(\tau, \alpha)$, contradicting that $(\tau, \alpha)$ is consumer-optimal. Then $\varsigma$ is on-off and funded, so by Step~2 it is pre-valvular. Let $\kappa \in \mathbb{P}^{\qcap}$ be the associated vector of cutoff prices.

Assume for contradiction that $(\varsigma, \kappa)$ violates $[\mathcal{V}_4]$. Then there is $q \in \llbracket 1, \qcap \rrbracket$ such that (i)~$\kappa_q < \sup \mathbb{P}_q$, (ii)~$\varsigma(\kappa_q) < \piavg(\kappa_q)$, and (iii)~for each $q^* \in \llbracket 1, \qcap \rrbracket$ such that $q^* > q$ and $\varsigma(\kappa_{q^*}) = \piavg(\kappa_{q^*})$, we have $\varsigma(\mathsf{S}(q^*)) > \varsigma(\kappa_q)$. Let $q_0$ be the maximum $q \in \llbracket 1, \qcap \rrbracket$ satisfying these three conditions. We consider two cases. First, if there is no $q^* \in \llbracket 1, \qcap \rrbracket$ such that $q^* > q_0$ and $\varsigma(\kappa_{q^*}) = \piavg(\kappa_{q^*})$, then it is straightforward to verify that $\varsigma$ is dominated by another valvular subsidy curve that has a greater cutoff price in $\mathbb{P}_{q_0}$, so by the \hyperlink{DominationLemma}{Domination~Lemma} $(\tau, \alpha)$ is consumer dominated by another endogenous shareholding auction, contradicting that $(\tau, \alpha)$ is consumer-optimal. Second, if there is $q^* \in \llbracket 1, \qcap \rrbracket$ such that $q^* > q_0$ and $\varsigma(\kappa_{q^*}) = \piavg(\kappa_{q^*})$, then let $q_1$ be the minimum $q^* \in \llbracket q_0+1, \qcap \rrbracket$ satisfying these conditions. Then (i)~$\varsigma(\mathsf{S}(q_1)) > \varsigma(\kappa_{q_0})$, and (ii)~by definition of $q_1$ and the \hyperlink{CeilingLemma}{Ceiling~Lemma}, for each $q' \in \llbracket q_0+1, q_1-1 \rrbracket$, $\varsigma(\kappa_{q'}) < \piavg(\kappa_{q'})$, so it is straightforward to verify that $\varsigma$ is dominated by another valvular subsidy curve that has a greater cutoff in $\mathbb{P}_{q_0}$, so by the \hyperlink{DominationLemma}{Domination~Lemma} $(\tau, \alpha)$ is consumer dominated by another endogenous shareholding auction, contradicting that $(\tau, \alpha)$ is consumer-optimal. Thus $(\varsigma, \kappa)$ satisfies $[\mathcal{V}_4]$, so $\varsigma$ is valvular, so $(\tau, \alpha)$ is a valvular auction, as desired.

\vspace{\baselineskip} \noindent \textsc{[$\Leftarrow$]} Let $(\tau, \alpha)$ be an endogenous subsidy auction, let $\varsigma$ be its subsidy curve, assume that $\varsigma$ is valvular, and let $\kappa \in \mathbb{P}^{\qcap}$ be the associated vector of cutoff prices. By Step~2, $\varsigma$ is funded, so $(\tau, \alpha)$ is an endogenous shareholding auction. If $\mathsf{S}(1) = \infty$, then we are done; thus let us assume $\mathsf{S}(1) < \infty$.

Assume, by way of contradiction, that $(\tau, \alpha)$ is not consumer-optimal. By the \hyperlink{DominationLemma}{Domination~Lemma}, $\varsigma$ is dominated by a funded subsidy curve $\varsigma''$. Moreover, (i)~if $\varsigma''$ is on-off, then by Step~2 it is pre-valvular, and (ii)~if $\varsigma''$ is not on-off, then by Step~1, $\varsigma''$ is dominated by a pre-valvular subsidy curve; thus in both cases $\varsigma$ is dominated by a pre-valvular subsidy curve $\varsigma'$ with associated vector of cutoff prices $\kappa' \in \mathbb{P}^{\qcap}$, and by Step~2, $\varsigma'$ is funded. Since $\varsigma'$ dominates $\varsigma$, thus there is $q \in \llbracket 1, \qcap \rrbracket$ such that $\kappa'_q > \kappa_q$. Then $\kappa_q < \sup \mathbb{P}_q$ and $\varsigma(\kappa_q) < \piavg(\kappa_q)$, so since $(\varsigma, \kappa)$ satisfies $[\mathcal{V}_4]$, thus there is $q^* \in \llbracket 1, \qcap \rrbracket$ such that (i)~$q^* > q$, (ii)~$\varsigma(\kappa_{q^*}) = \piavg(\kappa_{q^*})$, and (iii)~$\varsigma(\mathsf{S}(q^*)) = \varsigma(\kappa_q)$. Then by the \hyperlink{CeilingLemma}{Ceiling~Lemma}, $\piavg(\mathsf{S}(q^*)) \geq \piavg(\kappa_q) > \varsigma(\kappa_q) = \varsigma(\mathsf{S}(q^*))$, so since $\piavg(\kappa_{q^*}) = \varsigma(\kappa_{q^*})$ we have $\kappa_{q^*} > \mathsf{S}(q^*)$. Moreover, since $\varsigma'$ dominates $\varsigma$, thus $\piavg(\kappa_{q^*}) \geq \varsigma'(\kappa_{q^*}) \geq \varsigma(\kappa_{q^*}) = \piavg(\kappa_{q^*})$. Altogether, then, since $\varsigma'$ dominates $\varsigma$, thus by the \hyperlink{CeilingLemma}{Ceiling~Lemma}, both $\varsigma'$ and $\varsigma$ must increase with slope $1$ on the non-degenerate interval $[\mathsf{S}(q^*), \kappa_{q^*}]$ to reach the subsidy $\piavg(\kappa_{q^*})$. But then $\varsigma'(\mathsf{S}(q^*)) = \varsigma(\mathsf{S}(q^*)) = \varsigma(\kappa_q) < \varsigma'(\kappa'_q)$, contradicting that $\varsigma'$ is non-decreasing.~$\blacksquare$

\hypertarget{AppendixE}{}
\setcounter{secnumdepth}{0}
\section{Appendix E: Proofs for Section 5.4}

In this appendix, we prove the \hyperlink{CompactnessLemma}{Compactness~Lemma}, the \hyperlink{SummaryLemma}{Summary~Lemma}, and \hyperlink{Theorem6}{Theorem~6}. We begin with the \hyperlink{CompactnessLemma}{Compactness~Lemma}. Note that \hyperlink{Section6}{Section~6} contains a proof sketch for \hyperlink{Theorem6}{Theorem~6}.

\vspace{\baselineskip} \noindent \textsc{Proof of Compactness Lemma:} If $\mathsf{S}(1) = \infty$ then the result is trivial; thus let us assume $\mathsf{S}(1) \neq \infty$, so $\mathbb{P} \subseteq \mathbb{R}$. Let $d$ denote the usual Euclidean metric on $\mathbb{R}$. Throughout this proof, it should be understood that (i)~$\mathbb{P}$ is endowed with its Euclidean topology, (ii)~$\mathbb{R}$ is endowed with its Euclidean topology and the metric $d$, and (iii)~each subset of the function space $\mathbb{R}^\mathbb{P}$ is endowed with its topology of compact convergence; we suppress these details in our notation. The proof consists of four steps.

\vspace{\baselineskip} \noindent \textsc{Step 1:} Both $\mathbb{R}^\mathbb{P}$ and $\mathbb{S}$ are metrizable.

\vspace{\baselineskip} By Exercise 10 in $\S 46$ of \cite{Munkres2000}, (i)~since $\mathbb{P}$ is locally compact and second-countable, we have that it is $\sigma$-compact, and thus (ii)~since $\mathbb{P}$ is $\sigma$-compact and $\mathbb{R}$ is a metric space, we have that $\mathbb{R}^\mathbb{P}$ is metrizable. Then $\mathbb{S}$ is a subspace of a metrizable space, so $\mathbb{S}$ is metrizable, as desired.

\vspace{\baselineskip} \noindent \textsc{Step 2:} $\mathbb{S}$ is contained in a compact subspace of $\mathscr{C}(\mathbb{P}, \mathbb{R})$.

\vspace{\baselineskip} First, we claim that $\mathbb{S}$ is equicontinuous under $d$. Indeed, this is a direct consequence of the fact that each subsidy curve has Lipschitz constant $1$. Formally, we say {\it $\mathbb{S}$ is equicontinuous under $d$} if for each $p \in \mathbb{P}$ and each $\varepsilon > 0$, there is a neighborhood $\mathbb{P}'$ of $p$ such that for each $p' \in \mathbb{P}'$ and each $\varsigma \in \mathbb{S}$, $d(\varsigma(p), \varsigma(p')) < \varepsilon$. This is indeed the case: for each $p \in \mathbb{P}$ and each $\varepsilon > 0$, simply take $\mathbb{P}'$ to be $(p - \frac{\varepsilon}{2}, p + \frac{\varepsilon}{2}) \cap \mathbb{P}$.

Second, we claim that for each $p \in \mathbb{P}$, $\{\varsigma(p)|\varsigma \in \mathbb{S}\}$ has compact closure. Indeed, for each $p \in \mathbb{P}$, by the \hyperlink{CeilingLemma}{Ceiling~Lemma} we have $\varsigma \in \mathbb{S}$ implies $\varsigma(p) \in [0, \piavg(p)]$, so the closure of $\{\varsigma(p)|\varsigma \in \mathbb{S}\}$ is closed and bounded; thus by the Heine-Borel theorem, the closure of $\{\varsigma(p)|\varsigma \in \mathbb{S}\}$ is compact, as desired.

To conclude, we claim $\mathbb{S}$ is contained in a compact subspace of $\mathscr{C}(\mathbb{P}, \mathbb{R})$. Indeed, since (i)~$\mathbb{P}$ is a topological space, (ii)~$\mathbb{R}$ is a metric space with metric $d$, (iii)~$\mathbb{S} \subseteq \mathscr{C}(\mathbb{P}, \mathbb{R})$, (iv)~$\mathbb{S}$ is equicontinuous under $d$, and (v)~for each $p \in \mathbb{P}$, $\{\varsigma(p) | \varsigma \in \mathbb{S}\}$ has compact closure, thus by Theorem 47.1 of \cite{Munkres2000},\footnote{Munkres calls this Ascoli's theorem. It is one of several versions of the Arzel\`{a}-Ascoli theorem.} we have the desired conclusion.

\vspace{\baselineskip} \noindent \textsc{Step 3:} $\mathbb{S}$ is closed in $\mathbb{R}^\mathbb{P}$.

\vspace{\baselineskip} Since $\mathbb{R}^\mathbb{P}$ is metrizable, it is well-known that it suffices to show that $\mathbb{S}$ contains the limits of its convergent sequences; for example, this is an easy corollary of Lemma~21.1 in \cite{Munkres2000}. Thus let $(\varsigma_t) \in \mathbb{S}^\mathbb{N}$ be convergent in $\mathbb{R}^\mathbb{P}$. Since $\mathbb{R}^\mathbb{P}$ is metrizable, thus $(\varsigma_t)$ has a unique limit; denote it by $\varsigma_\infty \in \mathbb{R}^\mathbb{P}$. We must show $\varsigma_\infty \in \mathbb{S}$.

By Theorem~46.7 of \cite{Munkres2000}, the topology of compact convergence is finer than the topology of pointwise convergence, so $(\varsigma_t)$ converges to $\varsigma_\infty$ pointwise: for each $p \in \mathbb{P}$, $\lim \varsigma_t(p) = \varsigma_\infty(p)$. From here, it is straightforward to verify that $\varsigma_\infty$ satisfies each requirement for a funded subsidy curve by exploiting basic properties of limits. We demonstrate this with the argument that $\varsigma_\infty$ is a subsidy curve: for each pair $p, p' \in \mathbb{P}$ such that $p' > p$, we have
\begin{align*}
\frac{\varsigma_\infty(p) - \varsigma_\infty(p')}{p-p'} &= \frac{\lim \varsigma_t(p) - \lim \varsigma_t(p')}{p-p'}
\\ &= \lim \Big[ \frac{\varsigma_t(p) - \varsigma_t(p')}{p-p'} \Big]
\\ &\in [0, 1],
\end{align*}
as desired. By similar arguments, $\varsigma_\infty$ satisfies $[\mathcal{F}_1]$, $[\mathcal{F}_2]$, and $[\mathcal{F}_3]$; we omit the details.

\vspace{\baselineskip} \noindent \textsc{Step 4:} Conclude. 

\vspace{\baselineskip} By Step~1, $\mathbb{S}$ is metrizable. By Step~2 and Step~3, $\mathbb{S}$ is a closed subspace of a compact space, so by Theorem~26.2 of \cite{Munkres2000}, it is compact.~$\blacksquare$

\vspace{\baselineskip} Second, we prove the \hyperlink{SummaryLemma}{Summary~Lemma}.

\vspace{\baselineskip} \noindent \textsc{Proof of Summary Lemma:} Assume $\mathsf{S}(1) \neq \infty$. Throughout this proof, unless explicitly noted otherwise, it should be understood that (i)~$\mathbb{S}$ is endowed with its topology of compact convergence, (ii)~$V$, $\mathbb{U}$, and $\mathbb{P}$ are endowed with their Euclidean topologies, (iii)~$\mathbb{R}$ is endowed with its Euclidean topology and its Euclidean metric, and (iv)~any product of these sets is endowed with its associated product topology. We refer to these as the {\it implicit topologies}. In Step~3, we briefly consider explicit alternative topologies for $\mathbb{S}$ and $\mathbb{S} \times \mathbb{P}$.

For each $i \in N$, define $\mathbb{W}_i \equiv \{W \in \mathbb{W} | i \in W\}$ and $\mathbb{W}_{-i} \equiv \{W \in \mathbb{W} | i \not \in W\}$. The continuity argument involves the following functions:
\begin{itemize}
\item For each $\mathbb{W}' \subseteq \mathbb{W}$, define the associated optimal total surplus function, $\mathsf{OTS}_{\mathbb{W}'}: V \to \mathbb{R}$, to be the function $v \mapsto \max_{W \in \mathbb{W}'} \mathsf{TS}_v(W)$.

\item For each $i \in N$, define the associated Groves price function, $\pgro_i: V \to \mathbb{R}$, to be the function $v \mapsto \pgro_{v_{-i}}$.

\item Define the evaluation map, $\mathsf{ev}: \mathbb{S} \times \mathbb{P} \to \mathbb{R}$, to be the function $(\varsigma, p) \mapsto \varsigma(p)$.
\end{itemize}
The proof consists of two steps.

\vspace{\baselineskip} \noindent \textsc{Step 1:} For each $\varsigma \in \mathbb{S}$, each $(\tau, \alpha) \in \mathbb{A}$ that is supported by $\varsigma$, and each $v \in V$, $\mathcal{U}^{\mathbb{A} \times V}((\tau, \alpha), v) = \mathcal{U}^{\mathbb{S} \times V}(\varsigma, v)$.

\vspace{\baselineskip} Let $\varsigma$, $(\tau, \alpha)$, and $v$ satisfy the hypotheses. By \hyperlink{Theorem1}{Theorem~1}, for each $i \in N$ we have $(\tau_i(v), \alpha_i(v)) \in B^\delta_i(\varsigma(\pgro_{v_{-i}}), \pgro_{v_{-i}} | v_i)$ and thus
\begin{align*}
[\mathcal{U}^{\mathbb{A} \times V}((\tau, \alpha), v)]_i &= v_i \cdot \alpha_i(v) + \tau_i(v)
\\ &= \max\{v_i - \pgro_{v_{-i}}, 0\} + \varsigma(\pgro_{v_{-i}})
\\ &= [\mathcal{U}^{\mathbb{S} \times V}(\varsigma, v)]_i.
\end{align*}
Moreover, by {\it efficiency} of $(\tau, \alpha)$ and by part of the above equality,
\begin{align*}
[\mathcal{U}^{\mathbb{A} \times V}((\tau, \alpha), v)]_{i_0} &= [-\sum_{i \in N} \tau_i(v)] - \mathsf{C}(|\alpha(v)|)
\\ &= [\sum_{i \in N} v_i \cdot \alpha_i(v) - \mathsf{C}(|\alpha(v)|)] - \sum_{i \in N} [v_i \cdot \alpha_i(v) + \tau_i(v)] 
\\ &= \max_{W \subseteq N} [\sum_{i \in W} v_i - \mathsf{C}(|W|)] - \sum_{i \in N} [ \max\{v_i - \pgro_{v_{-i}}, 0\} + \varsigma(\pgro_{v_{-i}}) ]
\\ &= [\mathcal{U}^{\mathbb{S} \times V}(\varsigma, v)]_{i_0}.
\end{align*}
Altogether, then, $\mathcal{U}^{\mathbb{A} \times V}((\tau, \alpha), v) = \mathcal{U}^{\mathbb{S} \times V}(\varsigma, v)$, as desired.

\vspace{\baselineskip} \noindent \textsc{Step 2:} $\mathcal{U}^{\mathbb{S} \times V}$ is continuous.

\vspace{\baselineskip} First, we claim that for each $\mathbb{W}' \subseteq \mathbb{W}$, $\mathsf{OTS}_{\mathbb{W'}}$ is continuous. Indeed, let $\mathbb{W}' \subseteq \mathbb{W}$. For each $W \in \mathbb{W}'$, the function $v \to \mathsf{TS}_v(W) = \sum_{i \in W} v_i - \mathsf{C}(|W|)$ is continuous; thus $\mathsf{OTS}_{\mathbb{W}'}$ is continuous as a pointwise maximum (or ``upper envelope") of a finite family of continuous functions.

Second, we claim that for each $i \in N$, $\pgro_i$ is continuous. Indeed, let $i \in N$. Since $\mathsf{S}(1) \neq \infty$, thus by the \hyperlink{PriceLemma}{Price~Lemma}, $\pgro_i$ is the function $v \to v_i - [\mathsf{OTS}_{\mathbb{W}_i}(v) - \mathsf{OTS}_{\mathbb{W}_{-i}}(v)]$, so by the previous paragraph $\pgro_i$ is continuous.

Third, we claim that $\mathsf{ev}$ is continuous when $\mathbb{S} \times \mathbb{P}$ and $\mathbb{R}$ are endowed with their implicit topologies. Indeed, define the extended evaluation map $\mathsf{ev}^*: \mathscr{C}(\mathbb{P}, \mathbb{R}) \times \mathbb{P} \to \mathbb{R}$ to be the function $(f, p) \mapsto f(p)$. For $\mathscr{C}(\mathbb{P}, \mathbb{R}) \times \mathbb{P}$, let $\mathcal{T}^\mathsf{co}$ denote the product topology when $\mathscr{C}(\mathbb{P}, \mathbb{R})$ is endowed with the compact-open topology, and let $\mathcal{T}^\mathsf{cc}$ denote the product topology when $\mathscr{C}(\mathbb{P}, \mathbb{R})$ is endowed with the topology of compact convergence. First, since $\mathbb{P}$ is locally compact and Hausdorff, thus by Theorem~46.10 of \cite{Munkres2000}, $\mathsf{ev}^*$ is continuous when the domain and its topology are $(\mathscr{C}(\mathbb{P}, \mathbb{R}) \times \mathbb{P}, \mathcal{T}^\mathsf{co})$. Second, since $\mathbb{P}$ is a topological space and $\mathbb{R}$ is a metric space, thus by Theorem~46.8 of \cite{Munkres2000}, $(\mathscr{C}(\mathbb{P}, \mathbb{R}) \times \mathbb{P}, \mathcal{T}^\mathsf{co}) = (\mathscr{C}(\mathbb{P}, \mathbb{R}) \times \mathbb{P}, \mathcal{T}^\mathsf{cc})$. Third, since by definition $\mathbb{S}$ with its implicit topology is a subspace of $\mathscr{C}(\mathbb{P}, \mathbb{R})$ with its topology of compact convergence, thus by Theorem~16.3 of \cite{Munkres2000}, $\mathbb{S} \times \mathbb{P}$ with its implicit topology is a subspace of $(\mathscr{C}(\mathbb{P}, \mathbb{R}) \times \mathbb{P}, \mathcal{T}^\mathsf{cc})$. Altogether, then, since $\mathsf{ev}$ is the restriction of $\mathsf{ev}^*$ to $\mathbb{S} \times \mathbb{P}$, and since the restriction of a continuous function to a subset is continuous when the subset is endowed with the subspace topology, thus $\mathsf{ev}$ is continuous when $\mathbb{S} \times \mathbb{P}$ and $\mathbb{R}$ are endowed with their implicit topologies, as desired.

Fourth, we claim that for each $i \in N$, the component function $\mathbb{S} \times V \to \mathbb{R}$, $(\varsigma, v) \mapsto [\mathcal{U}^{\mathbb{S} \times V}(\varsigma, v)]_i$ is continuous. Indeed, let $i \in N$. Then
\begin{align*}
[\mathcal{U}^{\mathbb{S} \times V}(\varsigma, v)]_i &= \max\{v_i - \pgro_{v_{-i}}, 0\} + \varsigma(\pgro_{v_{-i}})
\\ &= \max\{v_i - \pgro_i(v), 0\} + \mathsf{ev}(\varsigma, \pgro_i(v)).
\end{align*}
By the second claim and third claim, $\pgro_i$ and $\mathsf{ev}$ are continuous. It then follows from well-known properties of continuous functions\footnote{Most notably, (i)~since $\pgro_i: V \to \mathbb{R}$ is continuous, thus $\mathbb{S} \times V \to \mathbb{S} \times \mathbb{P}$, $(\varsigma, v) \mapsto (\varsigma, \pgro_i(v))$ is continuous, and (ii)~the projection $\mathbb{S} \times V \to V$, $(\varsigma, v) \mapsto v$ is continuous.} that the given component function is continuous, as desired.

Fifth, we claim that the component function $\mathbb{S} \times V \to \mathbb{R}$, $(\varsigma, v) \mapsto [\mathcal{U}^{\mathbb{S} \times V}(\varsigma, v)]_{i_0}$ is continuous. Indeed,
\begin{align*}
[\mathcal{U}^{\mathbb{S} \times V}(\varsigma, v)]_{i_0} &= \max_{W \subseteq N} [\sum_{i \in W} v_i - C(|W|)] - \sum_{i \in N} [ \max\{v_i - \pgro_{v_{-i}}, 0\} + \varsigma(\pgro_{v_{-i}}) ]
\\ &= \mathsf{OTS}_\mathbb{W}(v) - \sum_{i \in N} [\max\{v_i - \pgro_i(v), 0\} + \mathsf{ev}(\varsigma, \pgro_i(v))].
\end{align*}
By the first three claims, each function in $\{\mathsf{OTS}_\mathbb{W}\} \cup \{\pgro_i\}_{i \in N} \cup \{\mathsf{ev}\}$ is continuous. It then follows from well-known properties of continuous functions that the given component function is continuous, as desired.

To conclude, by the fourth and fifth claim, each component function for $\mathcal{U}^{\mathbb{S} \times V}$ is continuous, so $\mathcal{U}^{\mathbb{S} \times V}$ is continuous, as desired.~$\blacksquare$

\vspace{\baselineskip} To conclude this appendix, we prove \hyperlink{Theorem6}{Theorem~6}.

\vspace{\baselineskip} \noindent \textsc{Proof of Theorem 6:} Let $\mu$ and $\mathcal{W}$ satisfy the hypotheses. Then there is $V^\mathsf{c}$ such that $\mu(V \backslash V^\mathsf{c}) = 0$. Throughout this proof, it should be understood that (i)~$\mathbb{S}$ is endowed with its topology of compact convergence, (ii)~$V$ and $\mathbb{U}$ are endowed with their Euclidean topologies, (iii)~any product of these sets is endowed with its associated product topology, and (iv)~any subset of these sets is endowed with its subspace topology. If $\mathsf{S}(1) = \infty$ then the result is trivial; thus let us assume $\mathsf{S}(1) \neq \infty$.

For each $\varsigma \in \mathbb{S}$, define the {\it profile summary given $\varsigma$}, $\mathcal{U}^{V|\varsigma}: V \to \mathbb{U}$, to be the function $v \mapsto \mathcal{U}^{\mathbb{S} \times V}(\varsigma, v)$. Define the functional $\mathscr{W}: \mathbb{S} \to \mathbb{R} \cup \{-\infty, \infty\}$ using the Lebesgue integral as follows: for each $\varsigma \in \mathbb{S}$,
\begin{align*}
\mathscr{W}(\varsigma) \equiv \int_{v \in V} \mathcal{W} \circ \mathcal{U}^{V|\varsigma}(v) d\mu.
\end{align*}
The proof consists of five steps.

\vspace{\baselineskip} \noindent \textsc{Step 1:} For each $\varsigma \in \mathbb{S}$, the function $\mathcal{W} \circ \mathcal{U}^{V|\varsigma}: V \to \mathbb{R}$ is continuous. Moreover, for each $\varsigma \in \mathbb{S}$, $\mathscr{W}(\varsigma)$ is a well-defined finite value.

\vspace{\baselineskip} Let $\varsigma \in \mathbb{S}$. We first claim that $\mathcal{W} \circ \mathcal{U}^{V|\varsigma}$ is sequentially continuous. Indeed, let $(v_t) \in (V)^\mathbb{N}$ be convergent. Since $V$ is metrizable, thus $(v_t)$ has a unique limit $v_\infty \in V$. By the \hyperlink{SummaryLemma}{Summary~Lemma} and hypothesis, both $\mathcal{U}^{\mathbb{S} \times V}$ and $\mathcal{W}$ are continuous; thus
\begin{align*}
\mathcal{W} \circ \mathcal{U}^{V|\varsigma}(v_\infty) &= \mathcal{W} \circ \mathcal{U}^{V|\varsigma}(\lim v_t)
\\ &= \mathcal{W} \circ \mathcal{U}^{\mathbb{S} \times V}(\varsigma, \lim v_t)
\\ &= \mathcal{W} \circ \lim \mathcal{U}^{\mathbb{S} \times V}(\varsigma, v_t)
\\ &= \lim \mathcal{W} \circ \mathcal{U}^{\mathbb{S} \times V}(\varsigma, v_t)
\\ &= \lim \mathcal{W} \circ \mathcal{U}^{V|\varsigma}(v_t),
\end{align*}
as desired. Since $V$ is metrizable and $\mathcal{W} \circ \mathcal{U}^{V|\varsigma}: V \to \mathbb{U}$ is sequentially continuous, thus by Theorem~21.3 of \cite{Munkres2000}, $\mathcal{W} \circ \mathcal{U}^{V|\varsigma}$ is continuous, as desired.

To conclude, let $\varsigma \in \mathbb{S}$, let $\mathcal{B}(V^\mathsf{c})$ denote the Borel $\sigma$-algebra of $V^\mathsf{c}$, and let $\mu^\mathsf{c}$ denote the restriction of $\mu$ to $V^\mathsf{c}$. Since (i)~$V^\mathsf{c}$ is compact, (ii)~$\mathcal{B}(V^\mathsf{c})$ contains the topology on $V^\mathsf{c}$, (iii)~$\mathcal{W} \circ \mathcal{U}^{V|\varsigma}$ is continuous (as is its restriction to $V^\mathsf{c}$), (iv)~$(V^\mathsf{c}, \mathcal{B}(V^\mathsf{c}), \mu^\mathsf{c})$ is a finite measure space, and (v)~$\mathscr{W}(\varsigma) = \int_{v \in V} \mathcal{W} \circ \mathcal{U}^{V|\varsigma}(v) d\mu = \int_{v \in V^\mathsf{c}} \mathcal{W} \circ \mathcal{U}^{V|\varsigma}(v) d\mu^\mathsf{c}$, thus by Corollary~16 in Chapter~18.3 of \citep{Royden-Fitzpatrick2010}, $\mathscr{W}(\varsigma)$ is a well-defined finite value.

\vspace{\baselineskip} \noindent \textsc{Step 2:} The family $\{\mathcal{W} \circ \mathcal{U}^{V|\varsigma} | \varsigma \in \mathbb{S} \}$ is uniformly pointwise bounded on $V^\mathsf{c}$.

\vspace{\baselineskip} By the \hyperlink{CompactnessLemma}{Compactness~Lemma} and construction, both $\mathbb{S}$ and $V^\mathsf{c}$ are compact, so by Tychonoff's theorem, $\mathbb{S} \times V^\mathsf{c}$ is compact. Then by the \hyperlink{SummaryLemma}{Summary~Lemma} and hypothesis, both $\mathcal{U}^{\mathbb{S} \times V}$ and $\mathcal{W}$ are continuous, so the image $\mathcal{W} \circ \mathcal{U}^{\mathbb{S} \times V}(\mathbb{S} \times V^\mathsf{c})$ is compact, so by the Heine-Borel theorem we have that $\mathcal{W} \circ \mathcal{U}^{\mathbb{S} \times V}(\mathbb{S} \times V^\mathsf{c})$ is bounded. Thus there is $M \geq 0$ such that for each $\varsigma \in \mathbb{S}$ and each $v \in V^\mathsf{c}$, $|\mathcal{W} \circ \mathcal{U}^{V|\varsigma}(v)| = |\mathcal{W} \circ \mathcal{U}^{\mathbb{S} \times V}(\varsigma, v)| \leq M$, so the family $\{\mathcal{W} \circ \mathcal{U}^{V|\varsigma} | \varsigma \in \mathbb{S} \}$ is uniformly pointwise bounded on $V^\mathsf{c}$, as desired.

\vspace{\baselineskip} \noindent \textsc{Step 3:} If $(\varsigma_t) \in \mathbb{S}^\mathbb{N}$ is convergent, then it has a unique limit $\varsigma_\infty \in \mathbb{S}$, and $(\mathcal{W} \circ \mathcal{U}^{V|\varsigma_t})$ converges to $\mathcal{W} \circ \mathcal{U}^{V|\varsigma_\infty}$ pointwise.

\vspace{\baselineskip} Let $(\varsigma_t) \in \mathbb{S}^\mathbb{N}$ be convergent. By the \hyperlink{CompactnessLemma}{Compactness~Lemma}, $\mathbb{S}$ is metrizable, so $(\varsigma_t)$ has a unique limit $\varsigma_\infty \in \mathbb{S}$. By the \hyperlink{SummaryLemma}{Summary~Lemma} and hypothesis, both $\mathcal{U}^{\mathbb{S} \times V}$ and $\mathcal{W}$ are continuous; thus for each $v \in V$,
\begin{align*}
\mathcal{W} \circ \mathcal{U}^{V|\varsigma_\infty}(v) &= \mathcal{W} \circ \mathcal{U}^{V|\lim \varsigma_t}(v)
\\ &= \mathcal{W} \circ \mathcal{U}^{\mathbb{S} \times V}(\lim \varsigma_t, v)
\\ &= \mathcal{W} \circ \lim \mathcal{U}^{\mathbb{S} \times V}(\varsigma_t, v)
\\ &= \lim \mathcal{W} \circ \mathcal{U}^{\mathbb{S} \times V}(\varsigma_t, v)
\\ &= \lim \mathcal{W} \circ \mathcal{U}^{V|\varsigma_t}(v),
\end{align*}
so $(\mathcal{W} \circ \mathcal{U}^{V|\varsigma_t})$ converges to $\mathcal{W} \circ \mathcal{U}^{V|\varsigma_\infty}$ pointwise, as desired.

\vspace{\baselineskip} \noindent \textsc{Step 4:} $\mathscr{W}$ is sequentially continuous.

\vspace{\baselineskip} Let $(\varsigma_t) \in \mathbb{S}^\mathbb{N}$ be convergent. By the \hyperlink{CompactnessLemma}{Compactness~Lemma}, $\mathbb{S}$ is metrizable, so $(\varsigma_t)$ has a unique limit $\varsigma_\infty \in \mathbb{S}$. By construction, $\mu(V \backslash V^\mathsf{c}) = 0$. By the \hyperlink{SummaryLemma}{Summary~Lemma} and hypothesis, both $\mathcal{U}^{\mathbb{S} \times V}$ and $\mathcal{W}$ are continuous. Moreover, (i)~by Step~1, $(\mathcal{W} \circ \mathcal{U}^{V|\varsigma_t})$ is a sequence of continuous functions and thus a sequence of measurable functions, (ii)~$\mu(V^\mathsf{c})$ is finite, (iii)~by Step~2, $\{\mathcal{W} \circ \mathcal{U}^{V|\varsigma_t} | t \in \mathbb{N} \}$ is uniformly pointwise bounded on $V^\mathsf{c}$, and (iv)~by Step~3, $(\mathcal{W} \circ \mathcal{U}^{V|\varsigma_t})$ converges to $\mathcal{W} \circ \mathcal{U}^{V|\varsigma_\infty}$ pointwise; thus by the bounded convergence theorem in Chapter 4.2 of \cite{Royden-Fitzpatrick2010},\footnote{This is an implication of Lebesgue's dominated convergence theorem.} we have that $\lim \int_{v \in V^\mathsf{c}} \mathcal{W} \circ \mathcal{U}^{V|\varsigma_t} = \int_{v \in V^\mathsf{c}} \mathcal{W} \circ \mathcal{U}^{V|\varsigma_\infty}$. Altogether, then,
\begin{align*}
\mathscr{W}(\varsigma_\infty) &= \int_{v \in V} \mathcal{W} \circ \mathcal{U}^{V|\varsigma_\infty}(v) d\mu
\\ &= \int_{v \in V^\mathsf{c}} \mathcal{W} \circ \mathcal{U}^{V|\lim \varsigma_t}(v) d\mu
\\ &= \int_{v \in V^\mathsf{c}} \mathcal{W} \circ \mathcal{U}^{\mathbb{S} \times V}(\lim \varsigma_t, v) d\mu
\\ &= \int_{v \in V^\mathsf{c}} \mathcal{W} \circ \lim \mathcal{U}^{\mathbb{S} \times V}(\varsigma_t, v) d\mu
\\ &= \int_{v \in V^\mathsf{c}} \lim \mathcal{W} \circ \mathcal{U}^{\mathbb{S} \times V}(\varsigma_t, v) d\mu
\\ &= \int_{v \in V^\mathsf{c}} \lim \mathcal{W} \circ \mathcal{U}^{V|\varsigma_t}(v) d\mu
\\ &= \int_{v \in V^\mathsf{c}} \mathcal{W} \circ \mathcal{U}^{V|\varsigma_\infty}(v) d\mu
\\ &= \lim \int_{v \in V^\mathsf{c}} \mathcal{W} \circ \mathcal{U}^{V|\varsigma_t}(v) d\mu
\\ &= \lim \mathscr{W}(\varsigma_t),
\end{align*}
as desired.

\vspace{\baselineskip} \noindent \textsc{Step 5:} Conclude.

\vspace{\baselineskip} By the \hyperlink{CompactnessLemma}{Compactness~Lemma} and Step~4, $\mathbb{S}$ is metrizable and $\mathscr{W}: \mathbb{S} \to \mathbb{R} \cup \{-\infty, \infty\}$ is sequentially continuous, so by Theorem~21.3 of \cite{Munkres2000}, $\mathscr{W}$ is continuous. Moreover, by the \hyperlink{CompactnessLemma}{Compactness~Lemma}, $\mathbb{S}$ is compact. Altogether, then, $\mathscr{W}$ is a continuous function with a compact domain, so it attains a maximum: there is $\varsigma^* \in \mathbb{S}$ such that for each $\varsigma \in \mathbb{S}$, $\mathscr{W}(\varsigma^*) \geq \mathscr{W}(\varsigma)$.

Let $(\tau^*, \alpha^*) \in \mathbb{A}$ be an endogenous shareholding auction supported by $\varsigma^*$, let $(\tau, \alpha) \in \mathbb{A}$ be any endogenous shareholding auction, and let $\varsigma \in \mathbb{S}$ be the funded subsidy curve that supports $(\tau, \alpha)$. By the \hyperlink{SummaryLemma}{Summary~Lemma},
\begin{align*}
\mathbb{E}_\mu \mathcal{W}(\tau^*, \alpha^*) &= \int_{v \in V} \mathcal{W} \circ \mathcal{U}^{\mathbb{A} \times V}((\tau^*, \alpha^*), v)d\mu
\\ &= \int_{v \in V} \mathcal{W} \circ \mathcal{U}^{\mathbb{S} \times V}(\varsigma^*, v)d\mu
\\ &= \int_{v \in V} \mathcal{W} \circ \mathcal{U}^{V|\varsigma^*}(v)d\mu
\\ &= \mathscr{W}(\varsigma^*)
\\ &\geq \mathscr{W}(\varsigma)
\\ &= \int_{v \in V} \mathcal{W} \circ \mathcal{U}^{V|\varsigma}(v)d\mu
\\ &= \int_{v \in V} \mathcal{W} \circ \mathcal{U}^{\mathbb{S} \times V}(\varsigma, v)d\mu
\\ &= \int_{v \in V} \mathcal{W} \circ \mathcal{U}^{\mathbb{A} \times V}((\tau, \alpha), v)d\mu
\\ &= \mathbb{E}_\mu \mathcal{W}(\tau, \alpha).
\end{align*}
Since $(\tau, \alpha) \in \mathbb{A}$ was arbitrary, thus (i)~by Step~1, $\mathbb{E}_\mu \mathcal{W}$ maps each endogenous shareholding auction to a well-defined finite value, and (ii)~$(\tau^*, \alpha^*)$ is $(\mu, \mathcal{W})$-optimal, as desired.~$\blacksquare$

\phantomsection

\end{document}